\RequirePackage{fix-cm} 
\documentclass[a4paper, twoside, reqno, 12pt]{amsart}
\usepackage{fixltx2e}   

\usepackage{etex}

\usepackage[latin1]{inputenc}
\usepackage[T1]{fontenc}

\usepackage{esint}
\usepackage{dsfont}
\usepackage{xspace}
\usepackage{amsgen}
\usepackage{amsthm}
\usepackage{amssymb}
\usepackage{amsmath}
\usepackage{wasysym}
\usepackage{upgreek}
\usepackage{amsfonts}
\usepackage{stmaryrd}
\usepackage{nicefrac}
\usepackage{mathtools}

\usepackage[mathcal, mathscr]{euscript}

\usepackage{mathrsfs}
\DeclareMathAlphabet{\mathscrbf}{OMS}{mdugm}{b}{n}

\usepackage[lined, boxed, linesnumbered, english]{algorithm2e}

\usepackage{multicol}
\usepackage{multirow}

\usepackage{a4wide}

\usepackage[dvipsnames, table]{xcolor}
\definecolor{bckg}{RGB}{20.8, 20.8, 20.8}
\definecolor{oneblue}{rgb}{0.0, 0.0, 0.85}
\definecolor{Lightblue}{RGB}{214, 214, 214}
\definecolor{bluepigment}{rgb}{0.2, 0.2, 0.6}
\definecolor{charcoal}{rgb}{0.21, 0.27, 0.31}
\definecolor{denimblue}{rgb}{0.08, 0.38, 0.74}
\definecolor{Lightgray}{rgb}{0.89, 0.89, 0.89}
\definecolor{darkgrey}{rgb}{0.273, 0.281, 0.30}
\definecolor{darkelectricblue}{rgb}{0.33, 0.41, 0.47}

\usepackage[sort&compress, comma, square, numbers]{natbib}

\usepackage{psfrag}
\usepackage{graphicx}
\usepackage{subfigure}
\usepackage{morefloats}
\usepackage{indentfirst}

\usepackage{acronym}
\usepackage{microtype}
\usepackage[labelsep=period,%
            labelfont={bf,sf,color=bluepigment},%
            justification=raggedright]{caption}

\usepackage[perpage, symbol]{footmisc}

\usepackage[usenames, pdf]{pstricks}
\usepackage{epsfig}
\usepackage{pst-grad} 
\usepackage{pst-plot} 

\usepackage{url}
\usepackage[colorlinks,
           urlcolor=oneblue,
           linkcolor=denimblue,
           citecolor=NavyBlue,
           bookmarksopen=false,
           pdfpagemode=UseNone,
           pagebackref]{hyperref}

\usepackage[explicit]{titlesec}

\titleformat{\section}[block]
  {\color{NavyBlue}\Large\sffamily\bfseries}
  {}
  {0.0em}
  {\colorbox{bckg!5}{\strut\parbox{\dimexpr\linewidth-2\fboxsep\relax}{\thesection. #1}}}
  [\vspace*{0.33em}]

\titleformat{name=\section,numberless}[block]
  {\color{NavyBlue}\Large\sffamily\bfseries}
  {}
  {0.0em}
  {\colorbox{bckg!5}{\strut\parbox{\dimexpr\linewidth-2\fboxsep\relax}{#1}}}
  [\vspace*{0.33em}]

\titleformat{\subsection}
  {\color{NavyBlue}\large\sffamily\bfseries}
  {}
  {0.0em}
  {\colorbox{bckg!5}{\parbox{\dimexpr\linewidth-2\fboxsep\relax}{\thesubsection. #1}}}
  [\vspace*{0.33em}]

\titleformat{name=\subsection,numberless}
  {\color{NavyBlue}\Large\sffamily\bfseries}
  {}
  {0em}
  {\colorbox{bckg!5}{\parbox{\dimexpr\linewidth-2\fboxsep\relax}{#1}}}
  [\vspace*{0.33em}]

\titleformat{\subsubsection}
  {\color{bluepigment}\sffamily\normalsize\bfseries}
  {\thesubsubsection}
  {0.5em}
  {#1}
  [\vspace*{0.33em}]

\titleformat{\paragraph}[runin]
  {\color{bluepigment}\sffamily\small\bfseries}
  {}
  {0em}
  {#1}

\titlespacing{\section}{0.0em}{1.5em plus 2pt minus 2pt}%
{1.0em plus 2pt minus 2pt}[0em]
\titlespacing{\subsection}{0.5em}{1.5em plus 2pt minus 2pt}%
{1.0em}[0em]
\titlespacing{\subsubsection}{0.5em}{1.5em plus 2pt minus 2pt}%
{1.0em plus 2pt minus 2pt}[0em]

\usepackage{titletoc}

\contentsmargin{0.5em}
\newlength{\tocsep} 
\titlecontents{section}[\tocsep]
  {\addvspace{10pt}\bfseries\sffamily}
  {\contentslabel[\thecontentslabel]{\tocsep}}
  {}
  {\ \titlerule*[0.75pc]{.}\ \thecontentspage}
  []
\titlecontents{subsection}[\tocsep]
  {\addvspace{8pt}\sffamily}
  {\contentslabel[\thecontentslabel]{\tocsep}}
  {}
  {\ \titlerule*[0.5pc]{.}\ \thecontentspage}
  []
\titlecontents*{subsubsection}[\tocsep]
  {\addvspace{2pt}\footnotesize\sffamily}
  {}
  {}
  {\ \titlerule*[0.35pc]{.}\ \thecontentspage}
  [\\*]

\makeatletter
\def\@setauthors{%
  \begingroup
  \def\thanks{\protect\thanks@warning}%
  \trivlist
  \centering\footnotesize \@topsep30\p@\relax
  \advance\@topsep by -\baselineskip
  \item\relax
  \author@andify\authors
  \def\\{\protect\linebreak}%
  \textsc{\normalsize\textcolor{darkelectricblue}{\authors}}%
  \ifx\@empty\contribs
  \else
    ,\penalty-3 \space \@setcontribs
    \@closetoccontribs
  \fi
  \endtrivlist
  \endgroup
}
\def\@settitle{\begin{center}%
  \baselineskip14\p@\relax
    \bfseries
    \textsc{\Large\textcolor{charcoal}{\@title}}
  \end{center}%
}
\makeatother

\usepackage{enumitem}
\setlist[description]{%
  topsep=30pt,               
  itemsep=5pt,               
  font={\bfseries\sffamily\color{NavyBlue}}, 
}

\usepackage{fancyhdr}
\usepackage{lastpage}

\newcommand*\Title{\textcolor{bluepigment}{Improved explicit scheme for whole-building}}
\newcommand*\Authors{\textcolor{bluepigment}{S.~Gasparin, J.~Berger, D.~Dutykh \& N.~Mendes}}
\newcommand*{\plogo}{\textcolor{gray}{{\texttt{arXiv.org} / \textsc{hal}}}} 

\numberwithin{equation}{section}

\newcommand{\DF}{\textsc{Dufort}--\textsc{Frankel}}
\newcommand{\Eu}{\textsc{Euler}}
\newcommand{\CN}{\textsc{Crank}--\textsc{Nicolson}}

\newcommand*\egal{\ = \ }
\newcommand*\plus{\ + \ }
\newcommand*\moins{\ - \ }
\newcommand*\egalb{\, =\, }

\renewcommand{\O}{\mathcal{O}}
\newcommand*{\Ox}{\Omega_{\, x}}

\newcommand{\const}{\mathrm{const}}

\newcommand{\caM}{\kappa_{\,M}}
\newcommand{\caTTzero}{\kappa_{\,TT,\,0}}
\newcommand{\caTTun}{\kappa_{\,TT,\,1}}
\newcommand{\caTT}{\kappa_{\,TT}}

\newcommand{\caTTref}{\kappa_{\,TT,\,0}}

\newcommand{\cM}{c_{\,M}}
\newcommand{\cMref}{c_{\,M,\,0}}
\newcommand{\cTM}{c_{\,TM}}
\newcommand{\cTMref}{c_{\,TM,\,0}}
\newcommand{\cTT}{c_{\,TT}}
\newcommand{\cTTref}{c_{\,TT,\,0}}
\newcommand{\cw}{c_{\,w}}

\newcommand{\cz}{c_{\,0}}

\newcommand{\hM}{h_{\,M}}
\newcommand{\hT}{h_{\,T}}
\newcommand{\kl}{k_{\,l}}
\newcommand{\kM}{k_{\,M}}
\newcommand{\kMref}{k_{\,M,\,0}}
\newcommand{\kTM}{k_{\,TM}}
\newcommand{\kTMref}{k_{\,TM,\,0}}
\newcommand{\kTT}{k_{\,TT}}
\newcommand{\kTTref}{k_{\,TT,\,0}}
\newcommand{\Lv}{L_{\,v}}
\newcommand{\Pa}{P_{\,a}}
\newcommand{\Pc}{P_{\,c}}
\newcommand{\Ps}{P_{\,s}}
\newcommand{\Pva}{P_{\,v, \, a}}
\newcommand{\Pv}{P_{\,v}}
\newcommand{\Pvi}{P_{\,v, \,i}}
\newcommand{\Pvref}{P_{\,v}^{\,\circ}}
\newcommand{\Pvs}{P_{\,v, \,s}}
\newcommand{\Pvinf}{P_{\,v, \, \infty}}
\newcommand{\Rv}{R_{\,v}}
\newcommand{\Ta}{T_{\,a}}
\newcommand{\Ts}{T_{\,s}}
\newcommand{\Ti}{T_{\,i}}
\newcommand{\Tinf}{T_{\, \infty}}
\newcommand{\tref}{t_{\,0}}
\newcommand{\wa}{w_{\, a}}

\newcommand{\rhol}{\rho_{\,l}}
\newcommand{\rhow}{\rho_{\,w}}

\newcommand{\rhoz}{\rho_{\,0}}

\newcommand{\BiMi}{\mathrm{Bi}_{\,M,\,i}}
\newcommand{\BiM}{\mathrm{Bi}_{\,M}}
\newcommand{\BiTT}{\mathrm{Bi}_{\,TT}}
\newcommand{\BiTTi}{\mathrm{Bi}_{\,TT,\,i}}
\newcommand{\BiTM}{\mathrm{Bi}_{\,TM}}
\newcommand{\BiTMi}{\mathrm{Bi}_{\,TM,\,i}}
\newcommand{\caTTuns}{\kappa_{\,aTT,\,1}^{\,\star}}

\newcommand{\cMs}{c_{\,M}^{\,\star}}
\newcommand{\cTTs}{c_{\,TT}^{\,\star}}
\newcommand{\cTMs}{c_{\,TM}^{\,\star}}
\newcommand{\dts}{\Delta t^{\,\star}}
\newcommand{\dxs}{\Delta x^{\,\star}}
\newcommand{\FoTT}{\mathrm{Fo}_{\,TT}}
\newcommand{\FoTM}{\mathrm{Fo}_{\,TM}}
\newcommand{\FoM}{\mathrm{Fo}_{\,M}}
\newcommand{\kMs}{k_{\,M}^{\,\star}}
\newcommand{\kTTs}{k_{\,TT}^{\,\star}}
\newcommand{\kTMs}{k_{\,TM}^{\,\star}}
\newcommand{\gsinf}{g^{\,\star}_{\,\infty}}
\newcommand{\qsinf}{q^{\,\star}_{\,\infty}}
\newcommand{\ts}{t^{\,\star}}
\newcommand{\thetaT}{\theta_{\,T}}
\newcommand{\thetaTi}{\theta_{\,T,\,i}}
\newcommand{\thetaM}{\theta_{\,M}}
\newcommand{\thetaMi}{\theta_{\,M,\,i}}
\newcommand{\ua}{u_{\,a}}
\newcommand{\uinf}{u_{\,\infty}}
\newcommand{\vinf}{v_{\,\infty}}
\newcommand{\va}{v_{\,a}}
\newcommand{\xs}{x^{\,\star}}

\newcommand{\scal}{\boldsymbol{\cdot}}
\newcommand*\pd[2]{\frac{\partial #1}{\partial #2}}
\newcommand*\od[2]{\frac{\mathrm{d} #1}{\mathrm{d} #2}}
\renewcommand{\div}{\grad\scal}
\newcommand{\grad}{\boldsymbol{\nabla}}
\newcommand{\eqdef}{\mathop{\stackrel{\,\mathrm{def}}{:=}\,}}

\renewcommand{\L}{\mathcal{L}}

\newcommand{\half}{{\textstyle{1\over2}}}

\newcommand{\dix}[1]{ \cdot 10^{\,#1}}

\newcommand{\unitfrac}[2]{\nicefrac{\mathsf{#1}}{\mathsf{#2}}}

\begin{document}

\title[\Title]{An improved explicit scheme for whole-building hygrothermal simulation}

\author[S.~Gasparin]{Suelen Gasparin$^*$}
\address{\textbf{S.~Gasparin:} LAMA, UMR 5127 CNRS, Universit\'e Savoie Mont Blanc, Campus Scientifique, F-73376 Le Bourget-du-Lac Cedex, France and Thermal Systems Laboratory, Mechanical Engineering Graduate Program, Pontifical Catholic University of Paran\'a, Rua Imaculada Concei\c{c}\~{a}o, 1155, CEP: 80215-901, Curitiba -- Paran\'a, Brazil}
\email{suelengasparin@hotmail.com}
\urladdr{https://www.researchgate.net/profile/Suelen\_Gasparin/}
\thanks{$^*$ Corresponding author}

\author[J.~Berger]{Julien Berger}
\address{\textbf{J.~Berger:} LOCIE, UMR 5271 CNRS, Universit\'e Savoie Mont Blanc, Campus Scientifique, F-73376 Le Bourget-du-Lac Cedex, France}
\email{Berger.Julien@univ-smb.fr}
\urladdr{https://www.researchgate.net/profile/Julien\_Berger3/}

\author[D.~Dutykh]{Denys Dutykh}
\address{\textbf{D.~Dutykh:} Univ. Grenoble Alpes, Univ. Savoie Mont Blanc, CNRS, LAMA, 73000 Chamb\'ery, France and LAMA, UMR 5127 CNRS, Universit\'e Savoie Mont Blanc, Campus Scientifique, F-73376 Le Bourget-du-Lac Cedex, France}
\email{Denys.Dutykh@univ-smb.fr}
\urladdr{http://www.denys-dutykh.com/}

\author[N.~Mendes]{Nathan Mendes}
\address{\textbf{N.~Mendes:} Thermal Systems Laboratory, Mechanical Engineering Graduate Program, Pontifical Catholic University of Paran\'a, Rua Imaculada Concei\c{c}\~{a}o, 1155, CEP: 80215-901, Curitiba -- Paran\'a, Brazil}
\email{Nathan.Mendes@pucpr.edu.br}
\urladdr{https://www.researchgate.net/profile/Nathan\_Mendes/}



\begin{titlepage}
\thispagestyle{empty} 
\noindent
{\Large Suelen \textsc{Gasparin}}\\
{\it\textcolor{gray}{Pontifical Catholic University of Paran\'a, Brazil}}\\
{\it\textcolor{gray}{LAMA--CNRS, Universit\'e Savoie Mont Blanc, France}}
\\[0.02\textheight]
{\Large Julien \textsc{Berger}}\\
{\it\textcolor{gray}{LOCIE--CNRS, Universit\'e Savoie Mont Blanc, France}}
\\[0.02\textheight]
{\Large Denys \textsc{Dutykh}}\\
{\it\textcolor{gray}{LAMA--CNRS, Universit\'e Savoie Mont Blanc, France}}
\\[0.02\textheight]
{\Large Nathan \textsc{Mendes}}\\
{\it\textcolor{gray}{Pontifical Catholic University of Paran\'a, Brazil}}
\\[0.10\textheight]

\colorbox{Lightblue}{
  \parbox[t]{1.0\textwidth}{
    \centering\huge\sc
    \vspace*{0.7cm}
    
    \textcolor{bluepigment}{An improved explicit scheme for whole-building hygrothermal simulation}

    \vspace*{0.7cm}
  }
}

\vfill 

\raggedleft     
{\large \plogo} 
\end{titlepage}


\newpage
\thispagestyle{empty} 
\par\vspace*{\fill}   
\begin{flushright} 
{\textcolor{denimblue}{\textsc{Last modified:}} \today}
\end{flushright}


\newpage
\maketitle
\thispagestyle{empty}


\begin{abstract}

Implicit schemes require important sub-iterations when dealing with highly nonlinear problems such as the combined heat and moisture transfer through porous building elements. The computational cost rises significantly when the whole-building is simulated, especially when there is important coupling among the building elements themselves with neighbouring zones and with HVAC (Heating Ventilation and Air Conditioning) systems. On the other hand, the classical \Eu ~explicit scheme is generally not used because its stability condition imposes very fine time discretisation. Hence, this paper explores the use of an improved explicit approach --- the \DF ~scheme --- to overcome the disadvantage of the classical explicit one and to bring benefits that cannot be obtained by implicit methods. The \DF ~approach is first compared to the classical  \Eu ~implicit and explicit schemes to compute the solution of nonlinear heat and moisture transfer through porous materials. Then, the analysis of the \DF ~unconditionally stable explicit scheme is extended to the coupled heat and moisture balances on the scale of a one- and a two-zone building models. The \DF ~scheme has the benefits of being unconditionally stable, second-order accurate in time $\O\,(\Delta t^{\,2})$ and to compute explicitly the solution at each time step, avoiding costly sub-iterations. This approach may reduce the computational cost by twenty, as well as it may enable perfect synchronism for whole-building simulation and co-simulation.


\bigskip
\noindent \textbf{\keywordsname:} heat and moisture transfer; numerical methods; finite-differences; explicit schemes; \DF ~scheme; whole-building simulation \\


\end{abstract}


\newpage
\tableofcontents
\thispagestyle{empty}


\newpage
\section{Introduction}

Models for the combined heat and moisture transfer through porous building elements have been implemented in building simulation tools since the 1990's in software such as Delphin \cite{BauklimatikDresden2011}, MATCH \citep{Rode2003}, MOIST \citep{Burch1993}, WUFI \cite{IBP2005} and UMIDUS \citep{Mendes1997, Mendes1999, DosSantos2006} among others. More recently, those models have been implemented in whole-building simulation tools and tested in the frame of the International Energy Agency Annex $41\,$, which reported on most of detailed models and their successful applications for accurate assessment of hygrothermal transfer in buildings \citep{Woloszyn2008}.

The \Eu ~and \CN ~implicit schemes have been used in many studies and implemented in building simulation tools, as reported in the literature \cite{Mendes2005, IBP2005, BauklimatikDresden2011, Steeman2009, Rouchier2013, Janssen2014, Janssen2007, VanGenuchten1982, Kalagasidis2007}, due to their numerical property of unconditional stability. Nevertheless, at every time step, one has to use a tridiagonal solver to invert the linear system of equations to determine the solution value at the following time layer. For instance in \cite{Mendes2005}, a multi-tridiagonal matrix algorithm has been developed to compute the solution of coupled equations of nonlinear heat and moisture transfer, using an \Eu ~implicit scheme. Furthermore, when dealing with nonlinearities, as when material properties are moisture content or temperature dependent, one has to perform \emph{sub-iterations} to linearise the system, increasing the total CPU time. In \cite{Janssen2014}, thousands of sub-iterations are reported to converge to the solution of a mass diffusion problem. Another disadvantage of implicit schemes appears when coupling the wall model, representing the transfer phenomena in porous building elements, to the room air model. The wall and the room air models must iterate within one time step until reaching a given tolerance \cite{Hensen1995}. If it does not impose any limitation on the choice of the time discretisation, it induces sub-iterations that increase the computational time of the simulation of the whole-building model. Moreover, it is valuable to decrease this computational cost knowing that the hygrothermal and energy building simulation is generally carried out for time scale periods as long as one year, or even more. However, the phenomena and particularly the boundary conditions evolve a time scale of seconds.

Recently, in \citep{Gasparin2017}, the improved \DF ~explicit scheme was explored for the solution of moisture diffusion equation highlighting that the standard stability limitation can be overcome, which inspired to investigate the use of the \DF ~scheme for the solution of the combined heat and moisture transfer through porous building elements coupled with room air models.

In this way, this paper first describes in Section~\ref{sec:HAM_transfer} the heat and moisture transfer model. In Section~ \ref{sec:numerical_schemes}, basics of the \DF ~explicit scheme is detailed before exploring the features of the scheme applied to linear and nonlinear cases, presented in Section~\ref{sec:AN2}. Then, the benefits of using an unconditionally stable explicit scheme are investigated to perform a whole-building hygrothermal simulation based on coupling the porous element model to a room air multizone model.


\section{Porous building element hygrothermal model}
\label{sec:HAM_transfer}

The physical problem considers one-dimensional heat and moisture transfer through a porous material defined by the spatial domain $\Ox \egal [\, 0, \, L \,]\,$. The following convention is adopted: $x \egal 0$ corresponds to the surface in contact with the inside room and, $x \egal L\,$, corresponds to the outside surface. The moisture transfer occurs due to capillary migration and vapour diffusion. The heat transfer is governed by diffusion and latent mechanisms. The physical problem can be formulated as \cite{Janssen2014,Tariku2010}:
\begin{subequations}\label{eq:HAM_equation}
  \begin{align}
  \label{eq:M_equation}
  \pd{\rhow}{t} 
  & \egal \pd{}{x} \Biggl( \, \kl \, \pd{\Pc}{x} \plus \delta_{\,v} \, \pd{\Pv}{x} \,  \Biggr)  \,, \\[3pt]
  \label{eq:H_equation}
  \bigl(\, \rhoz \ \cz \plus \rhow \ \cw \, \bigr) \ \pd{T}{t} 
  \plus \cw \ T \ \pd{\rhow}{t}
  & \egal \pd{}{x} \Biggl(\, \lambda \ \pd{T}{x} \plus \Lv \ \delta_{\,v} \ \pd{\Pv}{x} \, \Biggr) \,,
  \end{align}
\end{subequations}
where $\rhow$ is the volumetric moisture content of the material, $\delta_{\,v}$ and $\kl\,$, the vapour and liquid permeabilities, $\Pv\,$, the vapour pressure, $T\,$, the temperature, $\Rv\,$, the water vapour gas constant, $\Pc$ the capillary pressure, $\cz\,$, the material heat capacity, $\rhoz\,$, the material density, $\cw$ the water heat capacity, $\lambda$ the thermal conductivity, and, $\Lv$ the latent heat of evaporation. Eq.~\eqref{eq:M_equation} can be written using the vapour pressure $\Pv$ as the driving potential. For this, we consider the physical relation, known as the \textsc{Kelvin} equation, between $\Pv$ and $\Pc\,$, and the \textsc{Clausius}--\textsc{Clapeyron} equation:
\begin{align*}
  \Pc & \egal \rhol \, \Rv \, T \, \ln\left(\frac{\Pv}{\Ps(T)}\right)\,,\\
  \pd{\Pc}{\Pv} & \egal \frac{\rhol\, R_{\,v} \, T}{\Pv}\,.
\end{align*}
Neglecting the variation of the capillary pressure and the mass content with temperature \cite{Rouchier2013}, the partial derivative of $\Pc$ can be written as:
\begin{align*}
  \pd{\Pc}{x} \egal \pd{\Pc}{\Pv} \, \pd{\Pv}{x} \plus \pd{\Pc}{T} \, \pd{T}{x}\ \simeq \ \frac{\rhol\, R_{\,v} \, T}{\Pv} \, \pd{\Pv}{x}\,. 
\end{align*}

In addition, we have:
\begin{align*}
  & \pd{\rhow}{t} \egal \pd{\rhow}{\phi} \, \pd{\phi}{\Pv} \, \pd{\Pv}{t} \plus \pd{\rhow}{T} \, \pd{T}{t} \simeq \pd{\rhow}{\phi} \, \pd{\phi}{\Pv} \, \pd{\Pv}{t}\,.
\end{align*}
Considering the relation $\rhow \egal f\, (\phi)\,$, obtained from material properties, and from the relation between the vapour pressure $\Pv$ and the relative humidity $\phi\,$, we get: 
\begin{align*}
  & \pd{\rhow}{t} \egal \frac{f^{\,\prime}\, (\phi)}{\Ps} \ \pd{\Pv}{t}\,.
\end{align*}

We denote by:
\begin{align*}
  & \kM \ \eqdef \ \kl \, \dfrac{\rhol \, \Rv \, T}{\Pv} \plus \delta_{\,v} && \text{the total moisture transfer coefficient under vapour pressure gradient} \,,\\
  & \kTM \ \eqdef \ \Lv \ \delta_{\,v} && \text{the heat coefficient due to a  vapour pressure gradient} \,,\\
  & \kTT \ \eqdef \ \lambda && \text{the heat transfer coefficient under temperature gradient} \,, \\
  & \cM \ \eqdef \ \frac{f^{\,\prime}\, (\phi)}{\Ps(T)} && \text{the moisture storage coeficient}  \,,\\
  & \cTT \ \eqdef \ \rhoz \ \cz \plus f\, (\phi) \ \cw && \text{the energy storage coeficient} \,,\\
  & \cTM \ \eqdef \ \cw \ T \ \frac{f^{\,\prime}\, (\phi)}{\Ps(T)} && \text{the coupling storage coefficient} \,.\\
\end{align*}

Considering the previous notation, Eq.~\eqref{eq:HAM_equation} can be rewritten as:
\begin{subequations}\label{eq:HAM_equation2}
  \begin{align}
  \cM \ \pd{\Pv}{t} 
  & \egal \pd{}{x} \Biggl( \, \kM \, \pd{\Pv}{x} \,  \Biggr)  \,, \\[3pt]
  \cTT \ \pd{T}{t} 
  \plus \cTM \ \pd{\Pv}{t}
  & \egal \pd{}{x} \Biggl(\, \kTT \ \pd{T}{x} \plus \kTM \ \pd{\Pv}{x} \, \Biggr) \,.
  \end{align}
\end{subequations}

The boundary conditions at the interface between the porous material and the air are expressed as:
\begin{subequations}\label{eq:HAM_BC}
  \begin{align}
  \kM \, \pd{\Pv}{x} & 
  \egal \hM \, \bigl(\, \Pv - \Pvinf \,\bigr) \moins g_{\,\infty} \,, \\[3pt]
  \kTT \, \pd{T}{x} \plus \kTM \, \pd{\Pv}{x} & 
  \egal \hT \, \bigl(\, T- \Tinf \,\bigr) \plus \Lv \, \hM \, \bigl(\, \Pv - \Pvinf \,\bigr) \moins q_{\,\infty} \,,
  \end{align}
\end{subequations}
where $\Pvinf$ and $\Tinf$ stand for the vapour pressure and temperature of the air and $\hM$ and $\hT$ are the convective transfer coefficients. If the bounding surface is in contact with the outside building air, $g_{\,\infty} $ is the liquid flux from wind driven rain and $q_{\,\infty}$ is the total heat flux from radiation and the heat contribution from the inward liquid water penetration. If the bounding surface is in contact with the inside building air, $g_{\,\infty} \egal 0 $ and $q_{\,\infty}$  is the distributed short-wave radiative heat transfer rate $q_{\,\mathrm{rw}} $ in the enclosure and long-wave radiative heat exchanged among the room surfaces:
\begin{align*}
  q_{\,\mathrm{rw}} \egal \sum_{w=1}^{m} \mathrm{s} \, \epsilon \, \sigma \, \Biggl(\, \biggl(\, T_{\,w}(\, x = 0 \,) \, \biggr)^{\,4} \moins \biggl(\, T(\, x = 0 \,) \, \biggr)^{\,4} \, \Biggr) \,,
\end{align*} 
where $s$ is the view factor between two surfaces, $\sigma$ is the \textsc{Stefan}--\textsc{Boltzmann} constant, $\epsilon$ is the emissivity of the wall surface, $w$ represents the $m$ bounding walls. We consider a uniform vapour pressure and temperature distributions as initial conditions:
\begin{subequations}\label{eq:HAM_ic}
  \begin{align}
   \Pv &\egal \Pvi \,, && t\egal0 \,, \\
    T &\egal \Ti \,, && t\egal0 \,.
  \end{align}
\end{subequations}

The governing equations can be written in a dimensionless form as:
\begin{subequations}\label{eq:HAM_equation_dimless}
  \begin{align}\label{eq:heat_equation_dimless}
  \cMs \ \pd{v}{\ts} 
  & \egal \FoM \, \pd{}{\xs} \Biggl( \, \kMs \, \pd{v}{\xs} \,  \Biggr)  \,, \\[3pt]
  \label{eq:moisture_equation_dimless}
  \cTTs \ \pd{u}{\ts} 
  \plus \cTMs \ \gamma \ \pd{v}{\ts}
  & \egal \FoTT \, \pd{}{\xs} \Biggl(\, \kTTs \ \pd{u}{\xs} \, \Biggr) \plus \FoTM \ \gamma \, \pd{}{\xs} \Biggl(\,\kTMs \ \pd{v}{\xs}  \Biggr) \,.
  \end{align}
\end{subequations}
and the  boundary condition as:
\begin{subequations}\label{eq:HAM_BC_dimless}
  \begin{align}
  \kMs \ \pd{v}{\xs} & 
  \egal \BiM \, \bigl(\, v - \vinf \,\bigr) \moins \gsinf \,, \\[3pt]
  \FoTT \, \kTTs \ \pd{u}{\xs} \plus \kTMs \ \FoTM \ \gamma \ \pd{v}{\xs} & 
  \egal \BiTT \, \bigl(\, u- \uinf \,\bigr) \plus \BiTM \, \bigl(\, v - \vinf \,\bigr) \moins \qsinf \,,
  \end{align}
\end{subequations}
where the dimensionless quantities are defined as: 
\begin{align*}
  & u \ \eqdef \ \frac{T}{\Ti} \,,
  && v \ \eqdef \ \frac{\Pv}{\Pvi} \,,
  && \uinf \ \eqdef \ \frac{\Tinf}{\Ti} \,,
  && \vinf \ \eqdef \ \frac{\Pvinf}{\Pvi} \,, \\[3pt]
  & \xs \ \eqdef \ \frac{x}{L} \,, 
  && \ts \ \eqdef \ \frac{t}{\tref} \,,
  && \cMs \ \eqdef \ \frac{\cM }{\cMref} \,,
  && \cTTs \ \eqdef \ \frac{\cTT }{\cTTref} \,, \\[3pt]
  & \cTMs \ \eqdef \ \frac{\cTM }{\cTMref} \,, 
  && \kMs \ \eqdef \ \frac{\kM }{\kMref} \,, 
  && \kTTs \ \eqdef \ \frac{\kTT }{\kTTref} \,,
  && \kTMs \ \eqdef \ \frac{\kTM }{\kTMref} \,,  \\[3pt]
  & \FoM \ \eqdef \ \frac{\tref \cdot \kMref}{L^{\,2} \cdot \cMref} \,,
  && \FoTT \ \eqdef \ \frac{\tref \cdot \kTTref}{L^{\,2} \cdot \cTTref} \,,
  && \FoTM \ \eqdef \ \frac{\tref \cdot \kTMref}{L^{\,2}\cdot \cTMref} \,, 
  && \gamma \ \eqdef \ \frac{\cTMref \cdot \Pvi}{\cTTref \cdot \Ti} \,, \\[3pt]
  & \BiM \ \eqdef \ \frac{\hM \cdot L}{\kMref} \,, 
  && \BiTT \ \eqdef \ \frac{\hT \cdot L}{\kTTref} \,, 
  && \BiTM \ \eqdef \ \frac{\Lv \cdot \hM \cdot L \cdot \Pvi}{\kTTref \cdot \Ti} \,, 
  && \gsinf \ \eqdef \ \frac{L \cdot g_{\,\infty} }{\Pvi \cdot \kMref}  \,, \\[3pt] 
  & \qsinf \ \eqdef \ \frac{L \cdot q_{\,\infty}}{\Ti \cdot \kTTref}  \,.
\end{align*}

The dimensionless formulation enables to determine important scaling parameters (\textsc{Biot} and \textsc{Fourier} numbers for instance). Henceforth, solving one dimensionless problem is equivalent to solve a whole class of dimensional problems sharing the same scaling parameters. Then, dimensionless equations allow to estimate the relative magnitude of various terms, and thus, eventually to simplify the problem using asymptotic methods \citep{Nayfeh2000}. Finally, the floating point arithmetics is designed such as the rounding errors are minimal if you manipulate the numbers of the same magnitude \cite{Kahan1979}. Moreover, the floating point numbers have the highest density in the interval $(\, 0,\,1 \,)$ and their density decays exponentially when we move further away from zero. So, it is always better to manipulate numerically the quantities at the order of $\O\,(1)$ to avoid severe round-off errors and to likely improve the conditioning of the problem in hands.


\section{Numerical schemes}
\label{sec:numerical_schemes}

Let us consider a uniform discretisation of the interval $\Ox \ \rightsquigarrow\ \Omega_{\,L_{\,x}}\,$:
\begin{equation*}
  \Omega_{\,h}\ =\ \bigcup_{j\,=\,0}^{N-1} [\,x_{\,j},\;x_{\,j+1}\,]\,, \qquad
  x_{j+1}\ -\ x_{\,j}\ \equiv\ \Delta x\,, \quad \forall j\ \in\ \bigl\{0,\,1,\,\ldots,\,N-1\bigr\}\,.
\end{equation*}
The time layers are uniformly spaced as well $t^{\,n}\ =\ n\,\Delta t\,$, $\Delta t\ =\ \const\ >\ 0\,$, $n\ =\ 0,\,1,\,2,\,\ldots, \, N_{\,t}$\,. The values of the function $u\,(x,\,t)$ in discrete nodes will be denoted by $u_{\,j}^{\,n}\ \eqdef\ u\,(x_{\,j},\,t^{\,n}\,)\,$.

For the sake of simplicity and without loosing generality, simple diffusion equation is considered:
\begin{align}\label{eq:heat1d}
  \pd{u}{t} \egal \div \bigl( \, \nu \grad u \, \bigr) \,.
\end{align}
First, the numerical schemes are explained considering the linear case. Then, the extension to the nonlinear case is described.


\subsection{Improved explicit scheme: Dufort-Frankel method}
\label{sec:DF_scheme}

\begin{figure}
\begin{center}
  \includegraphics[scale=.3]{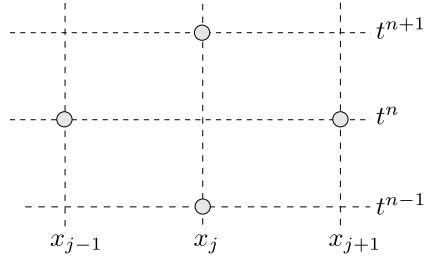}
  \caption{\small\em Stencil of the \textsc{Dufort--Frankel} numerical scheme.}
  \label{fig:stencil_Dufort}
\end{center}
\end{figure}

Using the so-called \textsc{Dufort}--\textsc{Frankel} method, the numerical scheme is expressed as:
\begin{align}\label{eq:dufort}
  & \frac{u_{\,j}^{\,n+1}\ -\ u_{\,j}^{\,n-1}}{2\,\Delta t} \egal \nu\;\frac{u_{\,j-1}^{\,n}\ -\ \bigl(u_{\,j}^{\,n-1}\ +\ u_{\,j}^{\,n+1}\bigr)\ +\ u_{\,j+1}^{\,n}}{\Delta x^{\,2}}\,, \qquad j\ =\ 1,\,\ldots,\,N-1\,, && n\ \geqslant\ 1\,,
\end{align}
where the term $2\,u_{\,j}^{\,n}\ $ from the explicit scheme is replaced by $u_{\,j}^{\,n-1}\ +\ u_{\,j}^{\,n+1}\,$. The Scheme~\eqref{eq:dufort} has the stencil depicted in Figure~\ref{fig:stencil_Dufort}. At a first glance, the scheme \eqref{eq:dufort} looks like an implicit scheme, however, it is not truly the case. Eq.~\eqref{eq:dufort} can be easily solved for $u_{\,j}^{\,n+1}$ to give the following discrete dynamical system:
\begin{align*}
  & u_{\,j}^{\,n+1}\ =\ \frac{1\ -\ \lambda}{1\ +\ \lambda}\;u_{\,j}^{\,n-1}\ +\ \frac{\lambda}{1\ +\ \lambda}\;\bigl(u_{\,j+1}^{\,n}\ +\ u_{\,j-1}^{\,n}\bigr) \,, && n\ \geqslant\ 1\,,
\end{align*}
where:
\begin{align*}
  \lambda\ \eqdef\ 2\,\nu\;\frac{\Delta t}{\Delta x^{\,2}} \,.
\end{align*}
The standard \textsc{von Neumann} stability analysis shows that the \textsc{Dufort}--\textsc{Frankel} scheme is \emph{unconditionally stable} \cite{Gasparin2017, Richtmyer1967, Taylor1970}. The consistency error analysis of the Scheme~\eqref{eq:dufort} shows the following result:
\begin{multline}\label{eq:expansion_dufort}
  \L_{\,j}^{\,n} \egal \nu\;\frac{\Delta t^{\,2}}{\Delta x^{\,2}} \;\pd{^2u}{t^2} \plus \pd{u}{t} \ -\ \nu\,\pd{^2u}{x^2} \plus   \frac{1}{6} \,\Delta t^{\,2}\,\pd{^3u}{t^3} \\
 \moins \frac{1}{12} \,\nu\,\Delta x^{\,2}\,\pd{^4u}{x^4} \ -\ \frac{1}{12} \,\nu\,\Delta t^{\,2}\,\Delta x\,\pd{^5u}{x^3 \, \partial t^2}\ +\ \O\Bigl(\frac{\Delta t^{\,4}}{\Delta x^{\,2}}\Bigr) \,,
\end{multline}
where:
\begin{align*}
  \L_{\,j}^{\,n}\ \eqdef\ \frac{u_{\,j}^{\,n+1}\ -\ u_{\,j}^{\,n-1}}{2\,\Delta t}\ -\ \nu\;\frac{u_{\,j-1}^{\,n}\ -\ \bigl(u_{\,j}^{\,n-1}\ +\ u_{\,j}^{\,n+1}\bigr)\ +\ u_{\,j+1}^{\,n}}{\Delta x^{\,2}} \,.
\end{align*}
So, from the asymptotic expansion for $\L_{\,j}^{\,n}$ one can see the \textsc{Dufort}--\textsc{Frankel} scheme is second-order accurate in time and:
\begin{itemize}
  \item First-order accurate in space if $\Delta t\ \propto\ \Delta x^{\,3/2}$\,;
  \item Second-order accurate in space if $\Delta t\ \propto\ \Delta x^{\,2}$\,.
\end{itemize}

In the nonlinear case, the ~numerical scheme can be derived as follows:
\begin{align}\label{eq:DF_NL}
  \frac{u_{\,j}^{\,n+1}\ -\ u_{\,j}^{\,n-1}}{2\,\Delta t} \egal \frac{1}{\Delta x} \Biggl[\, \left(\, \nu \pd{u}{x} \, \right)_{\,j+\half}^{\,n}   \moins \left(\, \nu \pd{u}{x} \, \right)_{\,j-\half}^{\,n}\,\Biggr]\,.
\end{align} 
The right-hand side term can be expressed as: 
\begin{align}\label{eq:DF_NL_RHS}
  & \frac{1}{\Delta x} \left[\, \left(\, \nu \pd{u}{x} \, \right)_{\,j+\half}^{\,n}   \moins \left(\, \nu \pd{u}{x} \, \right)_{\,j-\half}^{\,n}   \, \right] \egal \frac{1}{\Delta x^2} \left[\, \, \nu _{\,j+\half}^{\,n}  \, u_{\,j+1}^{\,n} \plus \, \nu _{\,j-\half}^{\,n}  \, u_{\,j-1}^{n} \moins \left(\, \nu _{\,j+\half}^{\,n}   \plus  \nu _{\,j-\half}^{\,n}  \, \right) u_{\,j}^{\,n} \,\right] \,.
\end{align}
Using the \DF ~stencil (see Figure~\ref{fig:stencil_Dufort}), the term $u_{\,j}^{\,n}$ is replaced by $\dfrac{u_{\,j}^{\,n+1} \plus u_{\,j}^{\,n-1}}{2}\,$.

Thus, considering Eq.~\eqref{eq:DF_NL}, the \DF ~schemes can be expressed as an explicit scheme: 
\begin{align*}
  & u_{\,j}^{\,n+1} \egal \frac{\lambda_{\,1}}{\lambda_{\,0} \plus \lambda_{\,3}} \cdot u_{\,j+1}^{\,n} \plus \frac{\lambda_{\,2}}{\lambda_{\,0} \plus \lambda_{\,3}} \cdot u_{\,j-1}^{\,n} \plus \frac{\lambda_{\,0} \moins \lambda_{\,3}}{\lambda_{\,0} \plus \lambda_{\,3}} \cdot u_{\,j}^{\,n-1}  \,, && n\ \geqslant\ 1\,,
\end{align*}
with 
\begin{align*}
  & \lambda_{\,0} \ \eqdef \ 1 \,,
  && \lambda_{\,1} \ \eqdef \  \frac{2 \, \Delta t}{\Delta x^2} \; \nu_{j+\half}^{\,n} \,, \\[3pt]
  & \lambda_{\,2} \ \eqdef \  \frac{2 \, \Delta t}{\Delta x^2} \; \nu_{j-\half}^{\,n} \,,
  && \lambda_{\,3} \ \eqdef \ \frac{\Delta t}{\Delta x^2} \, \left(\, \nu_{j+\half}^{\,n} + \nu_{j-\half}^{\,n} \, \right) \,.
\end{align*}

When dealing with the nonlinearities of the material properties, an interesting feature of explicit schemes is that it does not require any sub-iterations (using \textsc{Newton}--\textsc{Raphson} approach for instance). At the time layer  $n\,$, the material properties $\nu_{\,j+\half}\,$, $\nu_{\,j-\half}$ are \emph{explicitly} calculated at $t^{\,n}\,$. It should be noted that the material properties evaluated at $j \plus \half$ is formulated as:
\begin{align*}
  \nu^{\,n}_{\,j+\half} \egal \nu \ \Biggl(\, \frac{u^{\,n}_{\,j} \plus u^{\,n}_{\,j+1}}{2} \,\Biggr)\,.
\end{align*}

\subsection{Validation of the numerical solution}
\label{sec:validat}

All the numerical results in this paper were computed using \texttt{Matlab}. 
One possibility for comparing the computed solutions is by computing the $\mathcal{L}_{\,\infty}$ error between the solution $u_{\,\mathrm{num}}$ and a reference solution $u_{\, \mathrm{ref}}\,$:
\begin{align*}
  & \varepsilon \ \eqdef\ \Big| \Big| \, u_{\,\mathrm{ref}} \moins u_{\,\mathrm{num}}\,\Big| \Big|_{\,\infty}\,.
\end{align*}
The reference solution is computed using the \texttt{Matlab} open source package \texttt{Chebfun} \cite{Driscoll2014}. Using \texttt{pde} functions, it enables to compute a numerical solution of a partial derivative equation with the \textsc{Chebyshev} polynomials adaptive spectral methods.

The $\mathcal{L}_{\,\infty}$ error can be computed along the space or time domains, according to: 
\begin{align*}
  & \varepsilon(\,x\,) \ \eqdef\ \sup_{t \, \in \, \bigl[\,0 \,, \tau \,\bigr]} \,
  \biggl| \, u_{\,\mathrm{ref}} \, (\,x, \, t \,) \moins u_{\,\mathrm{num}} \, (\, x, \, t \,) \, \biggr| \,, \\
  & \varepsilon(\,t\,) \ \eqdef\ \sup_{x \, \in \, \bigl[\,0 \,, L \,\bigr]} \,
  \biggl| \, u_{\,\mathrm{ref}} \, (\,x, \, t \,) \moins u_{\,\mathrm{num}} \, (\, x, \, t \,) \, \biggr| \,.
\end{align*}


\section{Numerical application: porous wall transfer}
\label{sec:AN2}

\begin{figure}
  \begin{center}
  \subfigure[][\label{fig_AN1:BC_TL}]{\includegraphics[width=.4\textwidth]{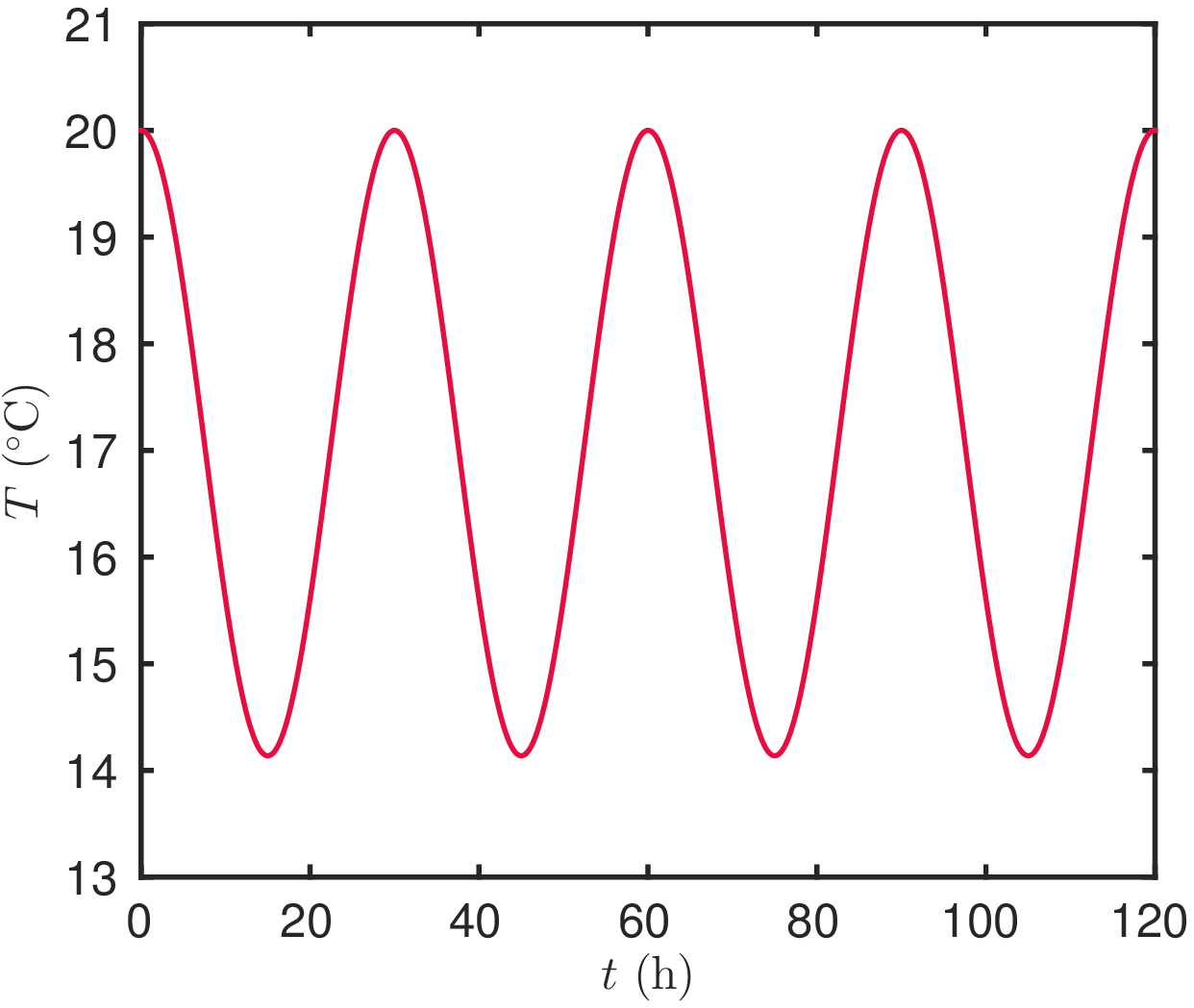}}
  \subfigure[][\label{fig_AN1:BC_TR}]{\includegraphics[width=.4\textwidth]{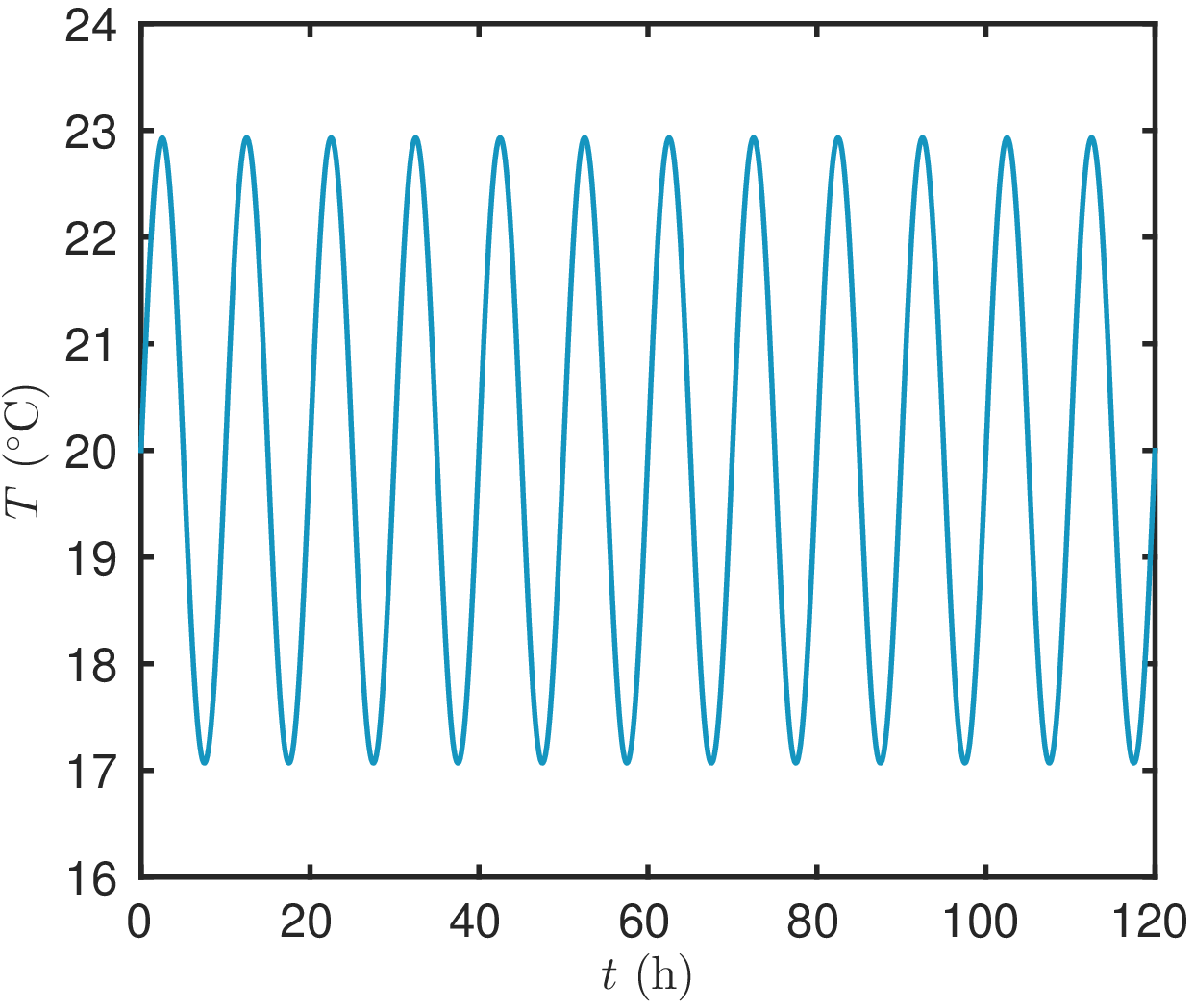}}
  \subfigure[][\label{fig_AN1:BCL_Phi}]{\includegraphics[width=.4\textwidth]{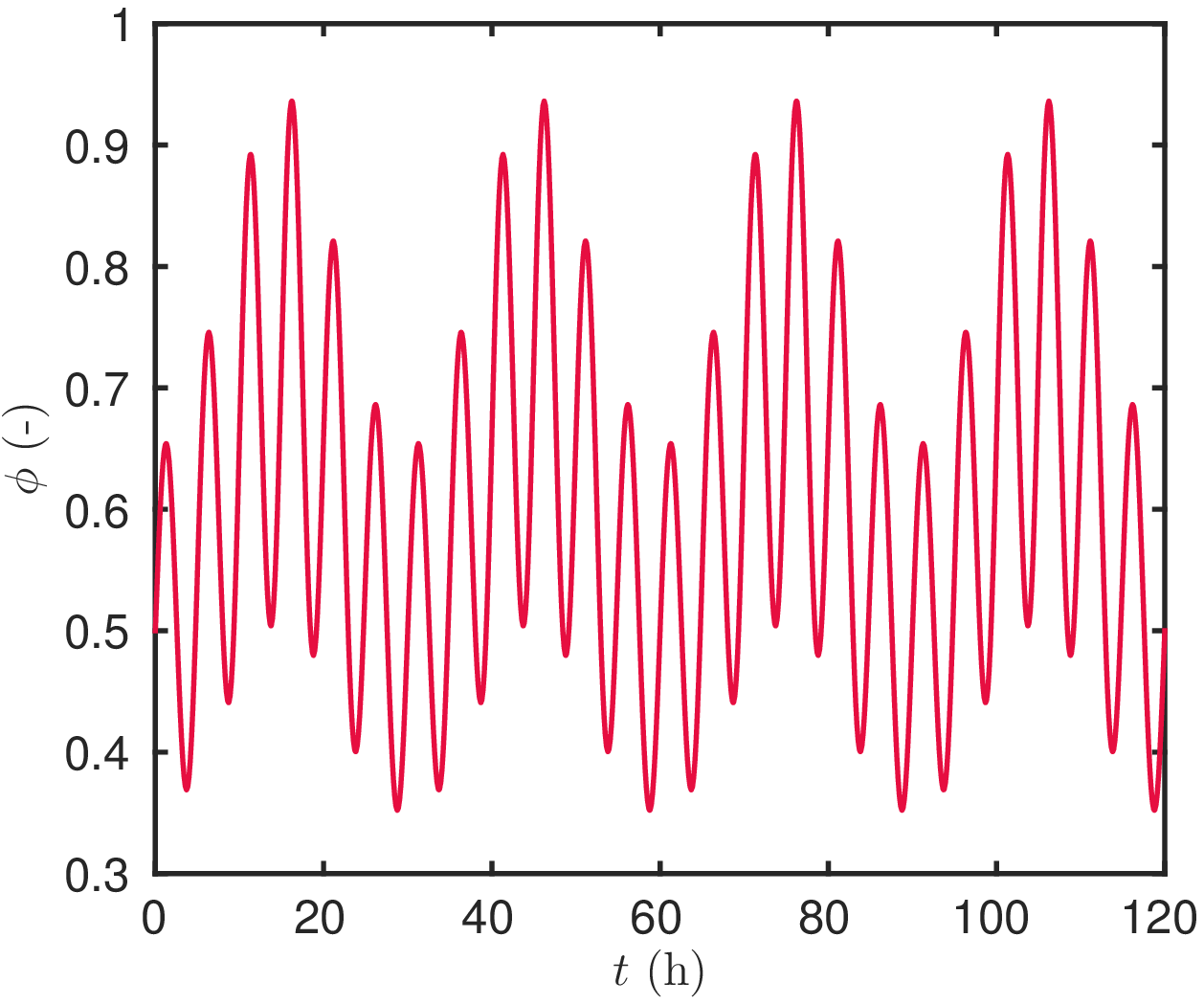}}
  \subfigure[][\label{fig_AN1:BCR_Phi}]{\includegraphics[width=.4\textwidth]{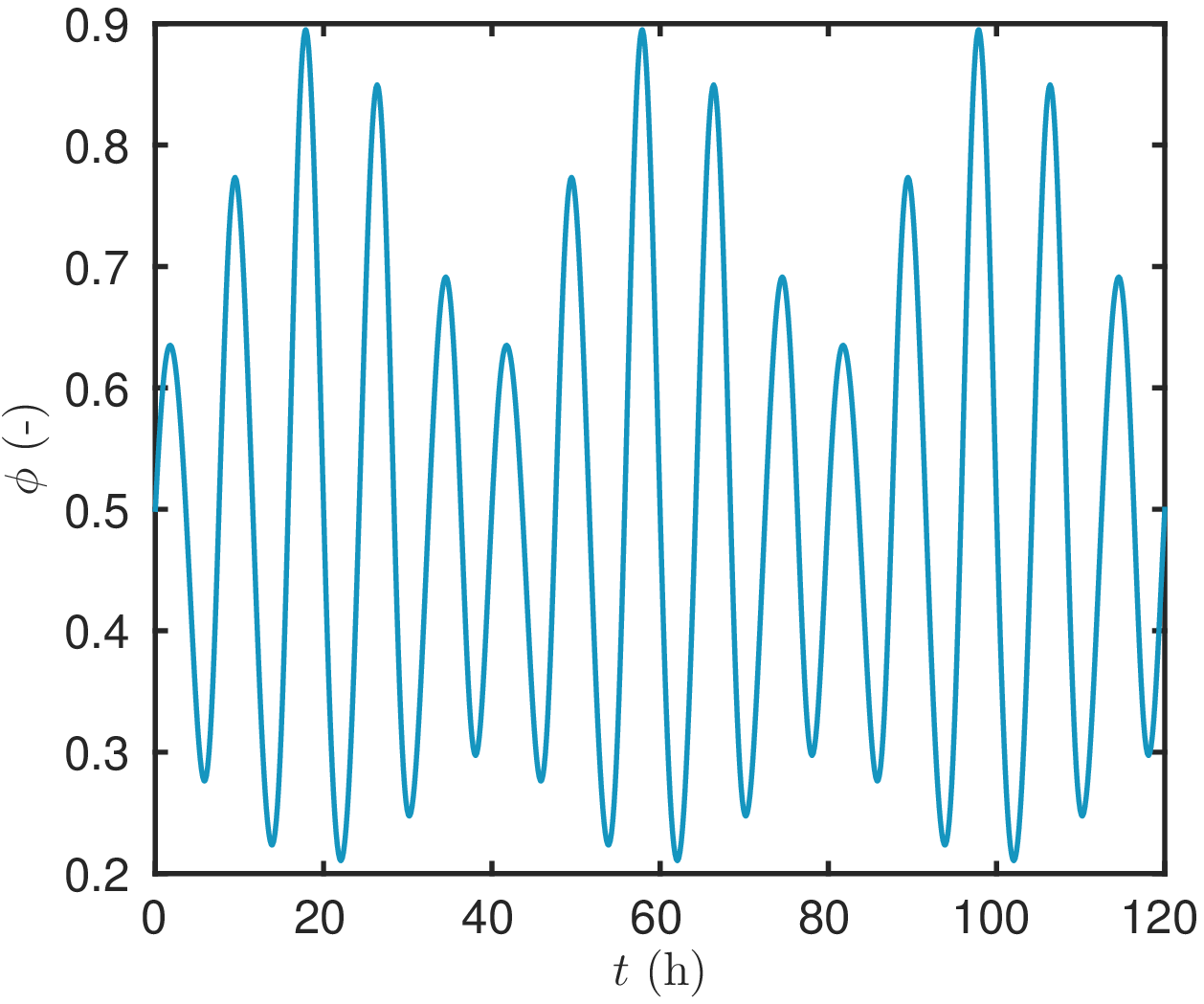}}
  \caption{\small\em Boundary conditions at $x \egal 0 \ \mathsf{m}$ (a,c) and $x \egal 0.1 \ \mathsf{m}$ (b,d).}
  \label{fig_AN1:BC}
  \end{center}
\end{figure}

Heat and moisture transfer are strongly nonlinear due to the variation of the material properties with the field. For this reason, this case study is devoted to investigated these effects by means of the \DF ~numerical scheme. 
A physical application is performed in a $10-\mathsf{cm}$ slab of bearing material. To test the robustness of the scheme, the properties of the material are gathered from \cite{Janssen2014, Hagentoft2004, Gasparin2017} and given in Table~\ref{table:properties_mat1}. At $t \egal 0 \ \mathsf{h}\,$, the material is considered with uniform fields, with  a temperature  of $\Ti \egal 20 \ \mathsf{^{\,\circ}C}$ and relative humidity of $\phi_{\,i} \egal 50 \ \mathsf{\%}\,$. The boundary conditions, represented by the relative humidity $\phi$ and temperature $T\,$, are given in Figure~\ref{fig_AN1:BC}. The convective mass and heat transfer coefficients are set to $\hM \egal 2 \cdot 10^{\,-7} \ \unitfrac{s}{m}\,$, $\hT \egal 25 \ \unitfrac{W}{(m^{\,2}\,.\,K)}$ and to $\hM \egal 3 \cdot 10^{\,-8} \ \unitfrac{s}{m}\,$, $\hT \egal 8 \ \unitfrac{W}{(m^{\,2}\,.\,K)}\,$, for the left and right boundary conditions, respectively. The dimensionless numerical values of this case are provided in \ref{sec:appendix}.

\begin{table}
\centering
\caption{\small\em Hygrothermal properties of the load bearing material.} \bigskip
\label{table:properties_mat1}
\footnotesize{
\begin{tabular}{l l}
\hline\hline
\textit{Property} & \textit{Value} \\
\hline\hline
Material density & $\rhoz \egal 790\quad [\unitfrac{kg}{m^3}]$  \\
Heat capacity & $\cz \egal 870\quad [\unitfrac{J}{(kg\cdot K)}]$\\
Water heat capacity & $\cw \egal 4180\quad [\unitfrac{J}{(kg\cdot K)}]$\\
Sorption isotherm &  $f \egal 47.1 \, \Bigl[1 + (-c_{\,1}\, \Pc)^{1.65}\Bigr]^{\,-0.39} + 109.9\, \Bigl[1 + (-c_{\,2}\, \Pc)^{\, 6}\Bigr]^{\,-0.83} \quad [\unitfrac{kg}{m^3}]$    \\
Permeability & $\delta_{\,v}\egal \dfrac{1.88\cdot 10^{\,-6}}{T}\cdot \dfrac{\Bigl(1-\frac{f}{157}\Bigr)}{0.503\, \Bigl(1-\frac{f}{157}\Bigr)^{\,2}+0.497} \quad [\mathsf{s}] $\\
Thermal conductivity & $\lambda\egal 0.2 \plus 0.0045\cdot f \quad [\unitfrac{W}{(m\cdot K)}]$\\
Moisture transfer coeff. & \parbox[t]{9cm}{ $\kM \egal  \exp \Bigl[-0.07 \cdot \log_{10} \Bigl(2\, \frac{\Pv}{\Ps}\Bigr)+1.44 \Bigr] \cdot 1.97\cdot 10^{\,-10}  + \\  \exp \Bigl[-8 \cdot \Bigl(2\, \frac{\Pv}{\Ps}\Bigr)-2 \Bigr]^{2}  \cdot 1.77\cdot 10^{\,-7}   \quad [\mathsf{s}]$ }\\
Moisture storage coeff. & \parbox[t]{9cm}{ $\cM \egal 
-30.62\cdot \Bigl(-c_1\,\frac{\Rv \, T}{\Pv} \Bigr)\cdot \Bigl(-c_1\, \Rv \, T \log_{10} \frac{\Pv}{\Ps} \Bigr)^{0.65} \cdot \\ 
\Bigl[1+\Bigl(-c_1\, \Rv \, T \, \log_{10} \frac{\Pv}{\Ps} \Bigr)^{1.65}\Bigr]^{\,-1.39} \moins 
549.5\cdot \Bigl(-c_2\,\frac{\Rv \, T}{\Pv} \Bigr)\cdot \\ \Bigl(-c_2\, \Rv \, T \log_{10} \frac{\Pv}{\Ps} \Bigr)^{5} \cdot
\Bigl[1+\Bigl(-c_2\, \Rv \, T \, \log_{10} \frac{\Pv}{\Ps} \Bigr)^{6}\Bigr]^{\,-1.83}\quad [\unitfrac{s^2}{m^2}]$ } \\
\hline \hline                 
$ ^{\star} c_1=1.25\cdot 10^{\,-5}$ and $c_2=1.8\cdot 10^{\,-5}\,$.
\end{tabular}}
\end{table}

The numerical solution was computed using three different schemes, considering a spatial discretisation parameter of $\Delta x^{\,\star} \egal 10^{\,-2}\,$. For the time domain, the discretisation parameter equals $\Delta t^{\,\star} \egal 10^{\,-3}$ for the \DF ~and \Eu ~implicit scheme. A tolerance $\eta \leqslant 10^{\,-2} \ \Delta t^{\,\star}$ has been used for the convergence of the sub-iterations of the implicit scheme, using a fixed-point algorithm. For the \Eu ~explicit scheme, a time discretisation $\Delta t^{\,\star} \egal 10^{\,-5}$ has been used, to respect the CFL stability condition. The latter has been computed for the heat transfer Equation~\eqref{eq:heat_equation_dimless}:
\begin{align}\label{eq:cfl_heat_Eq}
  & \Delta t^{\,\star} \ \leqslant \ 
  \frac{ \bigl(\,\Delta x^{\,\star}\, \bigr)^{\,2}}{2}  \, 
  \min_{\,u \,, v} \Biggl\{\,  \frac{\cTTs(\,u \,,v\,)}{\FoTT \, \kTTs(\,u \,,v\,)} \,,
  \frac{\cTTs(\,u \,,v\,)}{\FoTM \, \kTMs(\,u \,,v\,)} \,,
  \frac{\cTTs(\,u \,,v\,) \, \cMs(\,u \,,v\,)}{\cTMs(\,u \,,v\,) \ \gamma \, \FoM \, \kMs(\,u \,,v\,)} 
  \,\Biggr\} \,, 
\end{align}
and for the mass transfer Equation~\eqref{eq:moisture_equation_dimless}:
\begin{align}\label{eq:cfl_moisture_Eq}
  & \Delta t^{\,\star} \ \leqslant \ \frac{ \bigl(\,\Delta x^{\,\star}\, \bigr)^{\,2}}{2} \, \min_{\,u \,, v} \Biggl\{ \,  \frac{\cMs(\,u \,,v\,)}{\FoM \, \kMs(\,u \,,v\,)} \,\Biggr\}   \,,
\end{align}
which give the numerical values of $\Delta t^{\,\star} \ \leqslant \ 3 \dix{-4}$ and $\Delta t^{\,\star} \ \leqslant \ 7.4 \dix{-5}\,$, corresponding to physical values of $\Delta t^{\,\star} \ \leqslant \ 1.08 \ \mathsf{s}$ and $\Delta t^{\,\star} \ \leqslant \ 0.2 \ \mathsf{s}\,$,  respectively. It can be noted that the stability of the coupled equation is given by the moisture equation stability.

\begin{figure}
  \begin{center}
  \subfigure[][\label{fig_AN2:time_T_L}]{\includegraphics[width=.4\textwidth]{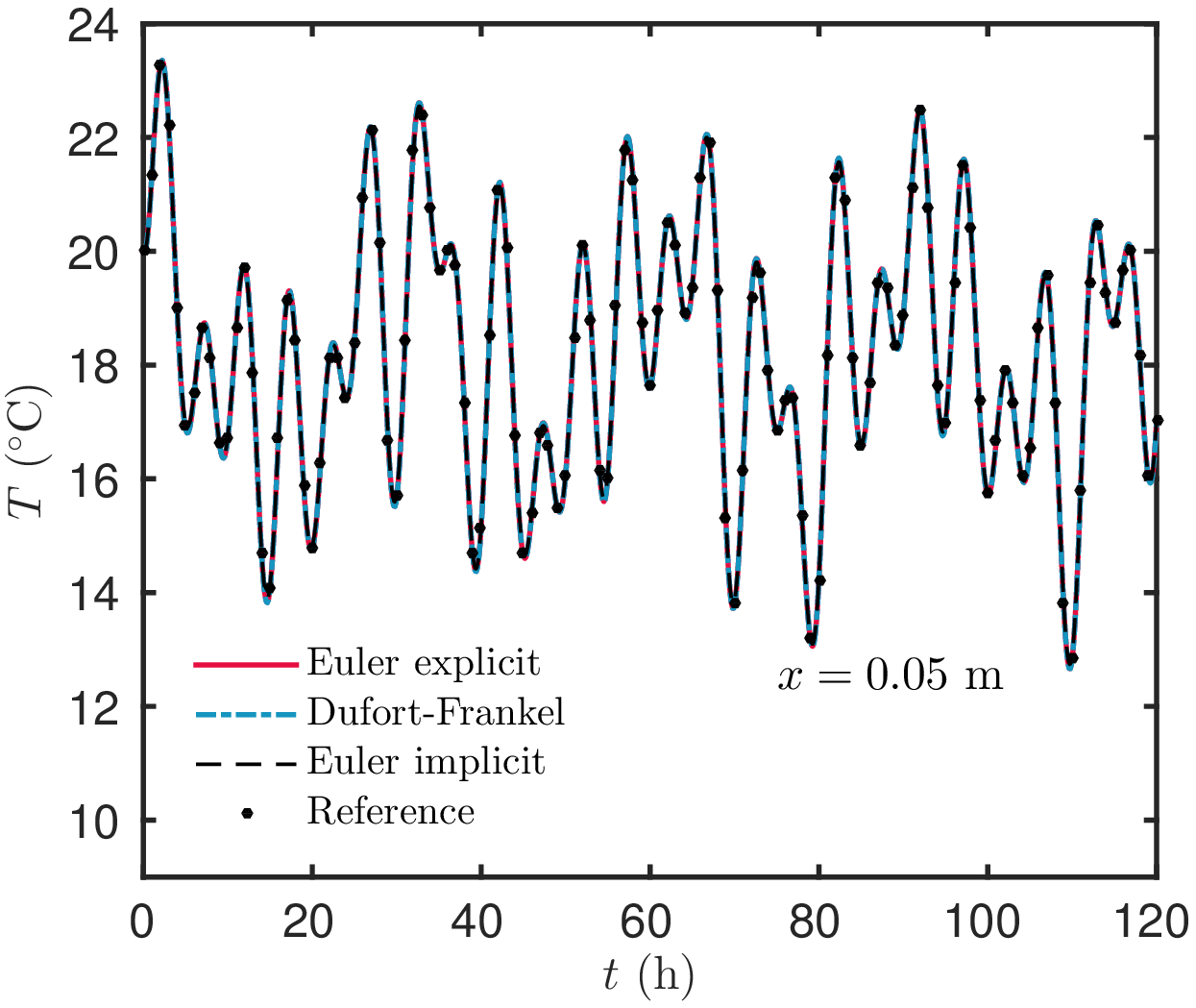}}
  \subfigure[][\label{fig_AN2:time_phi_L}]{\includegraphics[width=.4\textwidth]{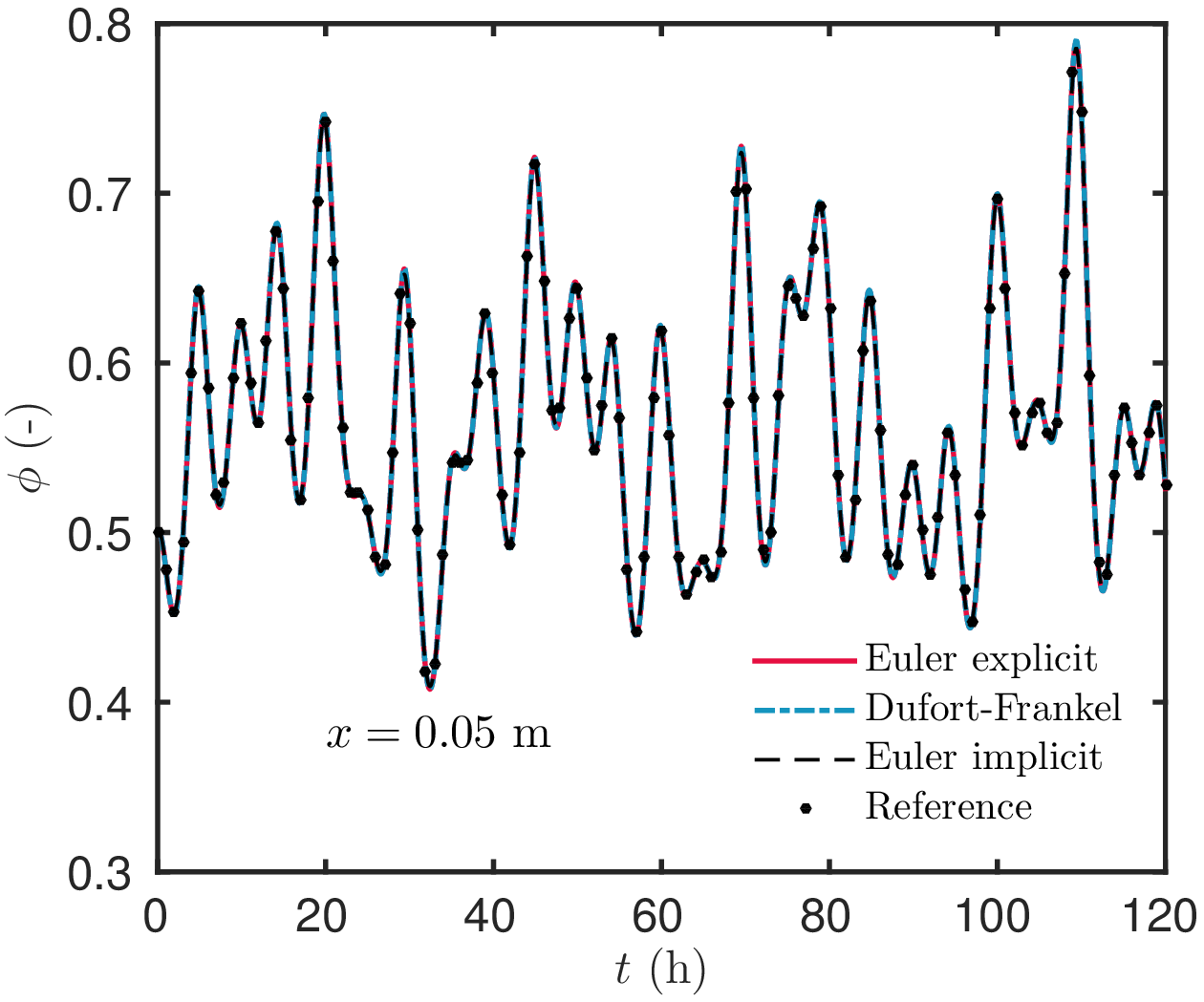}}
  \caption{\small\em Temperature (a) and relative humidity (b) evolution at the middle of the material $(x \egal 0.05 \ \mathsf{m})$.}
  \label{fig_AN2:times}
  \end{center}
\end{figure}

\begin{figure}
  \begin{center}
  \subfigure[][\label{fig_AN2:profil_phT}]{\includegraphics[width=.4\textwidth]{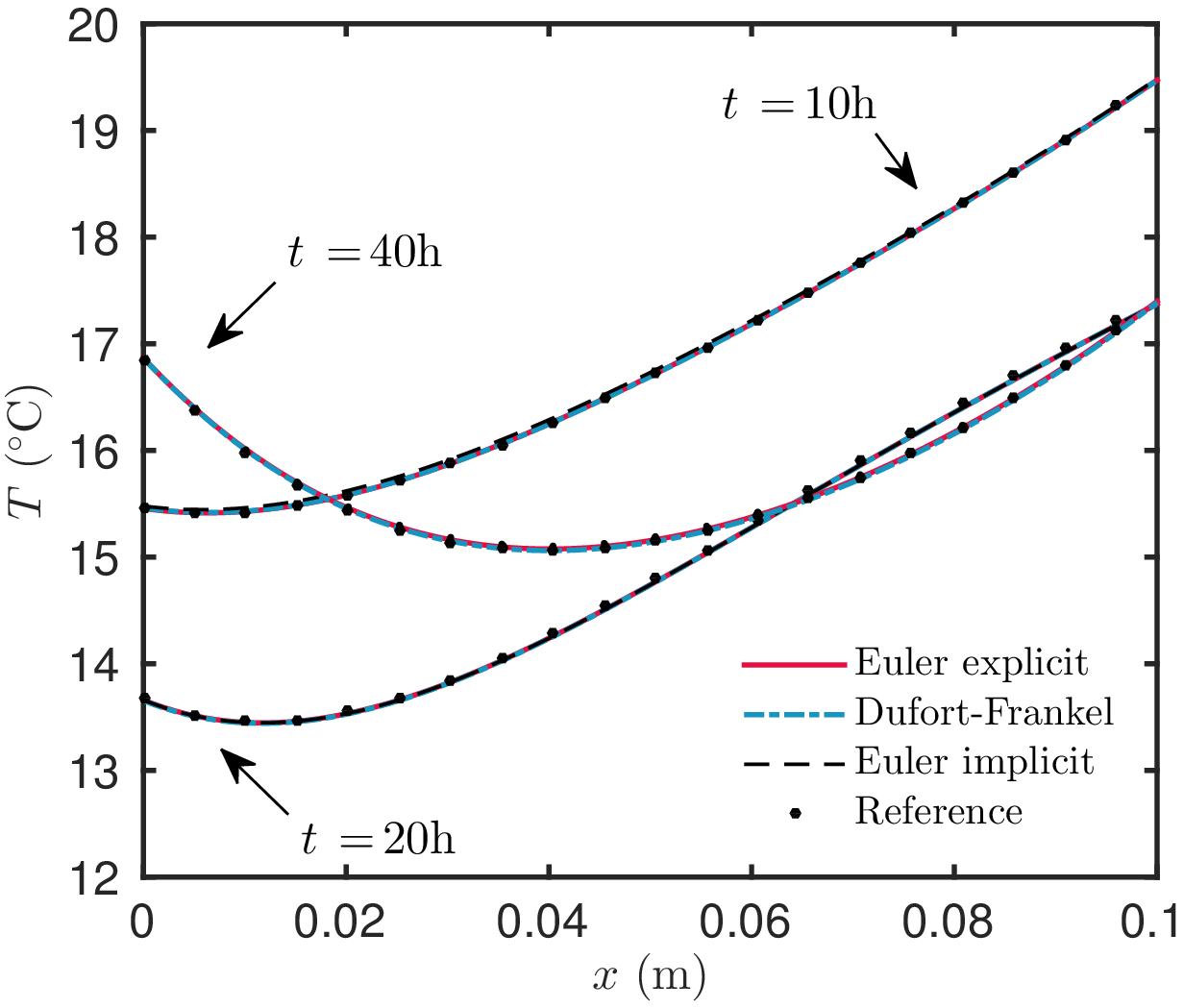}}
  \subfigure[][\label{fig_AN2:profil_phi}]{\includegraphics[width=.4\textwidth]{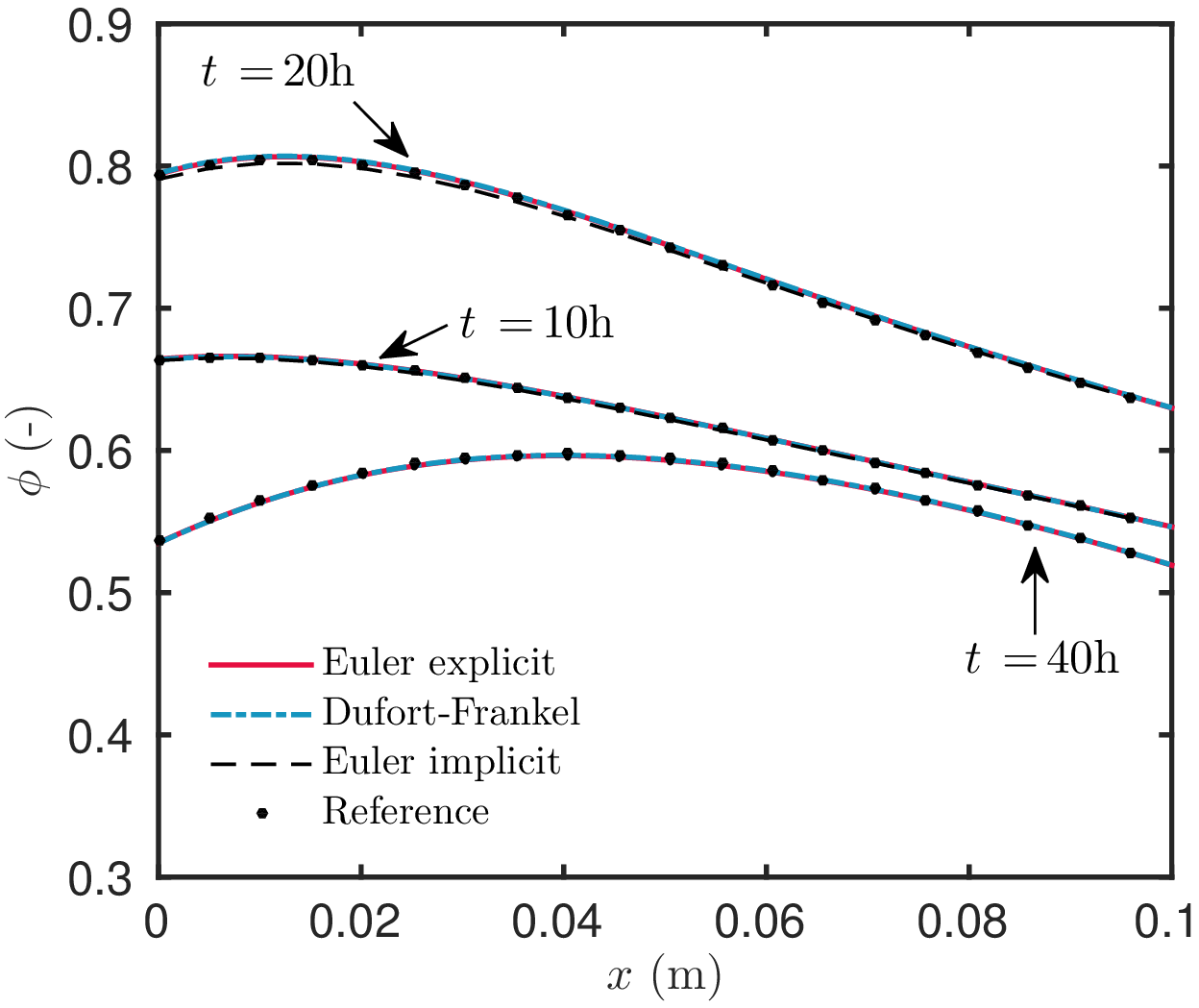}} 
  \caption{\small\em Temperature and relative humidity profiles at $t \egal 10 \ \mathsf{h}$, $t \egal 20 \ \mathsf{h}$  and $t \egal 40 \ \mathsf{h}\,$.}
  \label{fig_AN2:profils}
  \end{center}
\end{figure}

The time evolution of the fields in the middle of the material is presented in Figure~\ref{fig_AN2:times}, showing a good agreement between the solution of each numerical scheme. The coupling effect between heat and mass transfer can be specially noted in Figure~\ref{fig_AN2:time_T_L}. The temperature at $x \egal 0.05 \ \mathsf{m}$ varies according to both frequencies of the left side air temperature and relative humidity, illustrated in Figures~\ref{fig_AN1:BCL_Phi} and \ref{fig_AN1:BC_TL}. The relative humidity rises until $80 \ \mathsf{\%}$ at $t \egal 110 \ \mathsf{h}\,$, demonstrating that the material wont be saturated in the middle during the simulation. Figure~\ref{fig_AN2:profils} represents the field profiles at different time. We can observe that this material is non-hygroscopic but fairly permeable. It diffuses rapidly but do not retain the moisture.

\begin{figure}
  \begin{center}
  \subfigure[][\label{fig_AN2:errPhi_fx}]{\includegraphics[width=.4\textwidth]{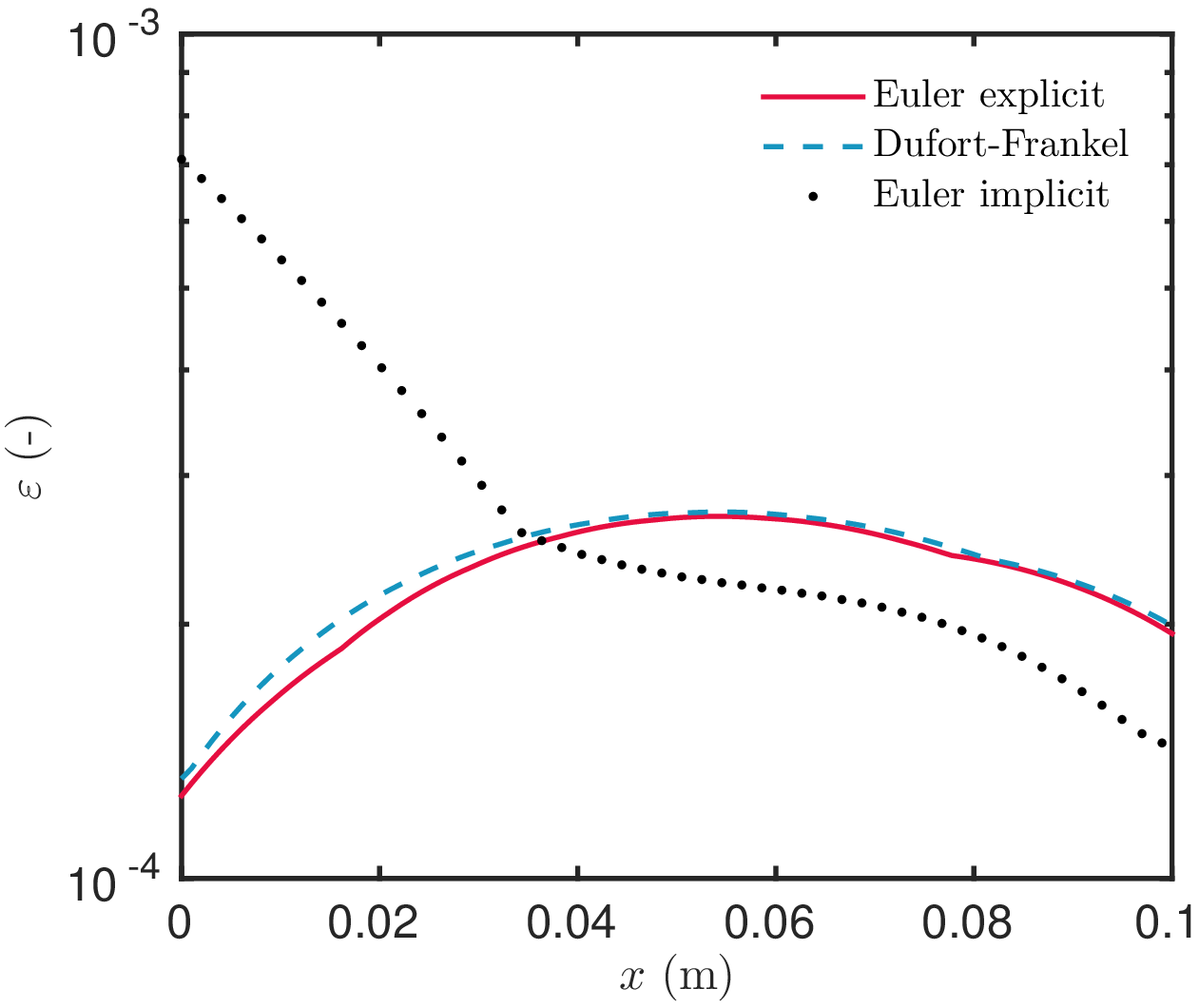}}
  \subfigure[][\label{fig_AN2:errT_fx}]{\includegraphics[width=.4\textwidth]{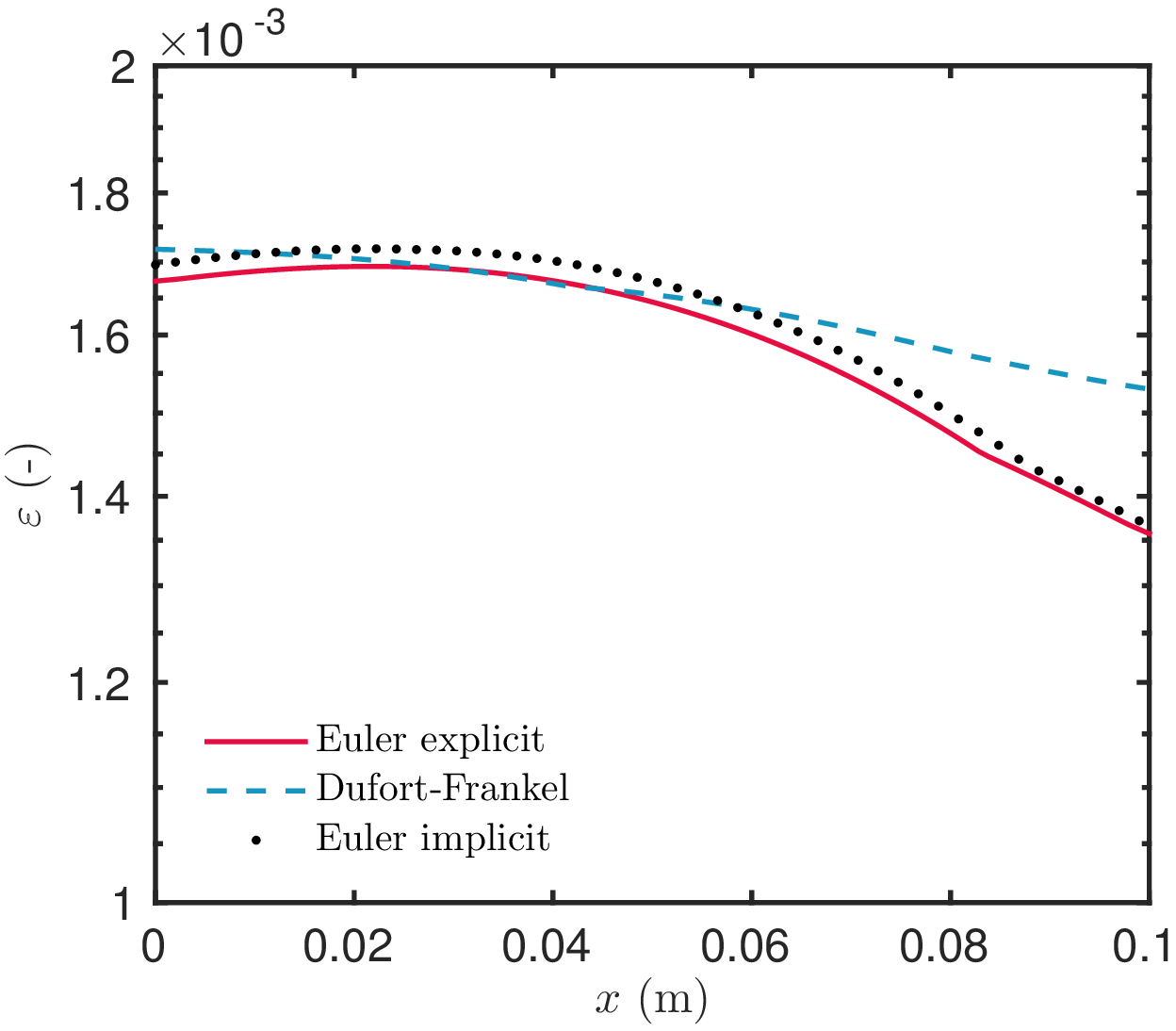}} 
  \caption{\small\em $\mathcal{L}_{\infty}$ error between the reference solution and the ones computed with the numerical schemes, for temperature (a) and relative humidity (b).}
  \label{fig_AN2:Erreur_fx}
  \end{center}
\end{figure}

The errors between the reference solution and the ones computed with the different numerical schemes are given in Figures~\ref{fig_AN2:errT_fx} and \ref{fig_AN2:errPhi_fx}. It confirms that all the numerical schemes enable to compute an accurate solution, at the order of $10^{\,-4}$ for both fields and considered spatial and temporal discretisations.

For the \Eu ~implicit and the \DF ~explicit schemes, the CPU time has been calculated, using \texttt{Matlab} platform on a computer with Intel i$7$ CPU and $32$GB of RAM, and shown in Table~\ref{tab:CPU_time}. The implicit scheme requires around $9$ sub-iterations per time step to treat the nonlinearities of the problem. The \DF ~approach computes directly the solution and therefore has a reduced computational costs, around $15\%$ of the \Eu ~implicit  scheme based algorithm.

\begin{figure}
  \begin{center}
  \subfigure[][\label{fig_AN2:errPhi_f_dt}]{\includegraphics[width=.4\textwidth]{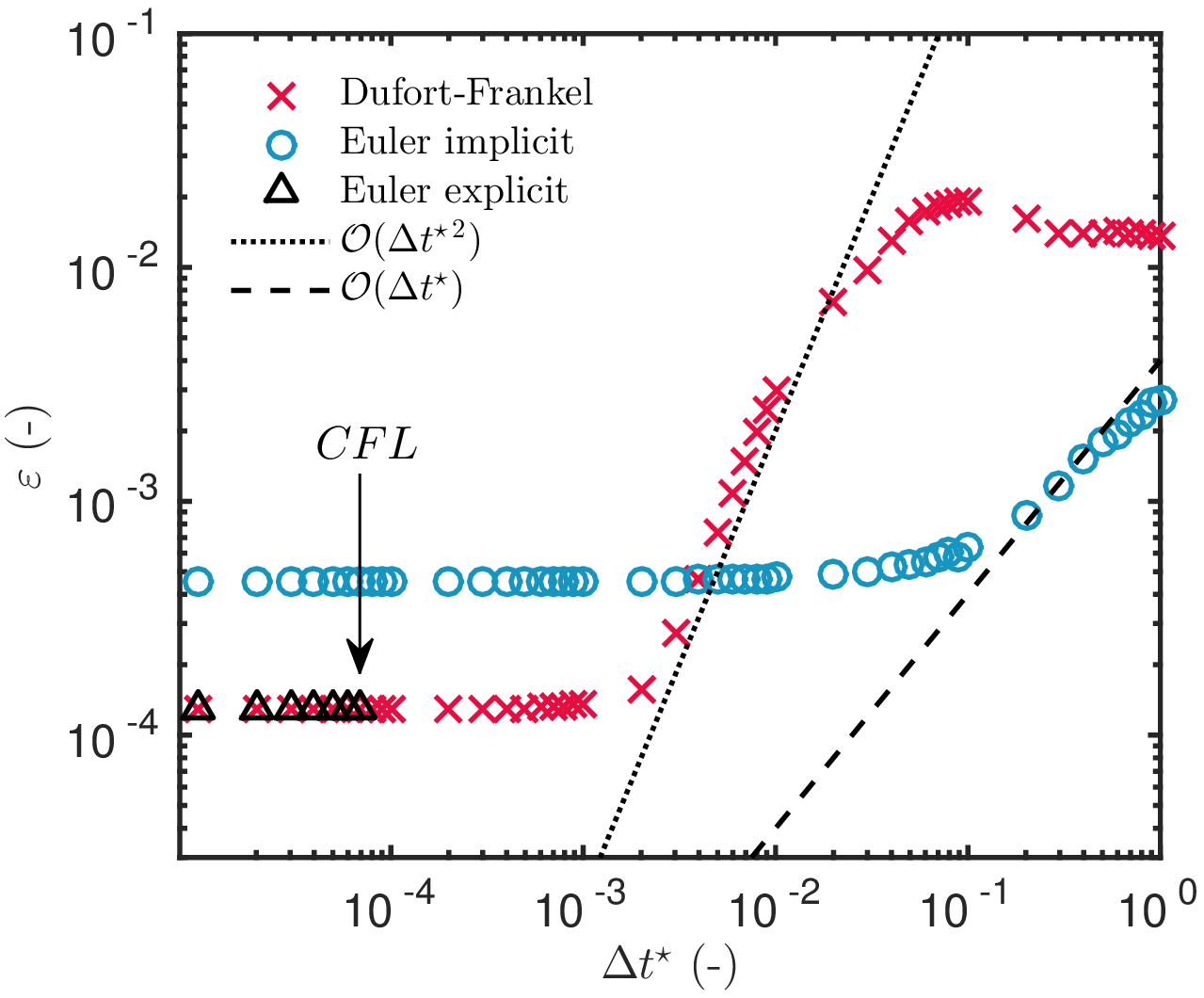}}
  \subfigure[][\label{fig_AN2:errT_f_dt}]{\includegraphics[width=.4\textwidth]{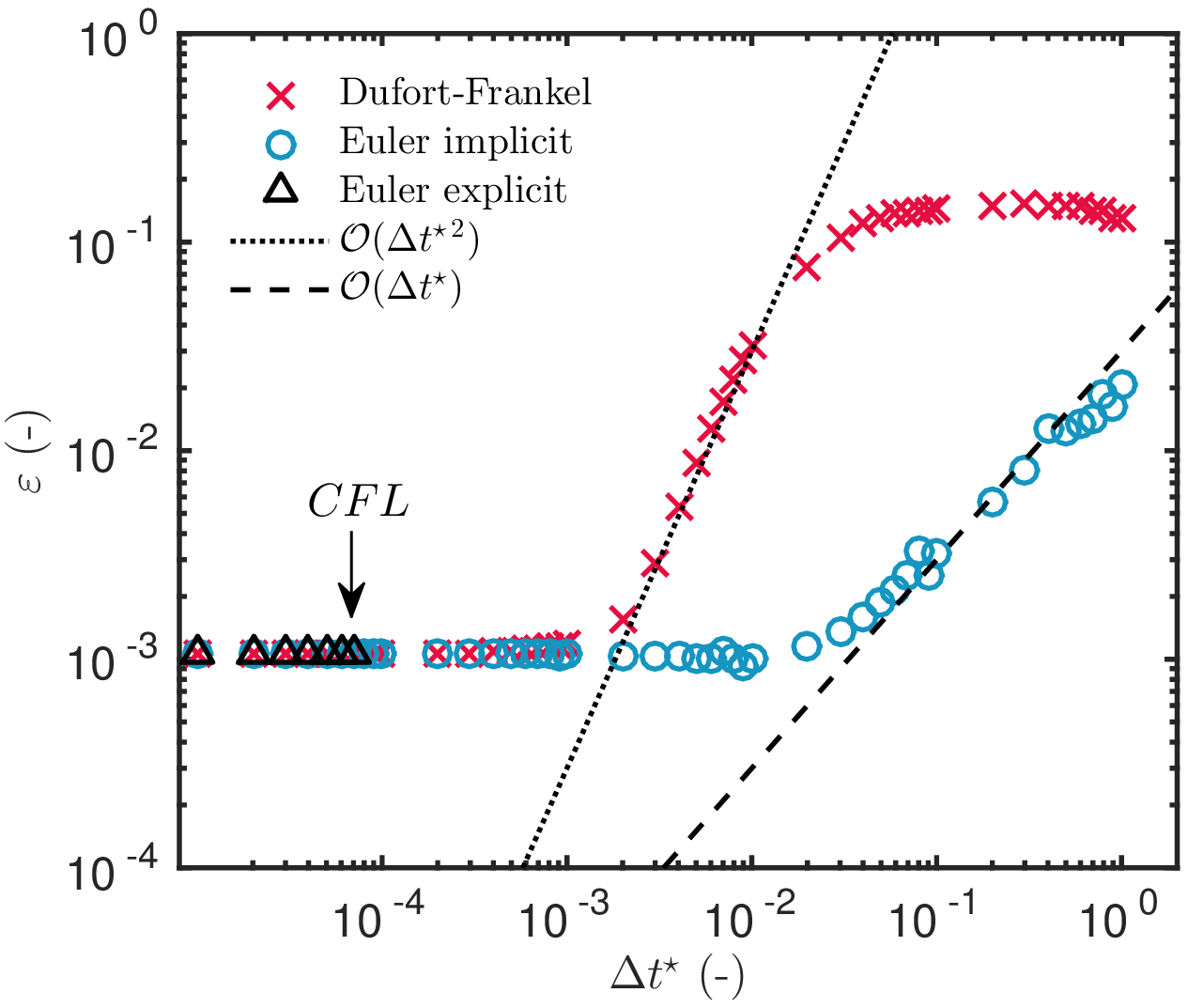}}
  \caption{\small\em $\mathcal{L}_{\infty}$ error between the reference and the computed solutions as a function of $\Delta \ts$ for temperature (a) and relative humidity (b).}
  \label{fig_AN2:err_f_dt}
  \end{center}
\end{figure}

A numerical analysis of the behaviour of the three numerical schemes has been carried out for different values of the temporal discretisation $\Delta \ts\,$. The spatial discretisation is maintained to $\Delta x^{\,\star} \egal 10^{\,-2}\,$. Results of the $\mathcal{L}_{\,\infty}$ error $\varepsilon$ are shown in Figure~\ref{fig_AN2:err_f_dt}. As expected, the \Eu ~explicit scheme was not able to compute a solution when the stability CFL condition is not respected (around $\dts \ \leqslant \ 8 \dix{-5}$). The values computed from Eq.~ \eqref{eq:cfl_heat_Eq} and \eqref{eq:cfl_moisture_Eq} are in accordance with the results from the convergence study. It also confirms that the \DF ~scheme is unconditionally stable, as it computes a solution for any discretisation parameter $\Delta t^{\,\star}\,$. The error of the numerical scheme is second-order accurate in time $\O\,(\Delta t^{\,2})\,$.

The choice of the discretisation parameters have strong influences on the solution computed. When analysing the error of the \DF ~scheme as a function of the time discretisation $\Delta \ts \,$, three regions have been highlighted. The first one corresponds to small discretisation parameter $\Delta \ts\, \leqslant\, 10^{\,-3}$, where the solution obtained with the \DF ~scheme reaches the constant error value which is lower than the \Eu ~implicit one. In this region, all schemes provide an accurate solution.  The second region is where the error is proportional to $\O\,(\Delta t^{\star \,2})$ but higher than the \Eu ~implicit one. The last region, when $\Delta \ts\, \geqslant \, 5 \cdot 10^{\,-2}\,$, includes the large time step discretisation. With this in mind, the first conclusion we can make is that until the second region the \DF ~scheme is more recommended, while in the third region, for bigger values of $\Delta \ts$ the \Eu ~implicit scheme is suggested. However, when the time discretisation $\Delta \ts$ is too large, both schemes do not succeed in representing the physical phenomena. As mentioned in \citep{Gasparin2017}, the time step has to be carefully chosen in accordance with the characteristic time of the physical phenomena. For the \DF ~scheme and the case studied, this condition is reached for time discretisation up to $180$-$\mathsf{s}\,$.

\section{Whole-building hygrothermal model}

\subsection{Coupling the lumped air multizone model with the transfer through porous walls}

The lumped multizone model divides the whole-building into $N_{\,z}$ perfectly mixed air zones. For each zone, the evolutions of the air temperature $\Ta$ and the humidity ratio $\wa$ are given by the equations \cite{Tariku2010, Berger2015b, Berger2016a}:
\begin{subequations}\label{eq:LAV_equation}
\begin{align}
\label{eq:LAV_equation_T}
\rho_{\,a} \ V \, \bigl(\, c_{\,p,a} \plus c_{\,p,v} \, \wa \, \bigr) \, \od{\Ta}{t} & \egal Q_{\,\mathrm{o}} \plus Q_{\,\mathrm{v}} \plus Q_{\,\mathrm{inz}} \plus \sum_{i=1}^{N_{\,w}} \, Q_{\,w,\,\,i} \,, \\[3pt]
\label{eq:LAV_equation_w}
\rho_{\,a} \ V \, \od{\wa}{t} & \egal G_{\,\mathrm{o}} \plus G_{\,\mathrm{v}} \plus G_{\,\mathrm{inz}} \plus \sum_{i=1}^{N_{\,w}} \, G_{\,w,\, i} \,.
\end{align}
\end{subequations}
Besides the dependence of the air heat capacity on the humidity ratio, the right-hand terms of the room air energy balance equation --- Eq.~\eqref{eq:LAV_equation_T} --- can also be strongly dependent on the humidity ratio, which may lead to a highly nonlinear problem.

Equation~\eqref{eq:LAV_equation_w} can be expressed using the vapour pressure of the air zone $\Pva\,$, by means of the following relation:
\begin{align*}
  \wa \egal \frac{M_{\,v}}{M_{\,a}} \, \frac{\Pva}{P_{\,a} \moins \Pva}\ \simeq\ \frac{\Pv}{\Pvref}\,,
\end{align*}
where $\Pvref \egal 1.61 \dix{5} \ \mathsf{Pa}\,$. In addition, according to the notation used before, we have: 
\begin{align*}
  & \caTTzero  \ \eqdef \  \frac{\rho_{\,a} \ V \ c_{\,p,a}}{\Pvref} \,, 
  && \caTTun \ \eqdef \ \frac{\rho_{\,a} \ V c_{\,p,v}}{\Pvref} \,, \\[3pt]
  & \caTT \ \eqdef \ \caTTzero \plus \caTTun \ \Pv \,, 
  && \caM \ \eqdef \ \rho_{\,a} \ V \,.
\end{align*}

The terms $G$ and $Q$ are associated to occupants and their activities, denoted with the subscript $o\,$, as well as air flow from ventilation systems or infiltration, denoted with the subscript $v\,$. The sources due to interzone airflow are designated by $Q_{\,\mathrm{inz}}$ and $G_{\,\mathrm{inz}}\,$. Interested readers may find a detailed list of these sources and their physical description in \citep{DosSantos2006}. The first can be expressed as:
\begin{align*}
  & Q_{\,\mathrm{o}} \egal \Lv \, g_{\,\mathrm{o}}(\,t\,)\,, 
  && G_{\,\mathrm{o}} \egal g_{\,\mathrm{o}}(\,t\,) \,,
\end{align*}
where $g_{\,\mathrm{o}}$ is the time variant vapour production. The sources due to ventilation system are written as:
\begin{align*}
  Q_{\,\mathrm{v}} & \egal g_{\,\mathrm{v}} \ \bigl(\, \cw \, \Tinf \moins \cw \, \Ta \,\bigr) \plus \Lv \, g_{\,\mathrm{v}} \ \bigl(\, w_{\,\infty} \moins \wa \,\bigr) \,, \\
  G_{\,\mathrm{v}} & \egal g_{\,\mathrm{v}} \ \bigl(\, w_{\,\infty} \moins \wa \,\bigr)  \,,
\end{align*}
where $g_{\,\mathrm{v}}$ stands for the air flow rate due to ventilation and infiltration, $\Tinf$ and $w_{\,\infty}$ are the outside temperature and humidity ratio, and $\cw$ the water heat capacity depending on the air temperature and humidity ratio. In a similar way, the sources due to an airflow $g_{\,\mathrm{inz}}$ between zones $i$ and $j$ are given by:
\begin{align*}
  Q_{\,\mathrm{inz}} & \egal g_{\,\mathrm{inz}} \ \bigl(\, c_{\,w\,j} \, T_{\,j} \moins c_{\,w\,i} \, T_{\,i}  \,\bigr) \plus \Lv \, g_{\,\mathrm{inz}} \ \bigl(\, w_{\,j} \moins w_{\,i} \,\bigr) \,, \\
  G_{\,\mathrm{inz}} & \egal g_{\,\mathrm{inz}} \ \bigl(\,  w_{\,j} \moins w_{\,i} \,\bigr)  \,.
\end{align*}
Here, a mean value of the latent heat of evaporation $\Lv\,$, evaluated at the considered temperatures, is used for numerical application. The terms $Q_{\,w}$ and $G_{\,w}$ represent the mass and heat quantities exchanged between the air room and the the $N_{\,w}$ bounding walls:
\begin{subequations}\label{eq:LAV_couplage}
\begin{align}
  Q_{\,w} & \egal \hT \, A \, \bigl(\, \Ts - \Ta \,\bigr) \plus \Lv \, \hM \, A \, \bigl(\, \Pvs - \Pva \,\bigr) \,, \\[3pt]
  G_{\,w} & \egal \hM \, A \, \bigl(\, \Pvs - \Pva \,\bigr) \,,
\end{align}
\end{subequations}
where $A$ is the wall surface area. The terms $\Ts$ and $\Pvs$ are the wall surface temperature and vapour pressure. Considering Eq.~\eqref{eq:HAM_equation2}, they correspond to $\Ts \egal T\,(x\ =\ 0)$ and $\Pvs \egal \Pv\,(x\ =\ 0)$ (with the adopted convention). Thus, the coupling between the wall model Eq.~\eqref{eq:HAM_equation2} and the lumped multizone model Eq.~\eqref{eq:LAV_equation} is operated by the sources $G_{\,w}$ and $Q_{\,w}$ from Eq.~\eqref{eq:LAV_couplage}.

The temperature and vapour pressure in the zone are initially at $T \egal \Ti$ and $\Pv \egal \Pvi\,$. To write a dimensionless formulation of Eq.~\eqref{eq:LAV_equation}, the following quantities are defined: 
\begin{align*}
  & \ua \ \eqdef \ \frac{\Ta}{\Ti} \,,
  && \va \ \eqdef \ \frac{\Pva}{\Pvi} \,,
  && \caTTuns \ \eqdef \ \frac{\caTTun}{\caTTzero} \, \Pvi \,,\\[3pt]
  & g_{\,o}^{\,\star} \ \eqdef \ \frac{\tref}{\Pvi \cdot \caM} \, g_{\,o}\,,
  && q_{\,o}^{\,\star} \ \eqdef \ \frac{\tref}{\Ti \cdot \caTTref} \, \Lv \, g_{\,o} \,, 
  && g_{\,v}^{\,\star} \ \eqdef \ \frac{g_{\,v} \, \tref}{\Pvref \, \caM} \,, \\[3pt]
  & q_{\,v,\,1}^{\,\star} \ \eqdef \ \frac{\cw \, g_{\,v} \, \tref \, \Pvi}{\Pvref \ \caTTref} \,,
  && q_{\,v,\,2}^{\,\star} \ \eqdef \ \frac{\Lv \, g_{\,v} \, \tref \, \Pvi}{\Pvref \ \caTTref \ \Ti} \,,
  && g_{\,inz}^{\,\star} \ \eqdef \ \frac{g_{\,inz} \, \tref}{\Pvref \, \caM} \,, \\[3pt]
  & q_{\,inz,\,1}^{\,\star} \ \eqdef \ \frac{\cw \, g_{\,inz} \, \tref \, \Pvi}{\Pvref \ \caTTref} \,,
  && q_{\,inz,\,2}^{\,\star} \ \eqdef \ \frac{\Lv \, g_{\,inz} \, \tref \, \Pvi}{\Pvref \ \caTTref \ \Ti} \,,
  &&\thetaT \ \eqdef \ \frac{\kTTref \cdot A \cdot \tref}{L \cdot \caTTref} \,, \\[3pt]
  & \thetaM \ \eqdef \ \frac{\kMref \cdot A \cdot \tref}{L \cdot \caM} \,.
\end{align*}
Thus, the dimensionless formulation of Eq.~\eqref{eq:LAV_equation} can be written as:
\begin{subequations}\label{eq:LAV_equation_dimless}
  \begin{align}
  \bigl(\, 1 \plus \caTTuns \, \bigr) \od{\ua}{\ts} & \egal q_{\,o}^{\,\star}
  \plus q_{\,v,\,1}^{\,\star} \, \bigl(\, \uinf \, \vinf \moins \ua \, \va \,\bigr)
  \plus q_{\,v,\,2}^{\,\star} \, \bigl(\, \uinf \moins \ua \,\bigr)  \nonumber  \\ 
  & \plus q_{\,inz,\,1}^{\,\star} \, \bigl(\, u_{\,a,2} \, v_{\,a,2} \moins \ua \, \va \,\bigr)
  \plus q_{\,inz,\,2}^{\,\star} \, \bigl(\, u_{\,a,2} \moins \ua \,\bigr) \nonumber \\ 
  & \plus \sum_{i=1}^{N_{\,w}} \, \BiTTi \, \thetaTi \, \bigl(\, u_{\,i}- \ua \,\bigr) \plus \BiTMi \, \thetaTi \, \bigl(\, v_{\,i} - \va \,\bigr)   \,, \\[3pt]
  \od{\va}{\ts} & \egal g_{\,o}^{\,\star} 
  \plus  g_{\,v}^{\,\star} \, \bigl(\, \vinf \moins \va \,\bigr) 
  \plus  g_{\,inz}^{\,\star} \, \bigl(\, v_{\,a,2} \moins \va \,\bigr) \nonumber \\ 
  & \plus \sum_{i=1}^{N_{\,w}} \, \BiMi \, \thetaMi  \, \bigl(\, \va \moins v_{\,i} \, \bigr)\,.
  \end{align}
\end{subequations}
The quantities $\uinf\,$, $\vinf$ and $u_{\,a,\,2}\,$, $v_{\,a,\,2}$ come from building outside (provided by weather data) and from the adjacent zone. The coupling between the wall model Eq.~\eqref{eq:HAM_equation_dimless} and the lumped multizone model Eq.~\eqref{eq:LAV_equation_dimless} is operated through the wall source terms. The dimensionless parameters $\BiTT\,$, $\BiTM$ and $\BiM$ qualify the penetration of the heat and moisture through the wall according to the physical mechanism. The parameters $\thetaTi$ and $\thetaMi$ depend on the wall surface on the room air, providing the weighted contribution of the wall to the energy and moisture balances.


\subsection{Implicit scheme for the whole-building energy simulation: problem statement}

For the sake of simplicity and without loosing the generality, the coupling procedure is explained considering \emph{only} the linear heat diffusion equation Eq.~\eqref{eq:heat1d} for one wall model with the  following boundary conditions for the surface in contact with the outside ($x \egal 0 $) and inside($x \egal 1$) air of the building:
\begin{subequations}
  \begin{align}\label{eq:wall_model_heatBC_L}
  & \pd{u}{x} \egal \, \BiTT \, \bigl(\, u \moins \uinf \, \bigr)\,, && x \egal 0 \,, \\ 
  \label{eq:wall_model_heatBC_R}
  & \pd{u}{x} \egal - \, \BiTT \, \bigl(\, u \moins \ua \, \bigr)\,, && x \egal 1 \,.
\end{align}
\end{subequations}
The room air energy conservation equation for the multizone model is expressed as: 
\begin{align}\label{eq:multizone_model_heat}
  & \frac{\mathrm{d} \ua}{\mathrm{d}t} \egal Q \plus \BiTT \, \Theta_{\,T} \, \bigl(\, u \moins \ua \, \bigr) \,,
\end{align}

Many whole-building models reported in the literature, such as MATCH \citep{Rode2003}, MOIST \citep{Burch1993} and DOMUS \citep{Mendes1997, Mendes1999,DosSantos2004}, use implicit (\Eu ~or \CN) approaches to solve these equations, mainly due to the unconditional stability property. Thus, using an \Eu ~implicit approach, the discretisations of Eqs.~\eqref{eq:heat1d}, \eqref{eq:wall_model_heatBC_L}, \eqref{eq:wall_model_heatBC_R} yield to:
\begin{align*}
  \frac{1}{\Delta t} \, \bigl(\, u^{\,n+1}_{\,j} \moins u^{\,n}_{\,j} \, \bigr) 
  &  \egal \frac{\nu}{\Delta x^2} \, \bigl(\, u^{\,n+1}_{\,j+1} \moins 2 \ u^{\,n+1}_{\,j} \plus u^{\,n+1}_{\,j-1} \, \bigr) \,,\\
   \frac{1}{\Delta x} \, \bigl( \, u^{\,n+1}_{\,1} - u^{\,n+1}_{\,0} \, \bigr) 
  & \egal \BiTT \, \bigl(\, u^{\,n+1}_{\,0} \moins \uinf^{\,n+1} \, \bigr) \,,\\
  \frac{1}{\Delta x} \, \bigl( \, u^{\,n+1}_{\,N} - u^{\,n+1}_{\,N-1} \, \bigr) 
  & \egal - \ \BiTT \, \bigl(\, u^{\,n+1}_{\,N} \moins \ua^{\,n+1} \, \bigr) \,.
\end{align*}
Therefore, at iteration $t^{\,n}\,$, solution $u^{\,n\,+\,1}\,$, of the wall model, is computed using the resolvent operator $\mathcal{R}_{\,\mathrm{imp}}^{\,\mathrm{w}}$ written as: 
\begin{align*}
  u^{\,n+1} \egal \mathcal{R}_{\,\mathrm{imp}}^{\,\mathrm{w}} \ \bigl(\, u^{\,n}, \, \ua^{\,n+1}, \, \uinf^{\,n+1} \,\bigr) \,,
\end{align*}
In the same way, the discretisation of Eq.~\eqref{eq:multizone_model_heat} yields to: 
\begin{align*}
  \frac{1}{\Delta t} \, \bigl(\, u_{\,a}^{\,n+1}\moins u^{\,n}_{\,a} \, \bigr) 
  & \egal Q^{\,n+1} \plus \BiTT \, \Theta_{\,T} \, \bigl(\, u^{\,n+1}_{N} \moins u_{\,a}^{\,n+1} \, \bigr) \,,
\end{align*} 
and, at iteration $t^{\,n}\,$, solution $\ua^{\,n\,+\,1}\,$, of the multizone model, is computed using the resolvent operator $\mathcal{R}_{\,\mathrm{imp}}^{\,\mathrm{a}}$ written as: 
\begin{align*}
  \ua^{\,n+1} \egal \mathcal{R}_{\,\mathrm{imp}}^{\,\mathrm{a}} \ \bigl(\, \ua^{\,n}, \, u^{\,n+1} \,\bigr) \,.
\end{align*}

By coupling the wall and the multizone models to perform a whole-building energy simulation, ones must solve at each time iteration the following system of equations:
\begin{subequations}\label{eq:wbse_tn_imp}
  \begin{align}
  u^{\,n+1} &\egal \mathcal{R}_{\,\mathrm{imp}}^{\,\mathrm{w}} \ \bigl(\, u^{\,n}, \, \ua^{\,n+1}, \, \uinf^{\,n+1} \,\bigr) \,, \\
  \ua^{\,n+1} & \egal \mathcal{R}_{\,\mathrm{imp}}^{\,\mathrm{a}} \ \bigl(\, \ua^{\,n}, \, u^{\,n+1} \,\bigr) \,.
  \end{align}
\end{subequations}
As the system Eq.~\eqref{eq:wbse_tn_imp} is nonlinear, it is not possible to solve it directly. Thus, $\bigl(\,u^{\,n\,+\,1},\, \ua^{\,n+1}\,\bigr)$ are computed using, for instance, a fixed point algorithm, until reaching a prescribed tolerance $\eta\,$, as illustrated in Alg.~\ref{alg:Fixed_point_wbse}. The number of subiterations should increase with the number of walls considered in the whole-building model, when dealing with nonlinear models and when considering interzone airflows. Interested readers may report to \cite{DosSantos2004} for more details on this problem statement. Indeed, authors investigate the influence of the time step on the computational time of a whole-building simulation model, considering only heat transfer, for different numerical schemes and analytical solution.

\IncMargin{1em}
\begin{algorithm}
  \SetKwData{Left}{left}\SetKwData{This}{this}\SetKwData{Up}{up}
  \SetKwFunction{Union}{Union}\SetKwFunction{FindCompress}{FindCompress}
  \SetKwInOut{Input}{input}\SetKwInOut{Output}{output}
  $ u^{\,k} \egal u^{\,n} $ \;
  $\ua^{\,k} \egal u^{\,n} $ \;
  \While{ $\big|\big| \, u^{\,k+1} \moins u^{\,k}, \, \ua^{\,k+1} \moins \ua^{\,k} \, \big|\big| \geqslant \eta$}%
  {
  $u^{\,k+1} \egal \mathcal{R}_{\,\mathrm{imp}}^{\,\mathrm{w}} \ \bigl(\, u^{\,n}, \, \ua^{\,k}, \, \uinf^{\,n+1} \,\bigr)$ \;
  $ \ua^{\,k+1} \egal \mathcal{R}_{\,\mathrm{imp}}^{\,\mathrm{a}} \ \bigl(\, \ua^{\,n}, \, u^{\,k+1} \,\bigr) $\;%
  $u^{\,k} \egal u^{\,k+1}$ \;
  $ \ua^{\,k} \egal \ua^{\,k+1} $\;%
  }
  $u^{\,n+1} \egal u^{\,k+1}$ \;
  $ \ua^{\,n+1} \egal \ua^{\,k+1} $\;%
  \bigskip 
  \caption{\small\em Fixed point algorithm to compute $\bigl(\, u^{\,n+1},\, \ua^{\,n+1} \,\bigr)\,$, using implicit numerical schemes, within the framework of whole-building simulation energy.}
  \label{alg:Fixed_point_wbse}
\end{algorithm}
\DecMargin{1em}


\subsection{Improved explicit schemes for the whole-building energy simulations}

Using the improved \DF ~explicit scheme, results from Section~\ref{sec:DF_scheme} have shown that the discretisation of Eqs.~\eqref{eq:heat1d}, \eqref{eq:wall_model_heatBC_L}, \eqref{eq:wall_model_heatBC_R} is given by: 
\begin{align*}
  u_{\,j}^{\,n+1}\ =\ \frac{1\ -\ \lambda}{1\ +\ \lambda}\;u_{\,j}^{\,n-1}\ +\ \frac{\lambda}{1\ +\ \lambda}\;\bigl(u_{\,j+1}^{\,n}\ +\ u_{\,j-1}^{\,n}\bigr) \,.
\end{align*}
Moreover, to solve Eq.~\eqref{eq:multizone_model_heat}, we get the scheme: 
\begin{align*}
  \frac{1}{\Delta t} \, \bigl(\, u_{\,a}^{\,n+1}\moins u^{\,n}_{\,a} \, \bigr)  
  & \egal Q^{\,n} \plus \BiTT \, \Theta_{\,T} \, \bigl(\, u^{\,n}_{N} \moins u_{\,a}^{\,n} \, \bigr) \,.
\end{align*} 
Therefore, by coupling the wall and the multizone models, at each time iteration, the solution of the equations is directly computed: 
\begin{subequations}\label{eq:wbse_tn_exp}
  \begin{align}
  u^{\,n+1} &\egal \mathcal{R}_{\,\mathrm{exp}}^{\,\mathrm{w}} \ \bigl(\, u^{\,n}, \, \ua^{\,n}, \, \uinf^{\,n} \,\bigr) \,, \\
  \ua^{\,n+1} & \egal \mathcal{R}_{\,\mathrm{exp}}^{\,\mathrm{a}} \ \bigl(\, \ua^{\,n}, \, u^{\,n} \,\bigr) \,.
\end{align}
\end{subequations}

Thus, the use of improved explicit schemes, such as the \DF ~one, is particularly advantageous to avoid subiterations at each time step to compute the dependent variable fields. In addition, the numerical property of being unconditionally stable avoids any limitation on the values of the time step $\Delta t \,$, which is chosen only according to the characteristic time of the physical phenomena \citep{Gasparin2017}.


\section{Numerical application: multizone transfer}

\subsection{One-zone case study}
\label{sec:AN3}

\begin{figure}
\begin{center}
\includegraphics[width=.5\textwidth]{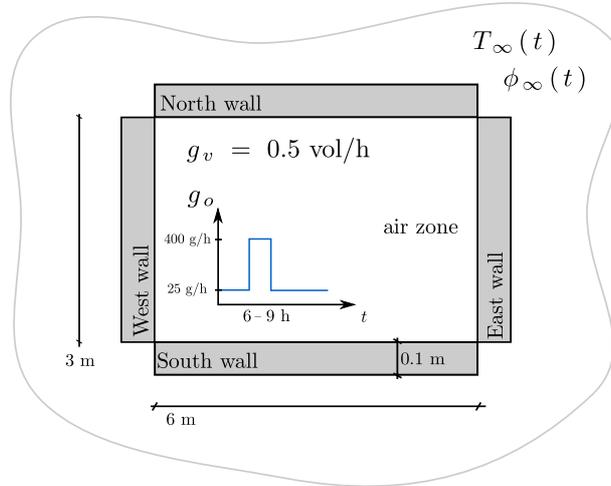}
\caption{\small\em Illustration of the one-zone case study.}
\label{fig_AN3:schema}
\end{center}
\end{figure}

\begin{figure}
\begin{center}
\subfigure[][\label{fig_AN3:BCL_T}]{\includegraphics[width=.4\textwidth]{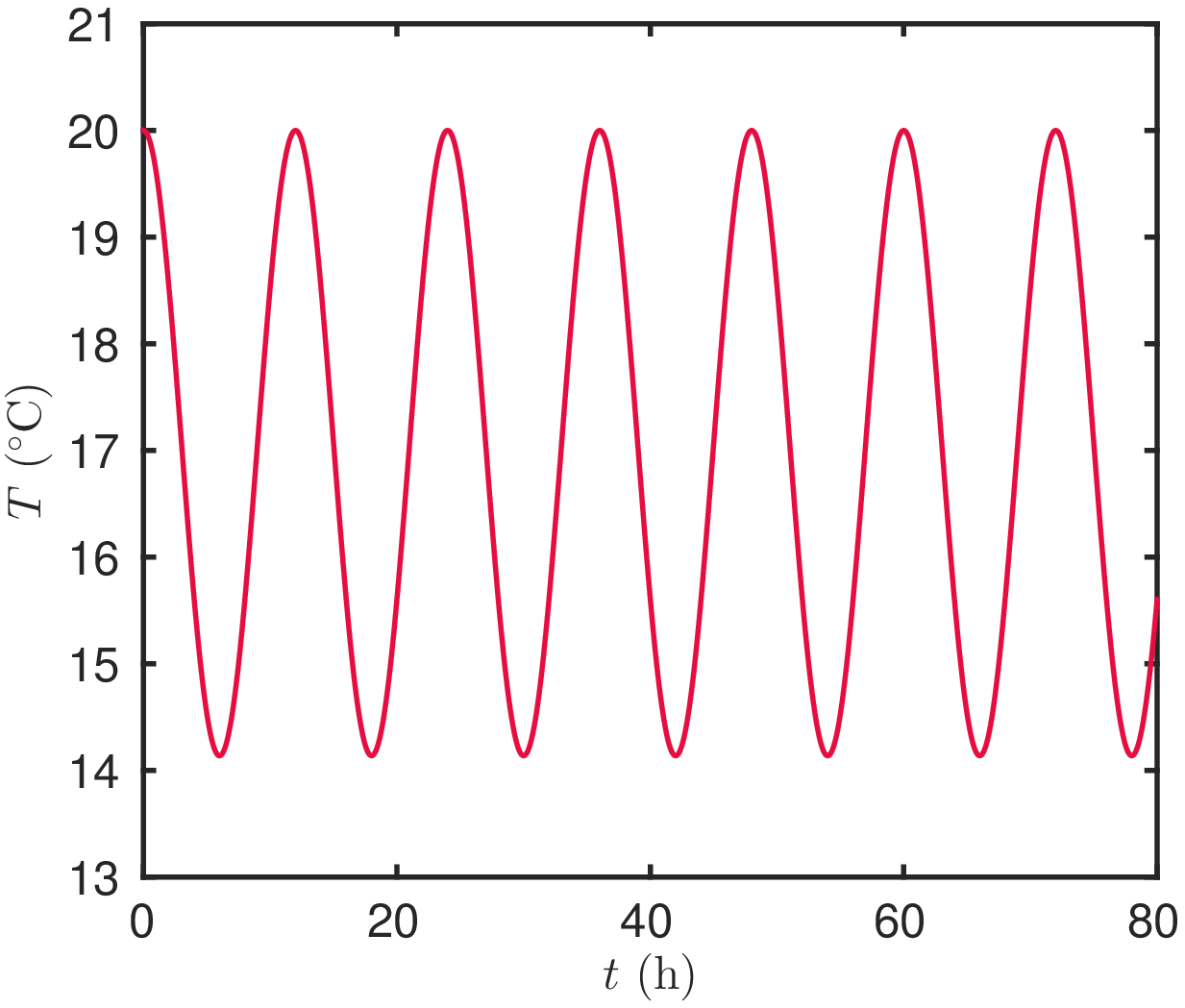}}
\subfigure[][\label{fig_AN3:BCL_Phi}]{\includegraphics[width=.4\textwidth]{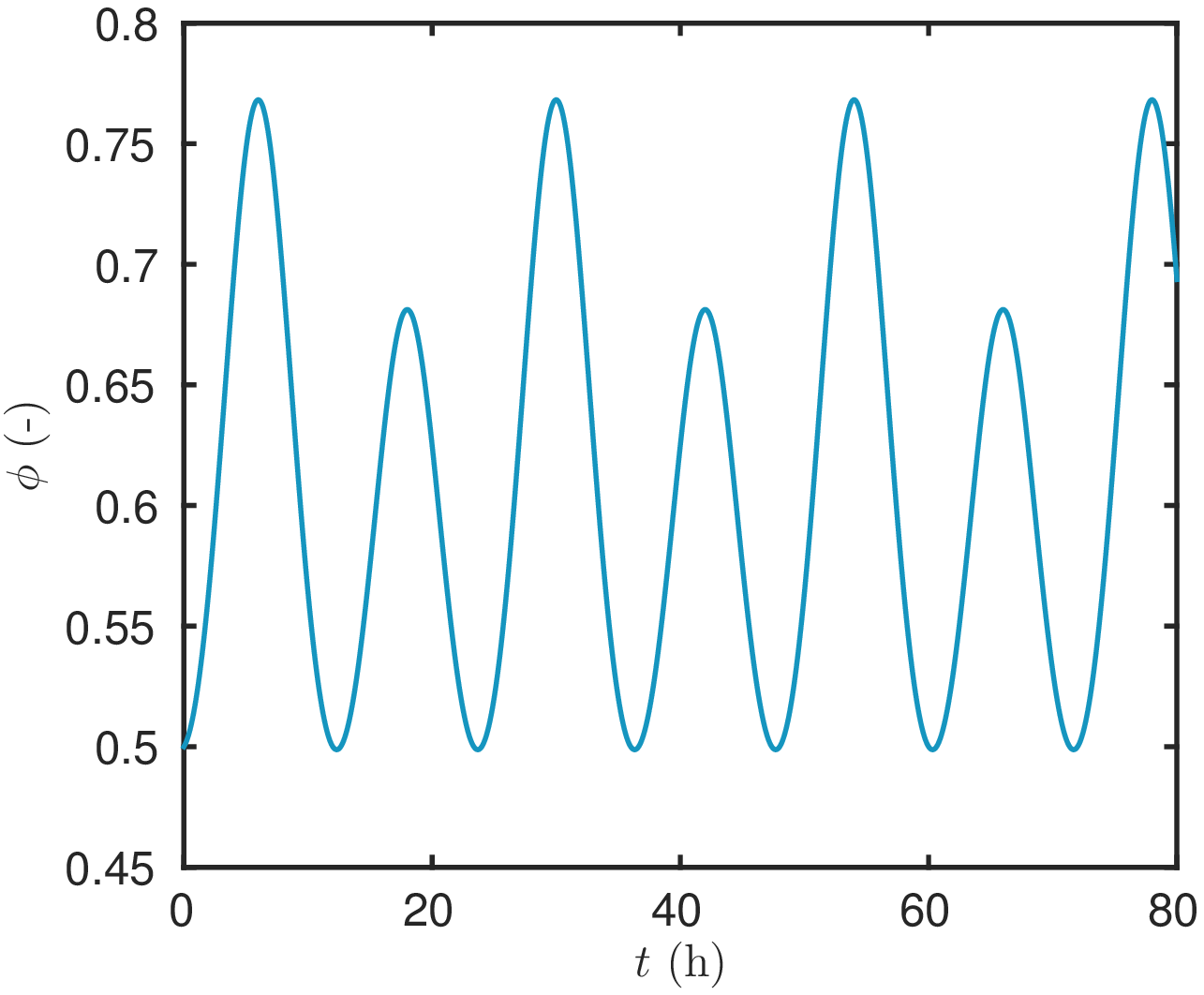}} 
\caption{\small\em Boundary conditions $\Tinf$ (a) and $\phi_{\,\infty}$ (b) for the one-zone case study.}
\label{fig_AN3:BC}
\end{center}
\end{figure}

A single zone, surrounded by four walls is considered, as illustrated in Figure~\ref{fig_AN3:schema}. All the walls have $0.1\ \mathsf{m}$ of length constituted by the same material, the load bearing material, whose properties are given in Table~\ref{table:properties_mat1}. Each wall is consider to have linear transfer and their coefficients are computed as follows: the \textit{north} wall with $23^{\,\circ}\mathsf{C}$ and $15\ \%\,$, the \textit{south} wall with $23^{\,\circ}\mathsf{C}$ and $90\ \%$ and the \textit{east} and \textit{west} walls with $23^{\,\circ}\mathsf{C}$ and $50\ \%\,$. The values of the linear coefficients are shown in Table~\ref{tab:wall_properties}. Their dimensionless values are provided in \ref{sec:appendix}. The convective mass and heat transfer coefficients are set to $\hM \egal 3 \cdot 10^{\,-8} \ \unitfrac{s}{m}$ and $\hT \egal 8 \ \unitfrac{W}{(m^{\,2}\,.\,K)}$ between the inside air zone and the walls. It is assumed that no transfer occurs through the ceiling and the floor. The zone has a floor area of $18 \ \mathsf{m^{\,2}}$ and a volume of $54 \ \mathsf{m^{\,3}}\,$. The room is subjected to a constant moisture load of $25 \ \mathsf{g/h}$ with an increase of $400 \ \mathsf{g/h}$ from $6 \ \mathsf{h}$ to $9 \ \mathsf{h}$ each day. External air enters the room through ventilation and infiltrations at a constant rate of $g_{\,v} \egal 0.5 \ \mathsf{h^{\,-1}}\,$. There are no radiative heat exchange among the walls. The initial temperature and relative humidity in the zone are $\Ti \egal 20 \ \mathsf{^{\,\circ}C}$ and $\phi_{\,i} \egal 50 \ \mathsf{\%}\,$, respectively. The outside boundary conditions $T_{\,\infty}$ and $\phi_{\,\infty}$ are given in Figures~\ref{fig_AN3:BCL_T} and \ref{fig_AN3:BCL_Phi}. The convective transfer coefficients between the walls and the outside conditions are also provided in Table~\ref{tab:wall_properties}. For this case, neither radiation or rain sources are considered for the boundary conditions.

\begin{table}
\centering
\caption{\small\em Parameter values for the one zone case.}\bigskip
\renewcommand*{\arraystretch}{.9}
\begin{tabular}{@{}>{\centering}m{.2\textwidth}
>{\centering}m{.2\textwidth}
>{\centering}m{.2\textwidth}
m{.2\textwidth}}
\hline
\hline
\textit{Parameter} & \textit{North Wall} & \textit{South Wall} & \textit{East and West walls} \\
\hline
\hline
$\cM \ [\,\unitfrac{s^{\,2}}{m^{\,2}} \,]$ &  $1.82 \dix{-2}$ & $1.18 \dix{-1}$ & $6.09 \dix{-2}$ \\
$\kM \ [\,\mathsf{s} \,]$ & $5.89 \dix{-9}$ & $2.92 \dix{-8}$ & $5.47 \dix{-9}$ \\
$\cTT \ \bigl[\,\unitfrac{W  \, . \,  s}{(K \, . \, m^{\,3})} \,\bigr]$ & $7.7 \dix{5} $ & $1.28 \dix{6}$ & $8.61 \dix{5}$ \\
$\kTT \bigl[\,\unitfrac{W}{(m \, . \, K)} \,\bigr]$ & $2.94 \dix{-1}$ & $8.41 \dix{-1}$ & $3.87 \dix{-1}$ \\ 
$\cTM \ \bigl[\,\unitfrac{W \, . \, s^{\,3}}{(kg \, . \, m^{\,2})} \,\bigr]$ & $1.52 \dix{3}$ & $9.88 \dix{3}$ & $5.09 \dix{3}$ \\ 
$\kTM \ \bigl[\,\unitfrac{m^{\,2}}{s}\,]$ & $1.59 \dix{-2}$ & $2.96 \dix{-3}$ & $1.53 \dix{-2}$\\ 
$A \ [\,\mathsf{m^{\,2}} \,]$ & $18$ & $18$ & $9$ \\ 
$\hT \ \bigl[\,\unitfrac{W}{(m^{\,2}\,.\,K)} \,\bigr]$ & $5$ & $25$ & $12$\\ 
$\hM \ [\,\unitfrac{s}{m} \,]$  & $2 \dix{-7}$ & $8 \dix{-7}$ & $4 \dix{-7}$\\
\hline
\hline
\end{tabular}
\label{tab:wall_properties}
\end{table}

The whole simulation is performed for $80 \ \mathsf{h}\,$. The walls and zone fields are computed using the \Eu ~implicit and \DF ~explicit schemes. The \Eu ~explicit scheme has  a stability restrictions imposing a very small time discretisation ($\Delta t^{\,\star} \ \leqslant \ 10^{\,-5}$ for the given space discretisation) and therefore an important computer run time. Moreover, to our knowledge, this scheme is not commonly used in building simulation programs. Thus, it was not used in this case study. The space and time discretisation parameters are $\Delta x^{\,\star} \egal 10^{\,-2}$ and $\Delta t^{\,\star} \egal 10^{\,-3}\,$, corresponding, from a physical point of view, to $\Delta x \egal 10^{\,-3} \ \mathsf{m}$ and $\Delta t \egal 3.6 \ \mathsf{s}\,$. Each solution is compared to a reference solution computed using the \texttt{Chebfun} open-source package for \texttt{Matlab} as explained in Section~\ref{sec:validat}.

\begin{figure}
\begin{center}
  \subfigure[][\label{fig_AN3:time_T1}]{\includegraphics[width=.4\textwidth]{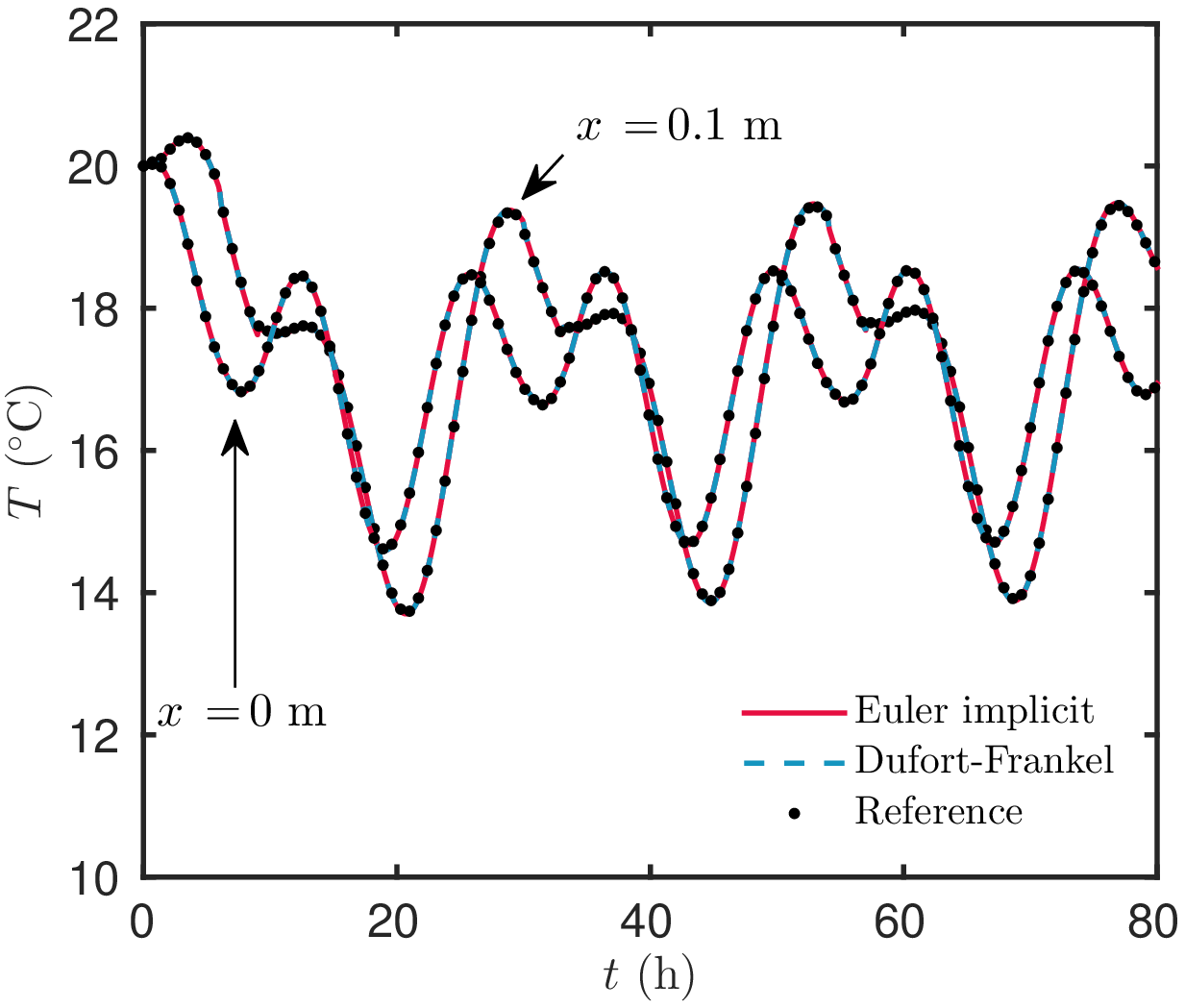}}
  \subfigure[][\label{fig_AN3:time_T2}]{\includegraphics[width=.4\textwidth]{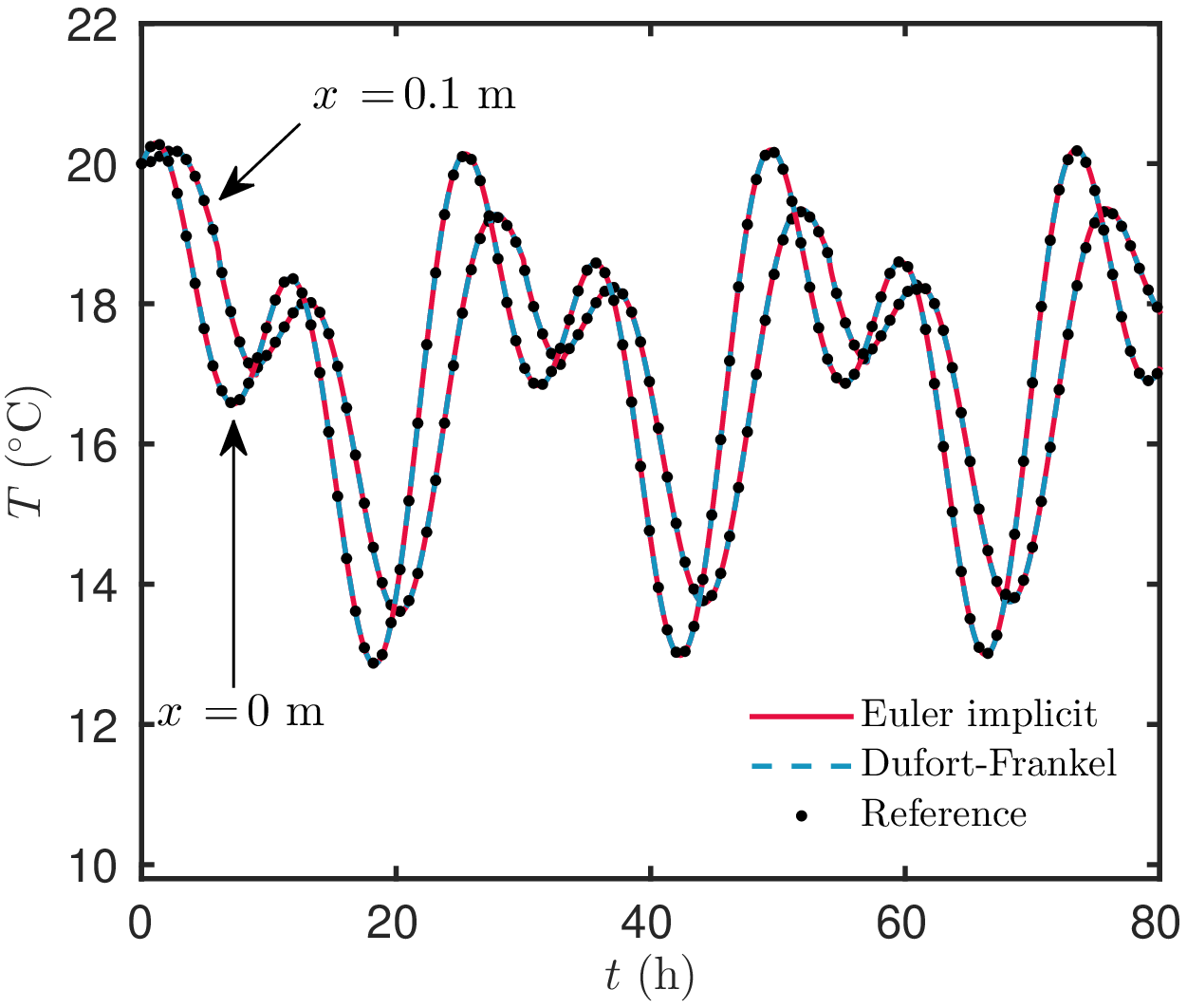}}
  \subfigure[][\label{fig_AN3:time_T34}]{\includegraphics[width=.4\textwidth]{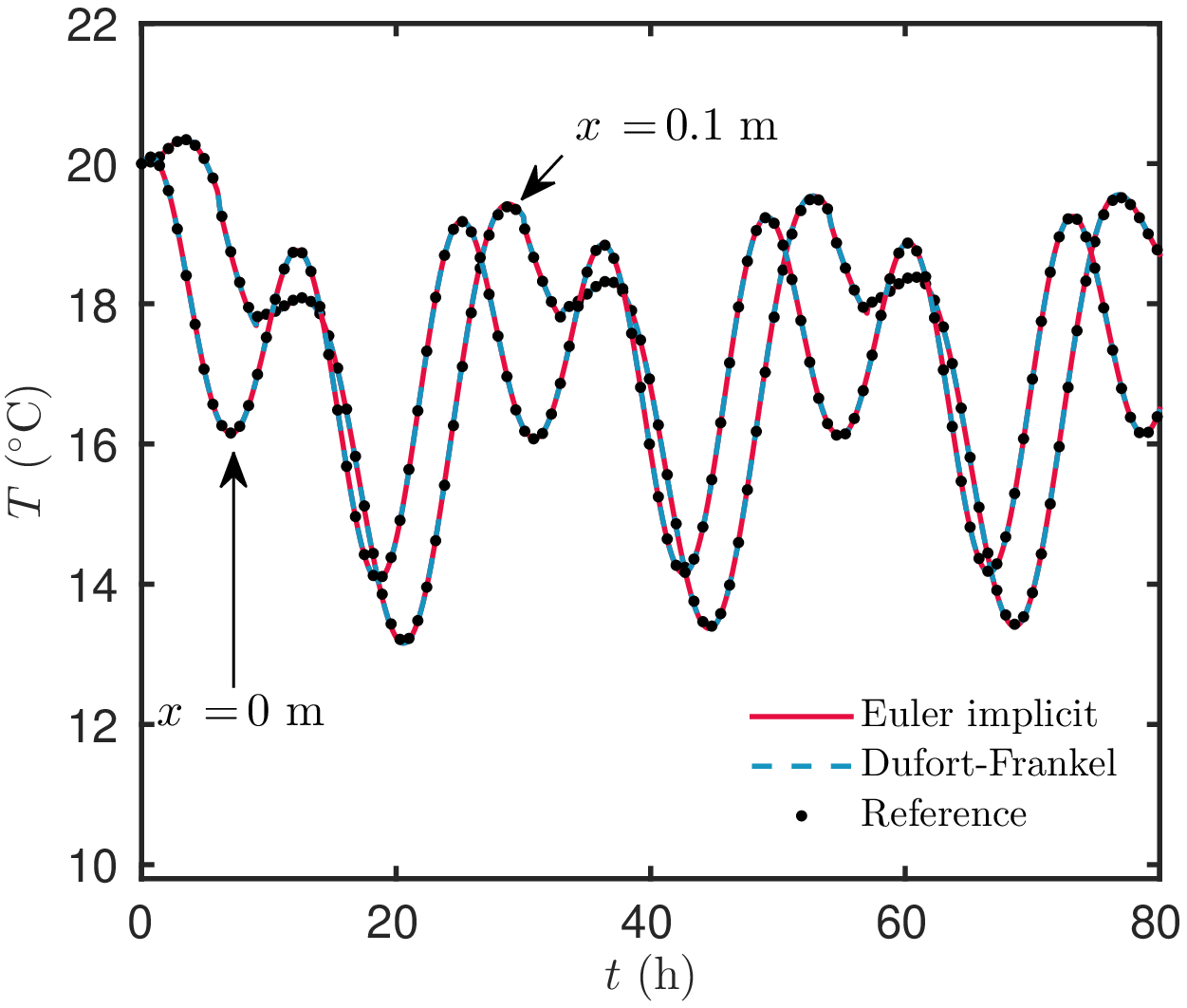}}
  \subfigure[][\label{fig_AN3:time_Tz}]{\includegraphics[width=.4\textwidth]{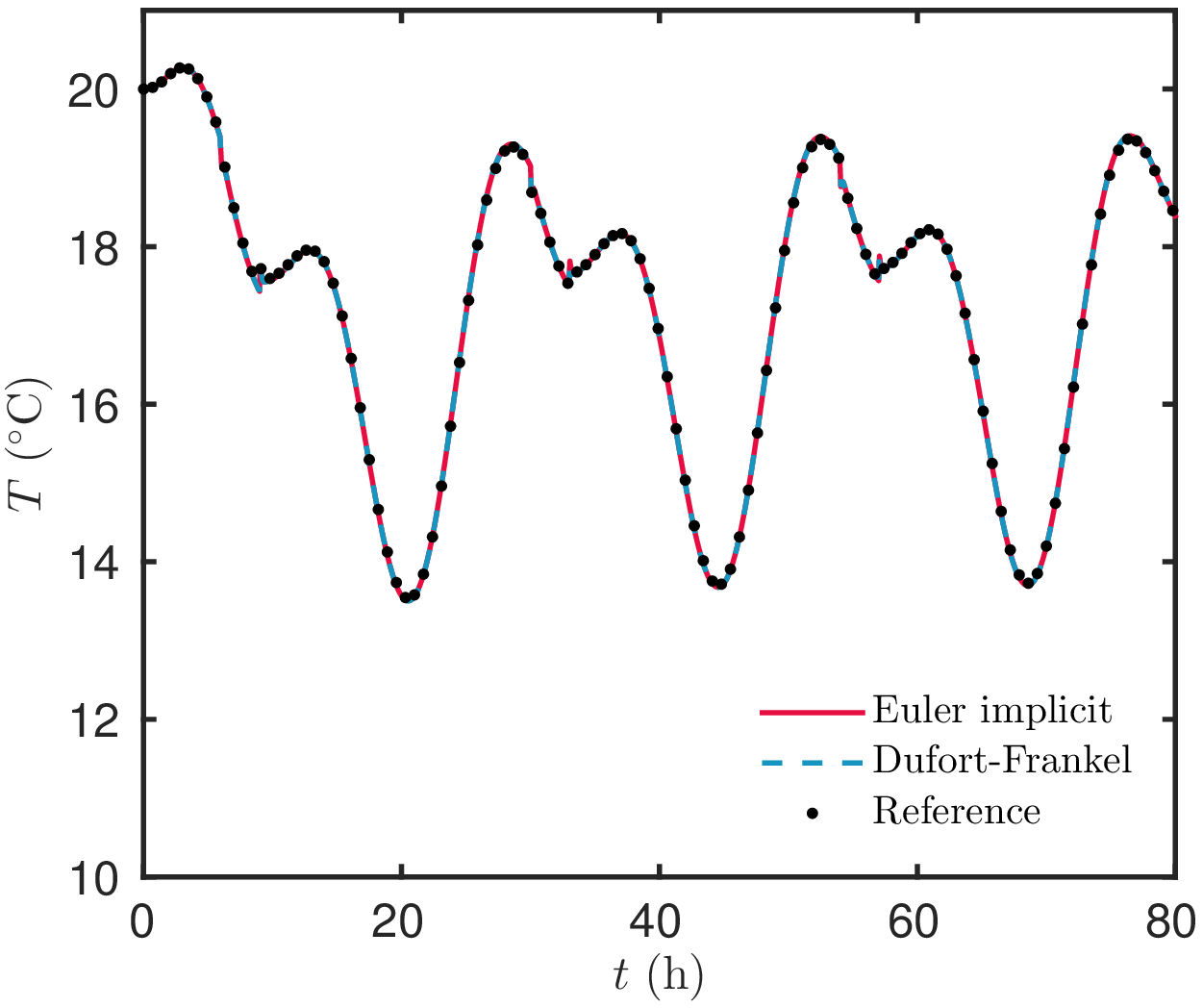}} 
  \caption{\small\em Temperature evolution for the North (a), South (b)  and East (c) walls and for the air zone (d) in the linear case.}
  \label{fig_AN3:time_T}
\end{center}
\end{figure}

\begin{figure}
\begin{center}
  \subfigure[][\label{fig_AN3:time_Phi1}]{\includegraphics[width=.4\textwidth]{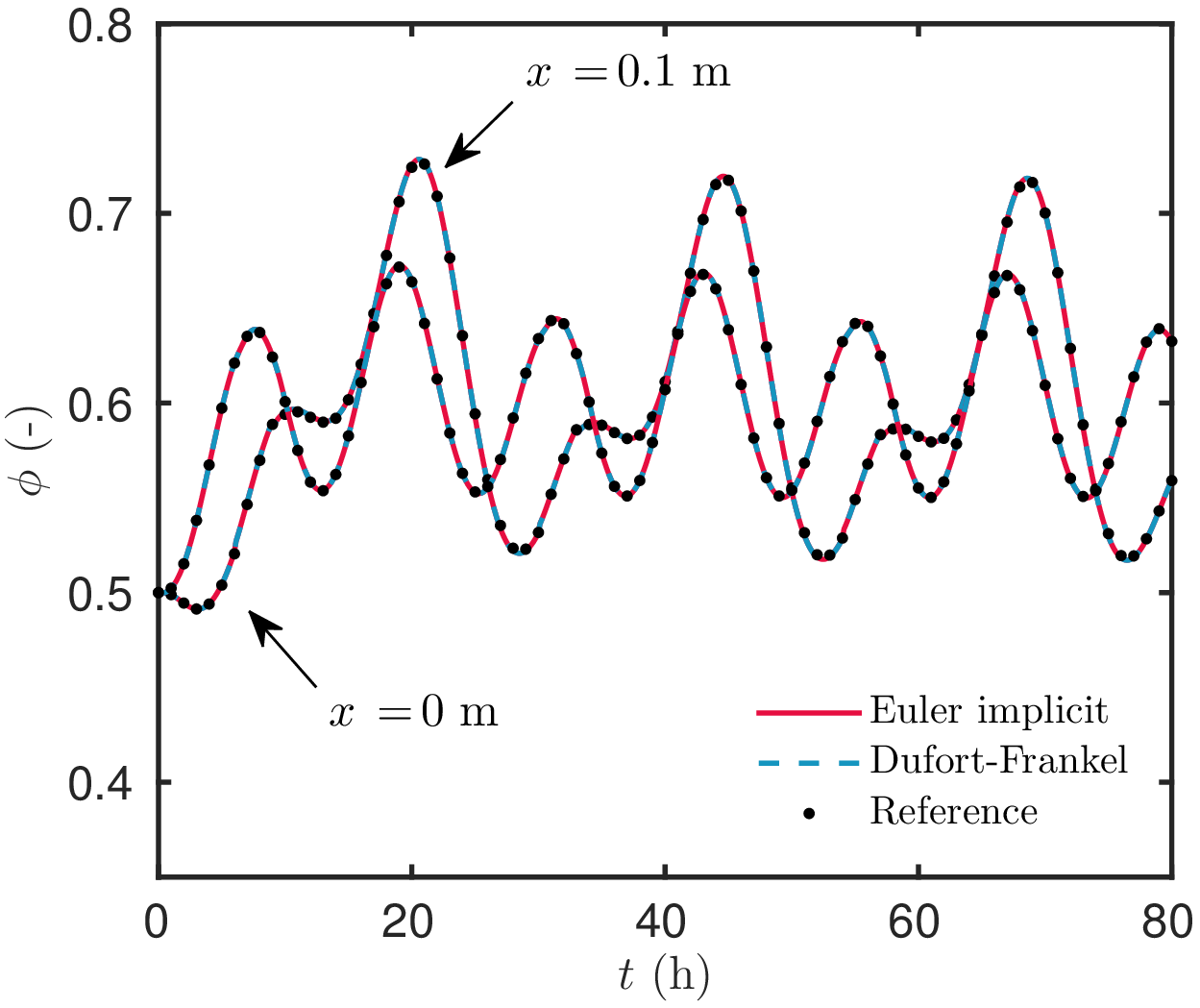}}
  \subfigure[][\label{fig_AN3:time_Phi2}]{\includegraphics[width=.4\textwidth]{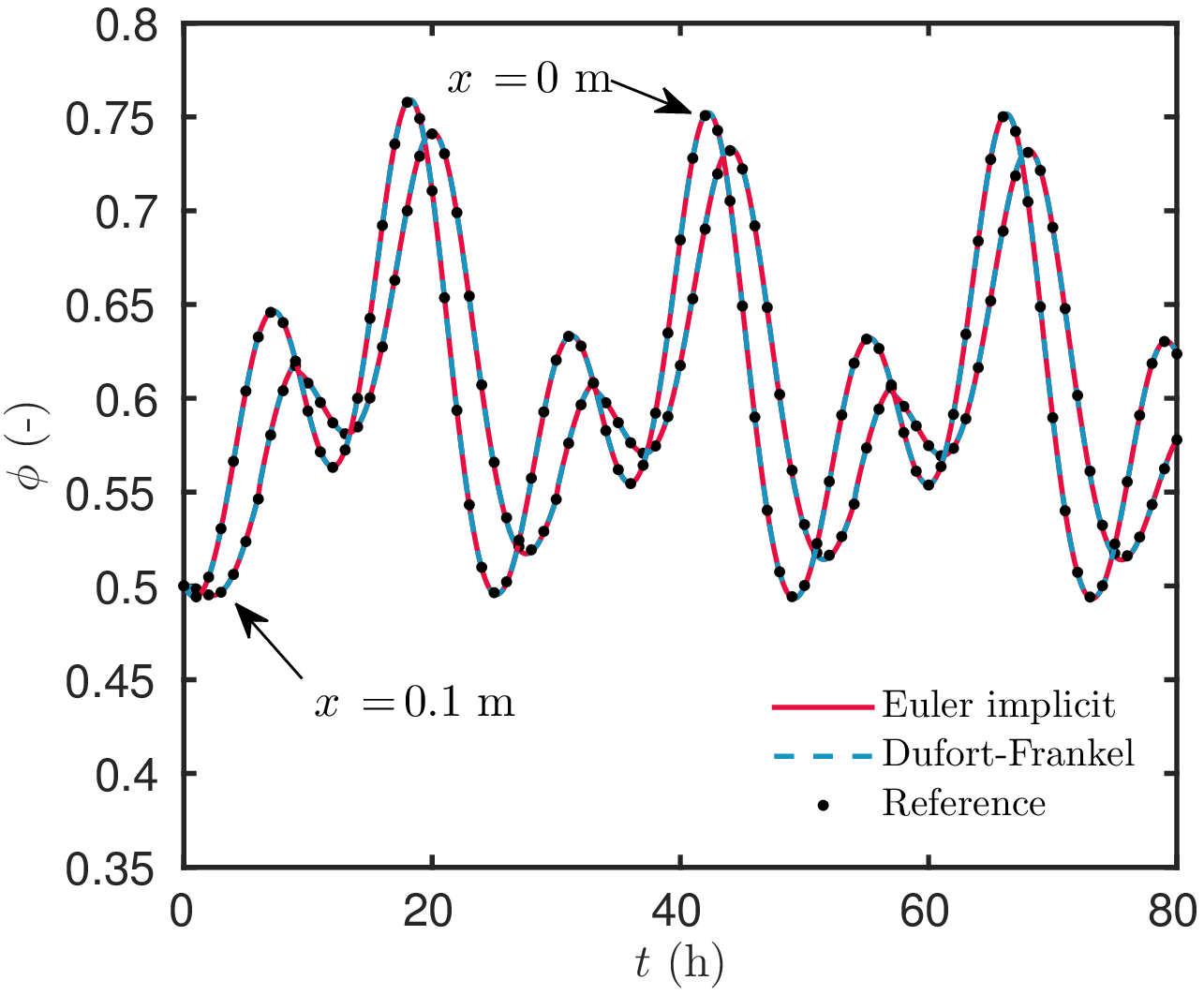}}
  \subfigure[][\label{fig_AN3:time_Phi34}]{\includegraphics[width=.4\textwidth]{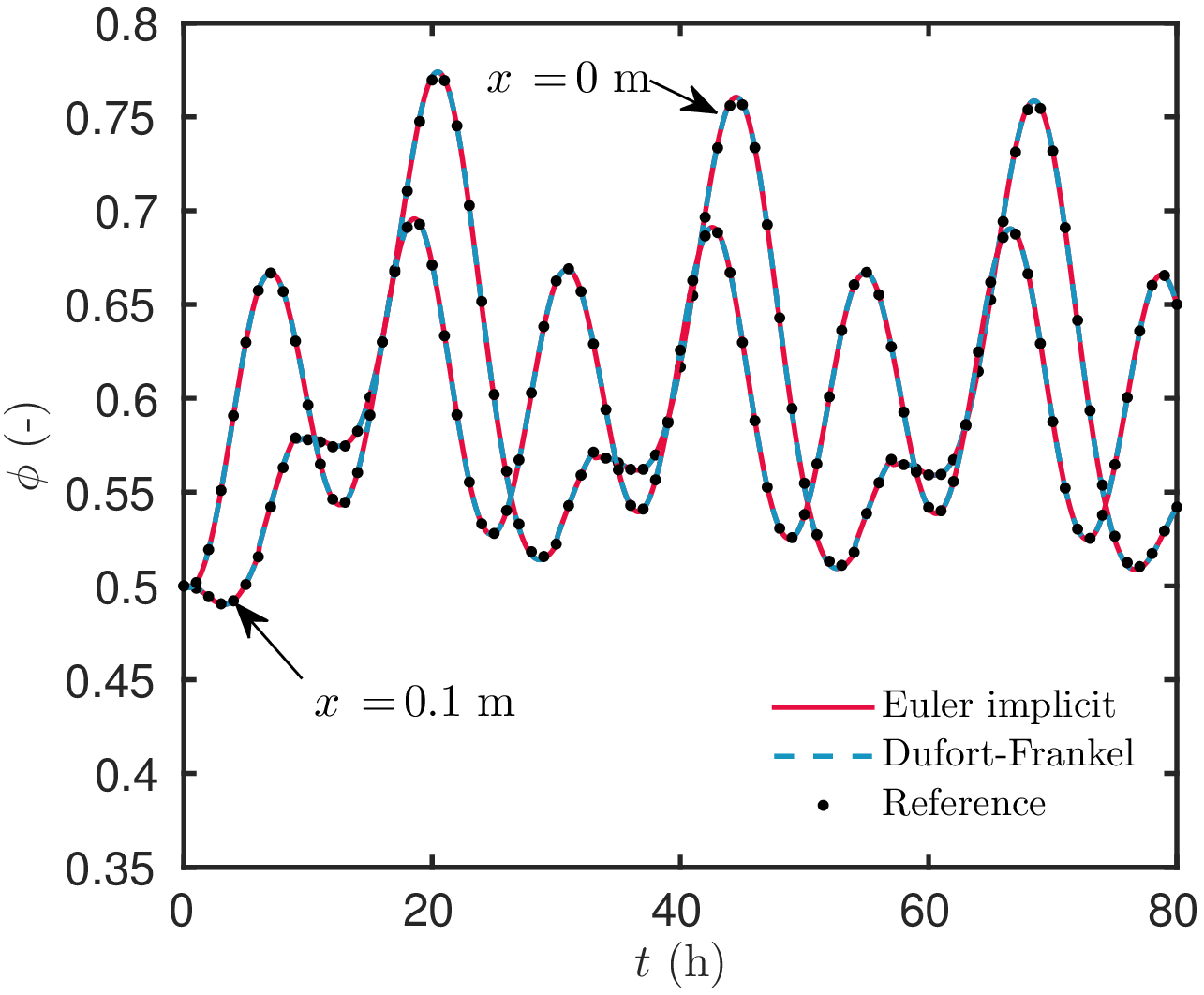}}
  \subfigure[][\label{fig_AN3:time_Phiz}]{\includegraphics[width=.4\textwidth]{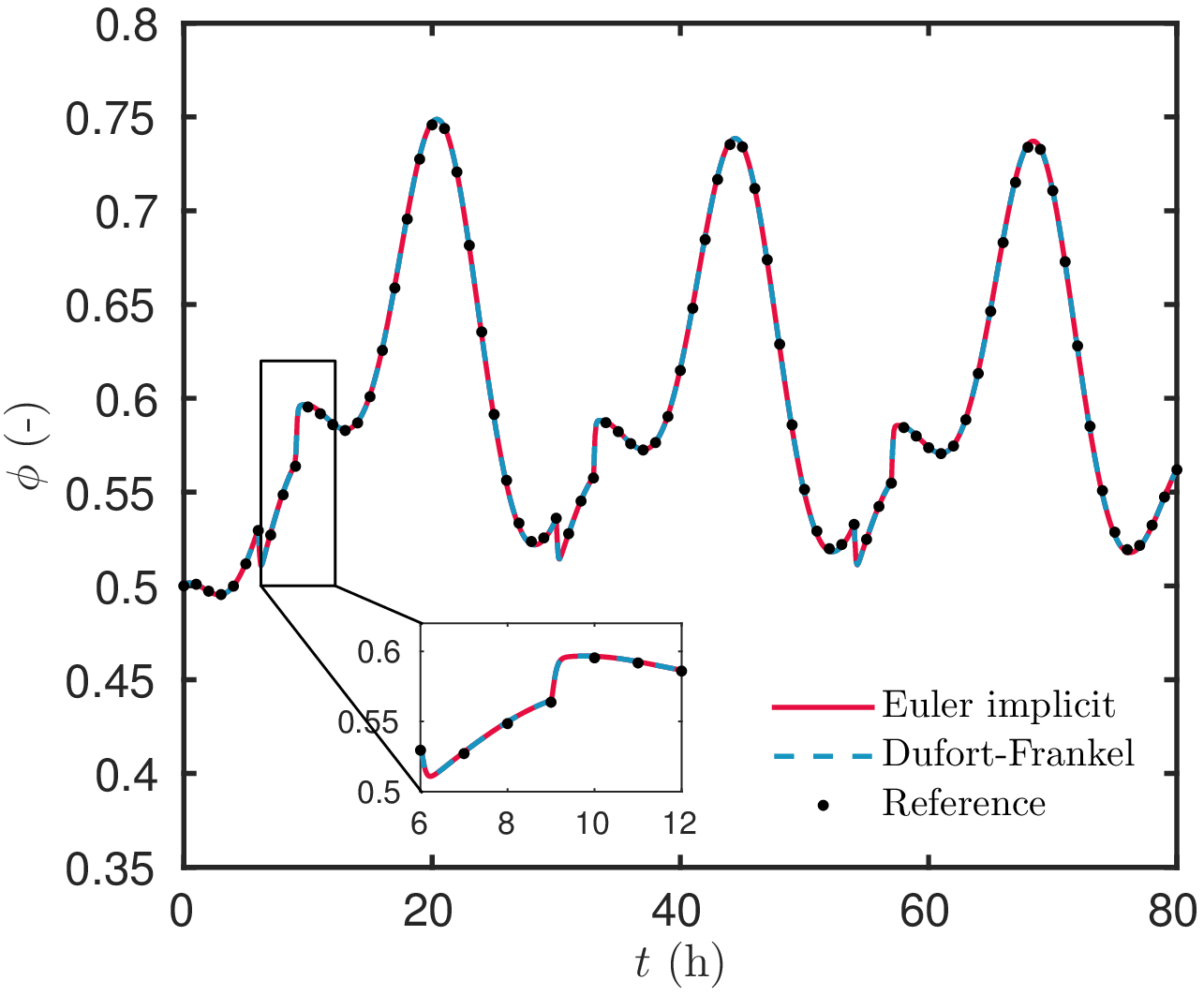}}
  \caption{\small\em Relative humidity evolution for the North (a), South (b) and East (c) walls and for the air zone (d) in the linear case.}
  \label{fig_AN3:time_Phi}
\end{center}
\end{figure}

The time evolution of the temperature and relative humidity for different walls and for the air zone are given in Figures~\ref{fig_AN3:time_T} and \ref{fig_AN3:time_Phi}. The solutions computed with the \Eu ~and \DF ~schemes are in very good agreement with the reference. As for the material properties of the wall, differences between the field evolution can be observed. Particularly, in Figures~\ref{fig_AN3:time_T1} and \ref{fig_AN3:time_T34}, the increase and decrease of temperature at $x \egal 0 \ \mathsf{m}$ are higher for the Eastern wall than for the Northern one. It is due to a higher \textsc{Biot} number for the latter: $\BiTT \egal 3.1$ against  $\BiTT \egal 1.7\,$. In Figures~\ref{fig_AN3:time_Phi1} and \ref{fig_AN3:time_Phi2} the increase and decrease of relative humidity are different at $x \egal 0 \ \mathsf{m}$ between the Northern and Southern walls. In Figure~\ref{fig_AN3:time_Phiz}, a slight increase of relative humidity due to moisture generation in the zone from $6$ to $9 \ \mathsf{h}$ can be observed.

\begin{figure}
\begin{center}
  \subfigure[][\label{fig_AN3:Erreur_fx_T}]{\includegraphics[width=.4\textwidth]{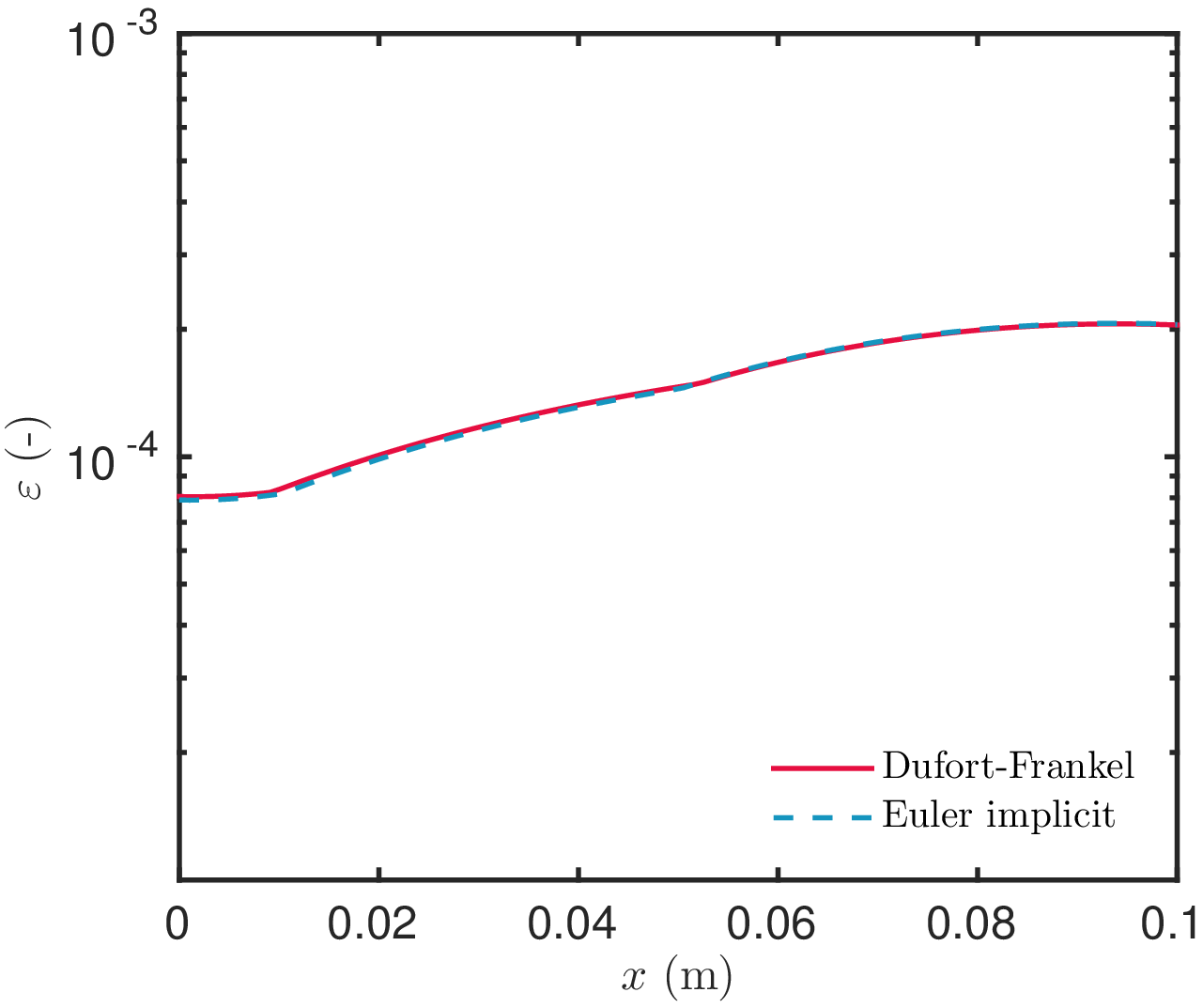}}
  \subfigure[][\label{fig_AN3:Erreur_ft_Phiz}]{\includegraphics[width=.4\textwidth]{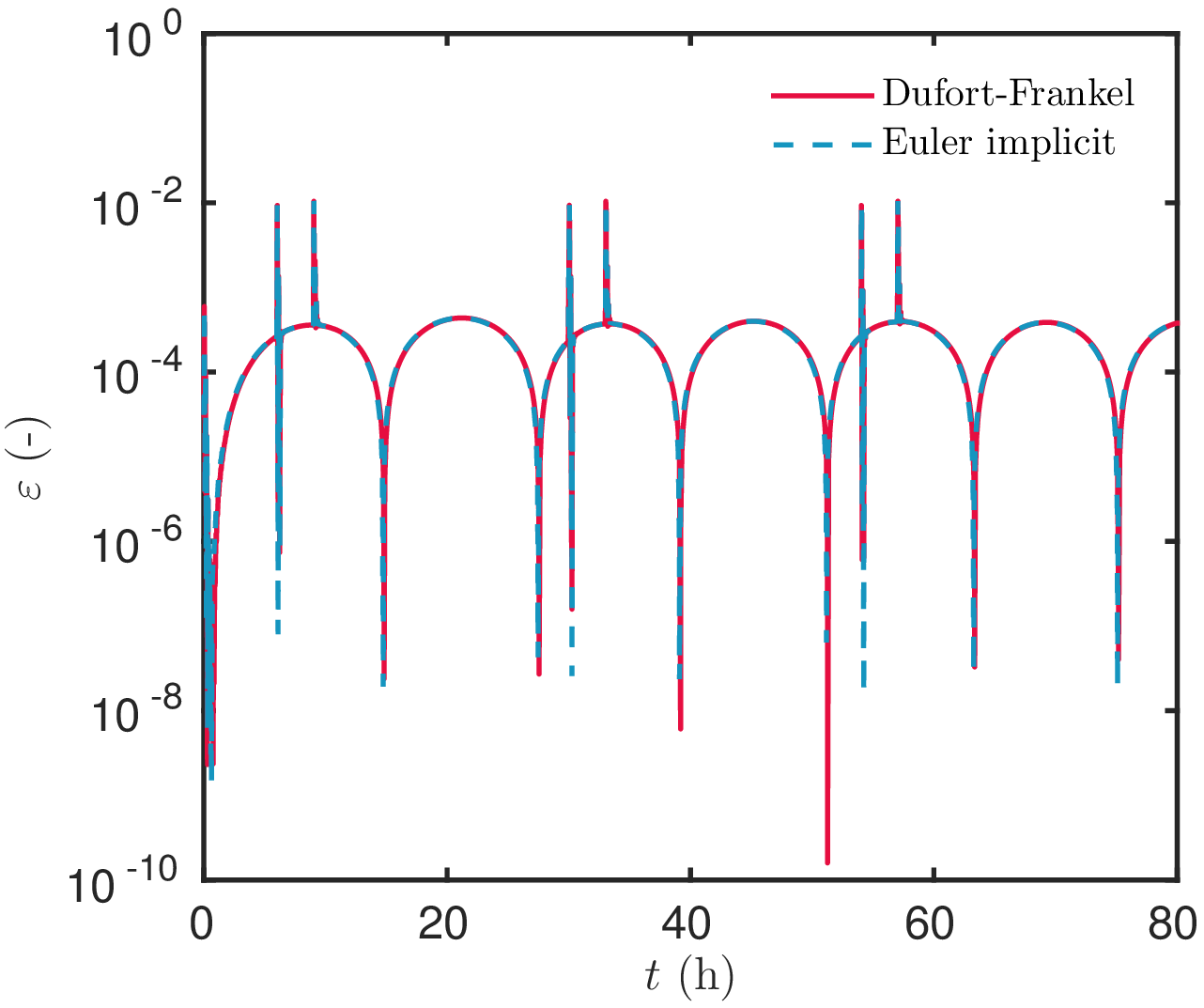}} 
  \caption{\small\em Error of the solutions computed with the \Eu ~implicit and \DF ~explicit schemes for the walls (a) and for the air zone (b) in the linear case.}
  \label{fig_AN3:error}
\end{center}
\end{figure}

The $\mathcal{L}_{\,\infty}$ error with respect to space $x$ has been computed for the fields in the four walls, as illustrated in Figure~\ref{fig_AN3:Erreur_fx_T}. Moreover, the $\mathcal{L}_{\,\infty}$ error with respect to $t$ for the fields in the zone is given in Figure~\ref{fig_AN3:Erreur_ft_Phiz}. The error is approximately $\O\,(10^{\,-4})$ for both models, proving the accuracy of the solution computed with the \Eu ~implicit and \DF ~explicit schemes.

\begin{figure}
\begin{center}
  \includegraphics[width=.4\textwidth]{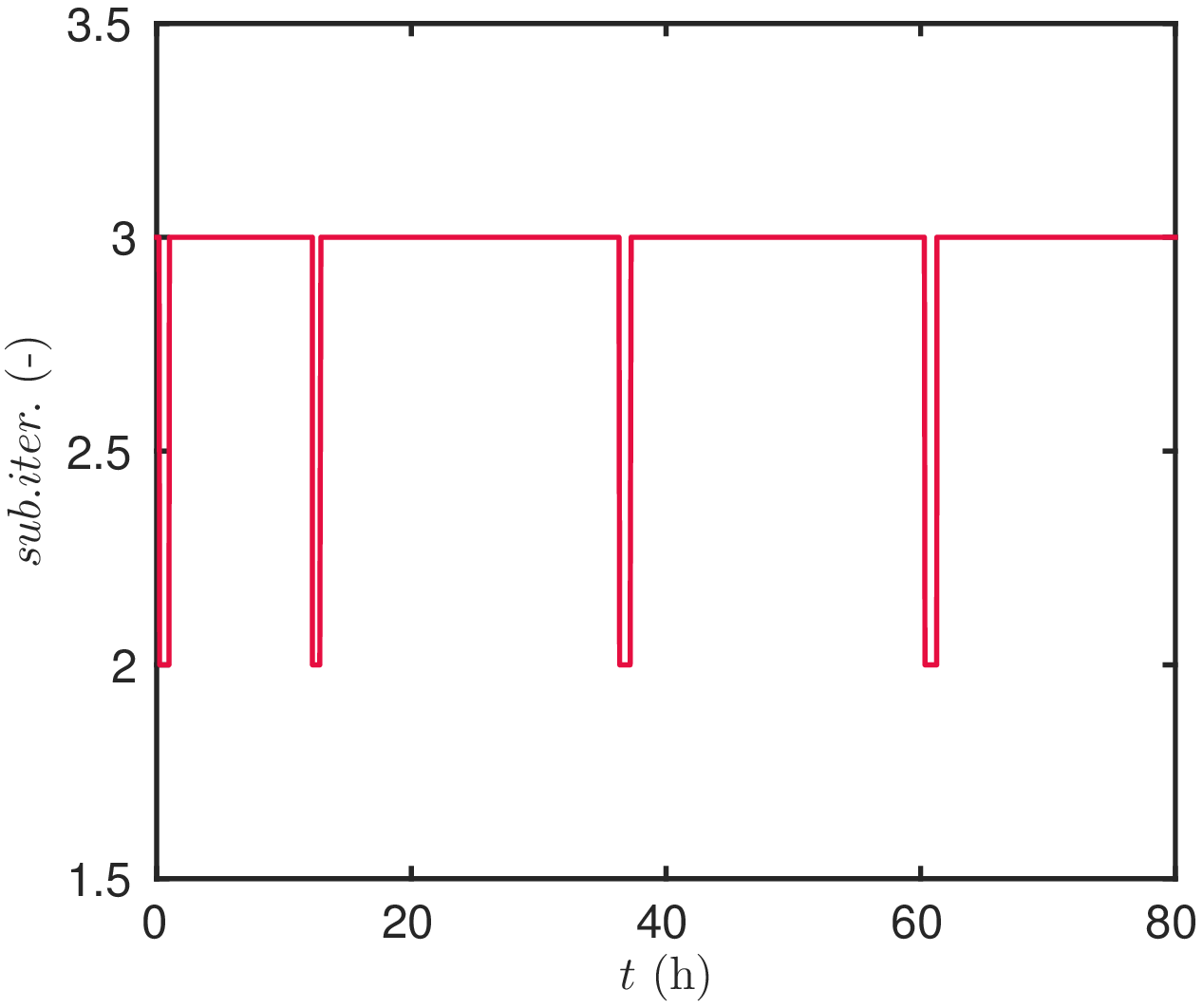}
  \caption{\small\em Sub-iterations required for the \Eu ~implicit scheme compute the solution of the whole-building energy model in the linear case.}
  \label{fig_AN3:subiter_ft}
\end{center}
\end{figure}

For each scheme, the CPU time has been calculated, using the \texttt{Matlab} platform on a computer with Intel i$7$ CPU and $32$GB of RAM, which is shown in Table~\ref{tab:CPU_time}. The \DF ~explicit scheme requires less than 10\% of the time needed for the \Eu ~implicit one, to compute the solution. This difference is due to the sub-iterations needed to solve the nonlinear system composed by equations of the wall and zone models. In this case, a fixed-point algorithm with tolerance parameter $\eta \egal 0.01 \cdot \Delta \ts\,$, as the one illustrated in Algorithm.~\ref{alg:Fixed_point_wbse}, has been used. As emphasized in Figure~\ref{fig_AN3:subiter_ft}, this algorithm required almost three sub-iterations to compute the solution at each time iteration even in the linear case.

\begin{figure}
\begin{center}
  \subfigure[][\label{fig_AN3:errT_f_dt}]{\includegraphics[width=.4\textwidth]{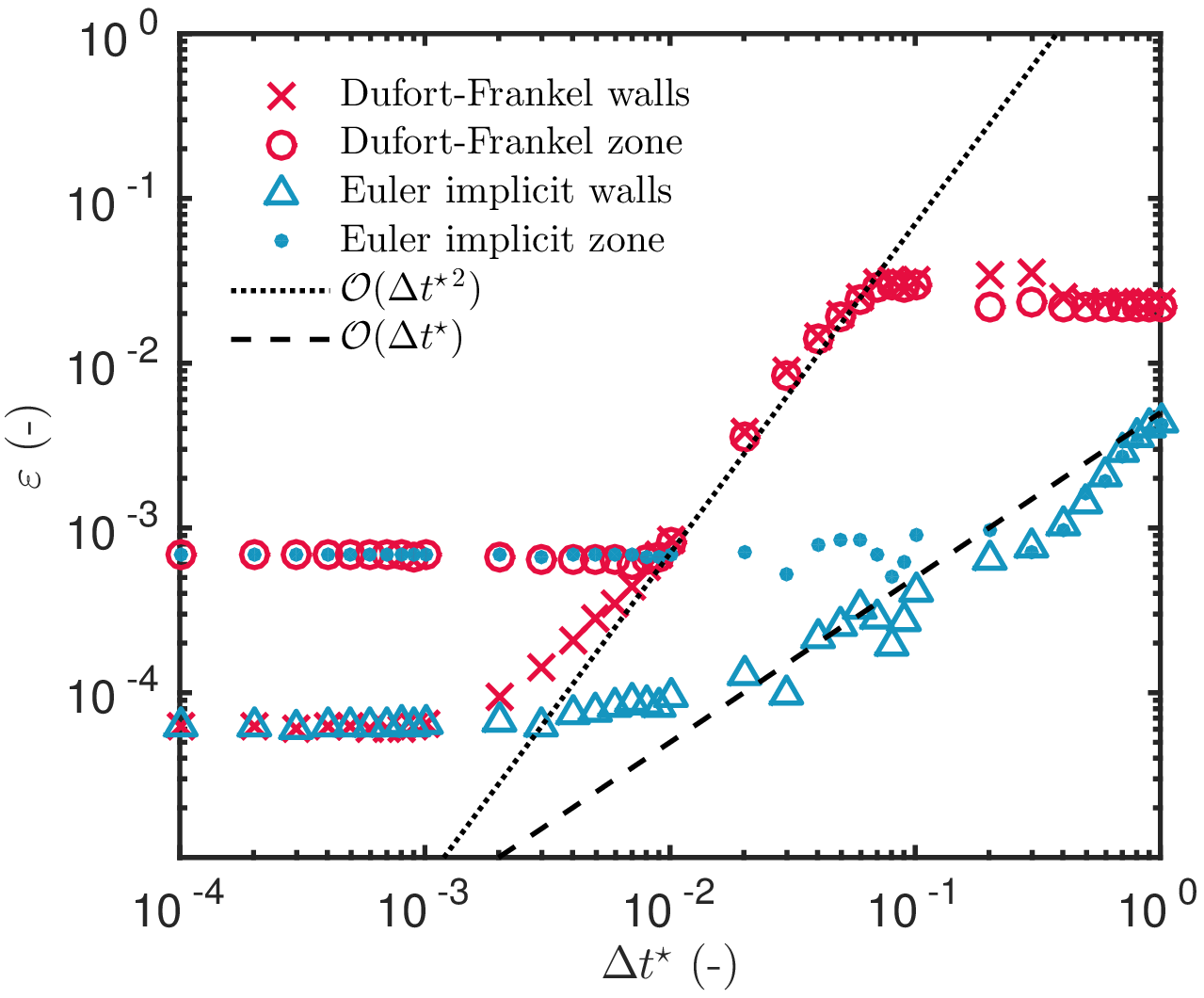}}
  \subfigure[][\label{fig_AN3:errPhi_f_dt}]{\includegraphics[width=.4\textwidth]{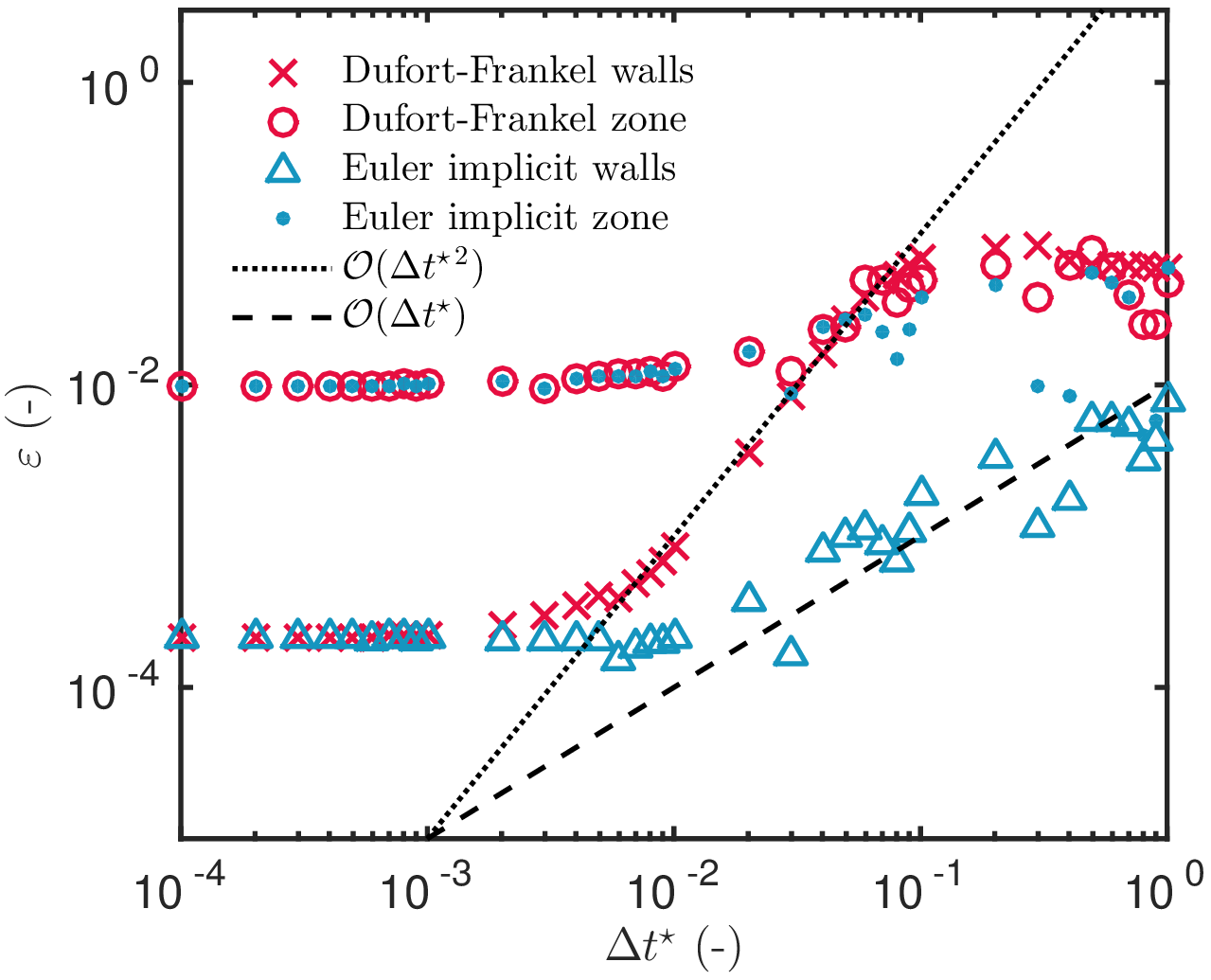}} 
  \caption{\small\em $\mathcal{L}_{\infty}$ error as a function of $\Delta \ts$ for the \Eu ~implicit and \DF ~explicit schemes, for temperature (a) and relative humidity (b) in the one-zone case study.}
  \label{fig_AN3:error_f_dt}
\end{center}
\end{figure}

A convergence study of the whole-building model has been carried out by fixing the space discretisation to $\dxs \egalb 10^{\,-2} $ and varying the time discretisation. Results are reported in Figures~\ref{fig_AN3:errT_f_dt} and \ref{fig_AN3:errPhi_f_dt}, for each field. The error with the reference solution has been computed for each field (temperature and relative humidity) and for each model (wall and zone). It can be noted that the wall model reach a constant absolute accuracy lower than the one for the zone. As discussed in the previous case, the \DF ~scheme is preferable to use for $\dts\, \leqslant\, 5\cdot 10^{\,-2}$ the equivalent to $180$-$\mathsf{s}\,$. For bigger time step values the implicit scheme would be more advisable at the strict condition that it respects the characteristic time of the physical phenomena \cite{Gasparin2017}. Moreover, in building simulation, the recommended time step is about $100$-$\mathsf{s}$ \cite{DosSantos2004} which makes the \DF ~scheme suitable for this kind of application.

\subsection{Two-zones case study}

\begin{figure}
\begin{center}
  \includegraphics[width=.8\textwidth]{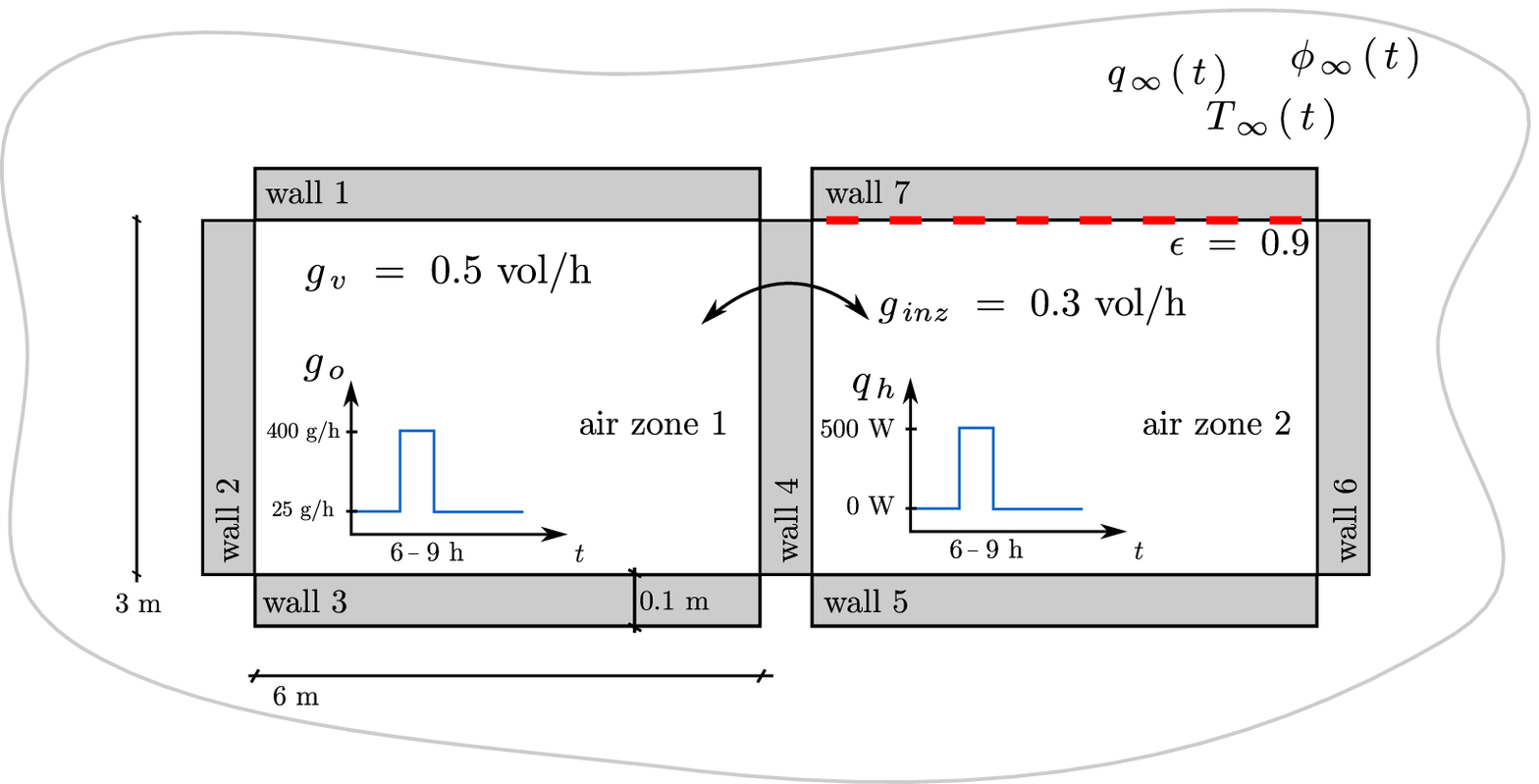}
  \caption{\small\em Illustration of the two-zones nonlinear case.}
  \label{fig_AN4:schema_bat}
\end{center}
\end{figure}

\begin{figure}
\begin{center}
  \includegraphics[width=.4\textwidth]{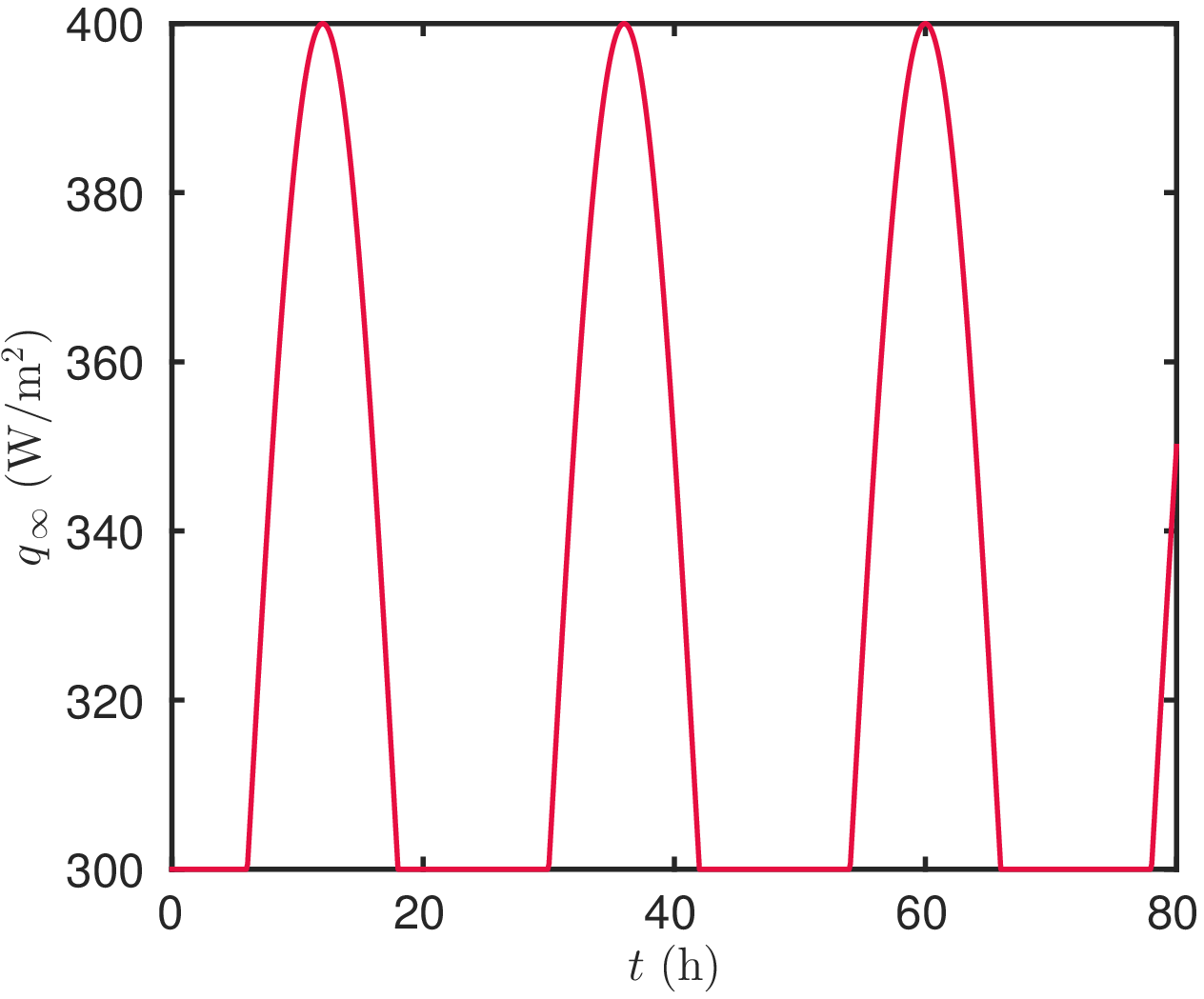}
  \caption{\small\em Outside radiative heat flux for the two-zones nonlinear case.}
  \label{fig_AN4:time_qinf}
\end{center}
\end{figure}

Previous case study considered a single zone building hygrothermal simulation with linear wall material properties. It enabled to enhance the sub-iterations required using \Eu ~implicit scheme when coupling the wall and zone models. The present case focus on a bi-zone building, as shown in Figure~\ref{fig_AN4:schema_bat}, and take into account the variation of the wall material properties with temperature and relative humidity, such as the ones used in Section~\ref{sec:AN2}. Furthermore, the radiative heat exchanged among the room air surfaces are included. All the inside walls have an emissivity $\epsilon \egal 0.5$ except the wall $7\,$, which have a higher value $\epsilon \egal 0.9$. The view factor is set to $s \egal 0.2\,$. Incident radiation is considered for the outside boundary condition, shown in Figure~\ref{fig_AN4:time_qinf}. An airflow of $g_{\,\mathrm{inz}} \egal 0.3 \ \mathsf{h^{\,-1}}$ occurs between both zones. Only zone $1$ is subjected to moisture sources due to occupants and to air ventilation. Zone $2$ receives a heat source $q_{\,h} \egal 500 \ \mathsf{W}$ from $3$ to $4 \ \mathsf{h}\,$. The other parameters have the same numerical values as in the previous case study. This case study was designed in order to enforce the nonlinearities of the whole-building model, to increase the sub-iterations of the \Eu ~implicit approach and thus enhance the efficiency of the \DF ~explicit scheme. The solution is computed using both schemes, with a time and space discretisations $\Delta t^{\,\star} \egal 10^{\,-3}$ and $\Delta x^{\,\star} \egal 10^{\,-2}\,$, equivalent to $\Delta x \egal 10^{\,-3} \ \mathsf{m}$ and $\Delta t \egal 3.6 \ \mathsf{s}\,$.

\begin{figure}
\begin{center}
  \subfigure[][\label{fig_AN4:time_T1}]{\includegraphics[width=.4\textwidth]{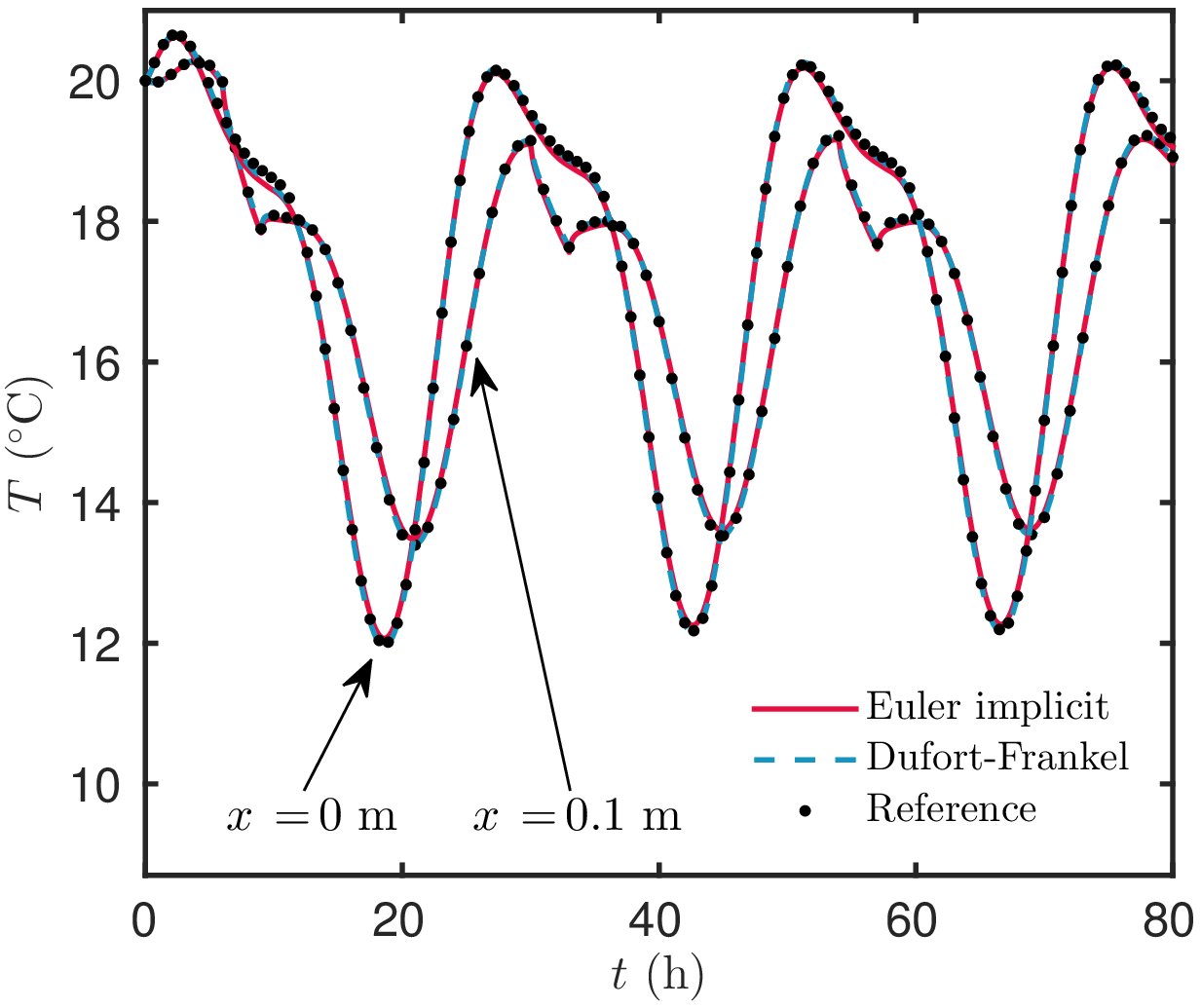}}
  \subfigure[][\label{fig_AN4:time_Phi1}]{\includegraphics[width=.4\textwidth]{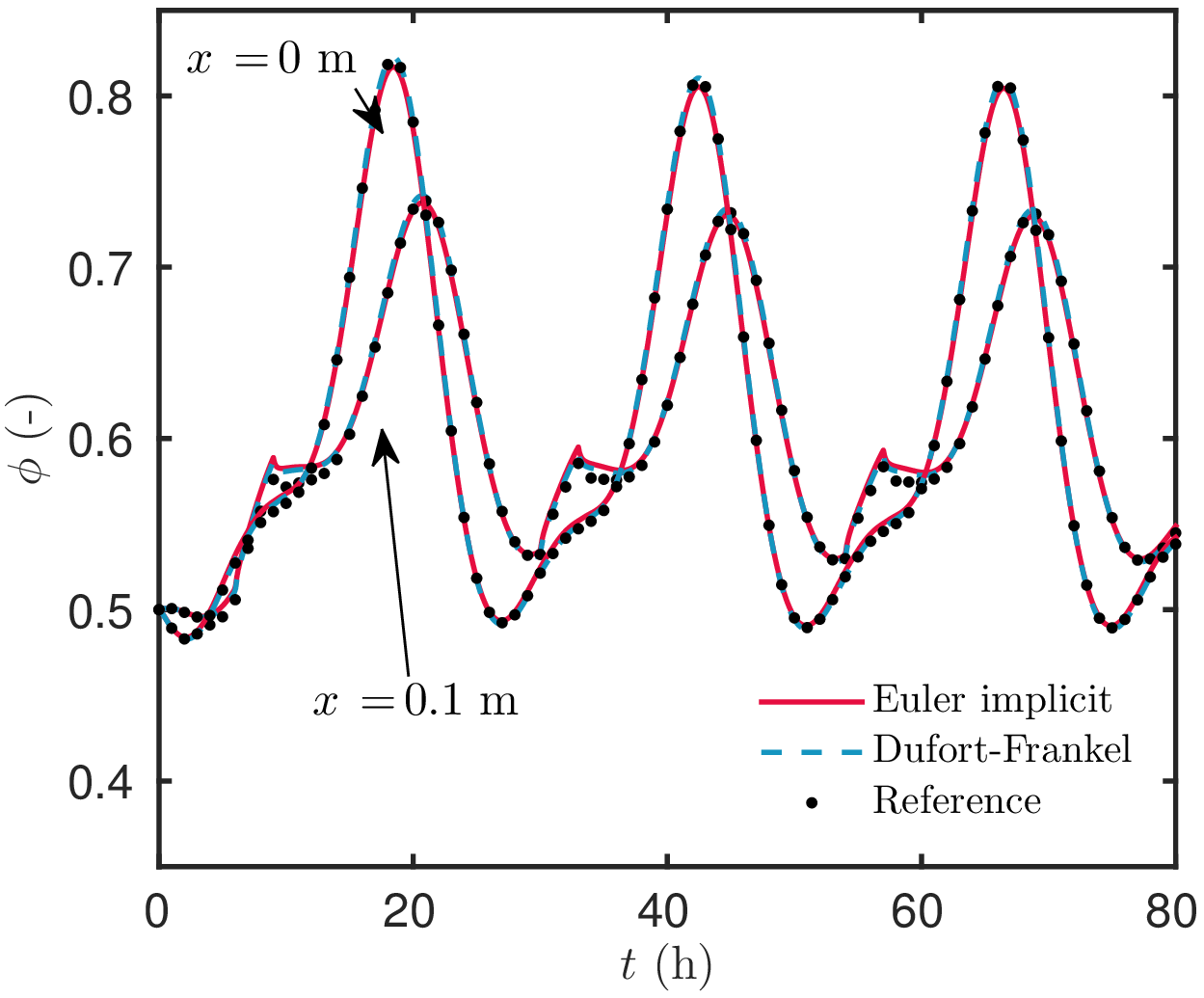}} 
  \caption{\small\em Evolution of the temperature (a) and relative humidity (b) for the wall $1$ in the nonlinear case.}
  \label{fig_AN4:time_fields_wall}
\end{center}
\end{figure}

\begin{figure}
\begin{center}
  \subfigure[][\label{fig_AN4:time_Tz1}]{\includegraphics[width=.4\textwidth]{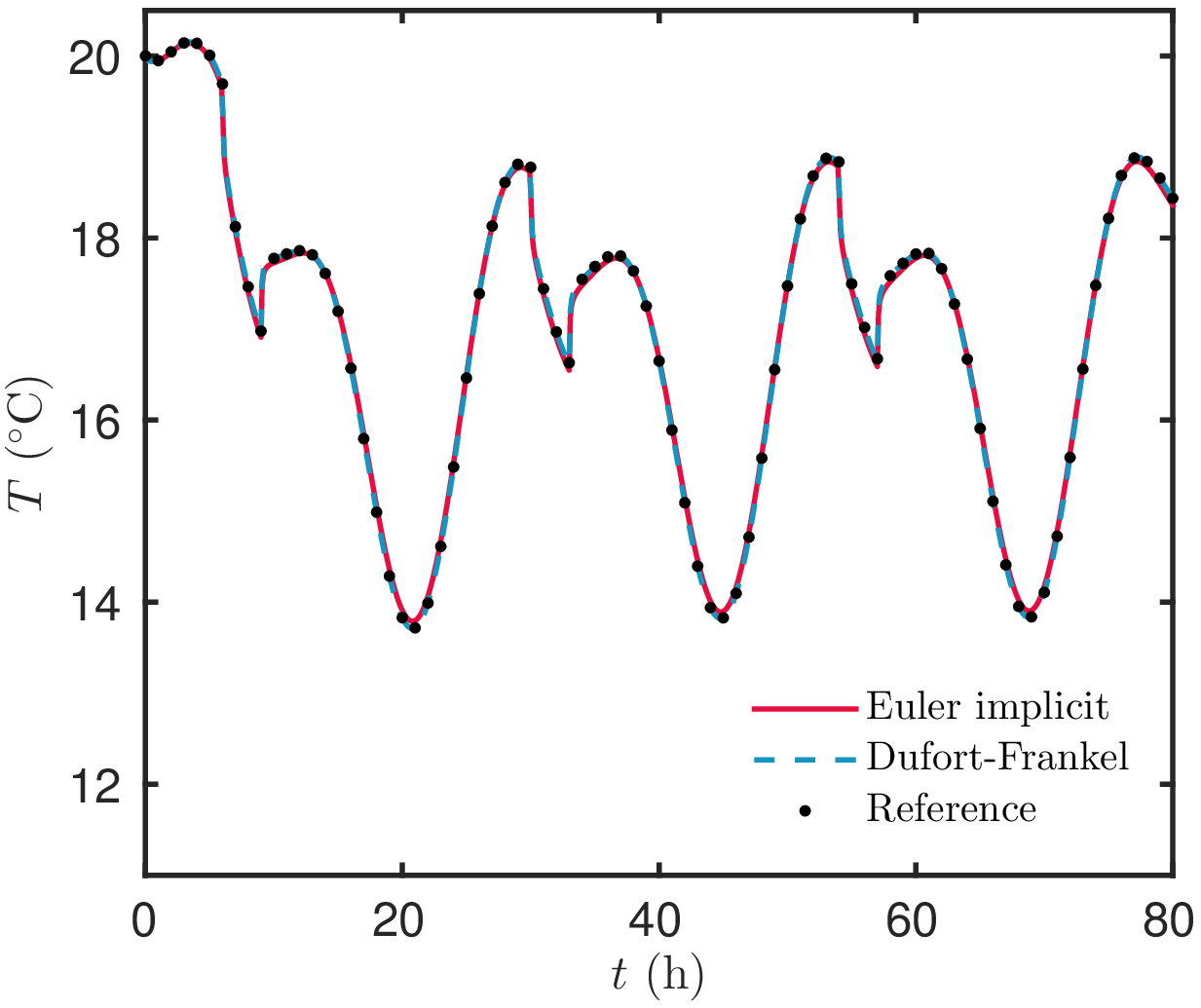}}
  \subfigure[][\label{fig_AN4:time_Phiz1}]{\includegraphics[width=.4\textwidth]{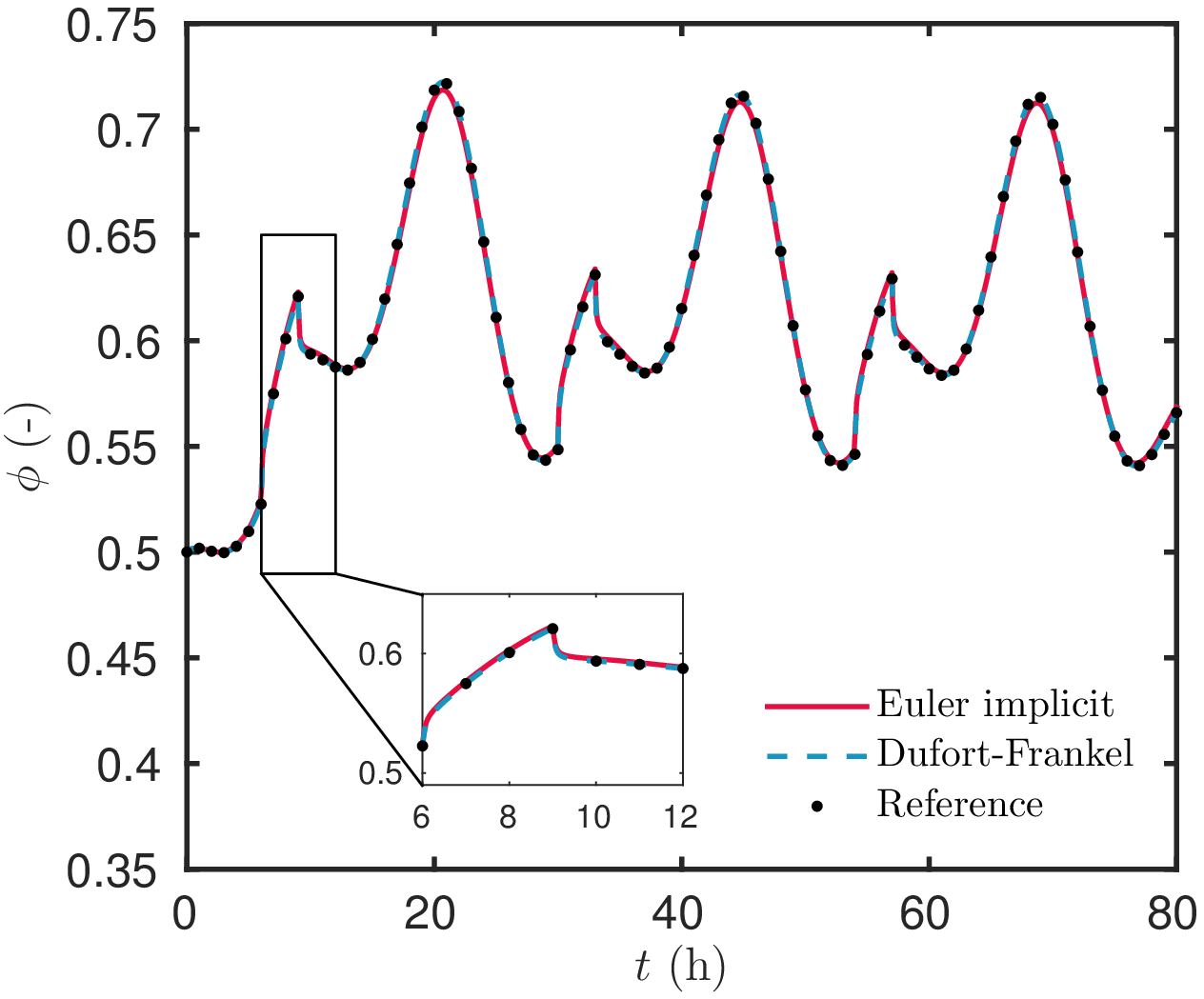}}
  \subfigure[][\label{fig_AN4:time_Tz2}]{\includegraphics[width=.4\textwidth]{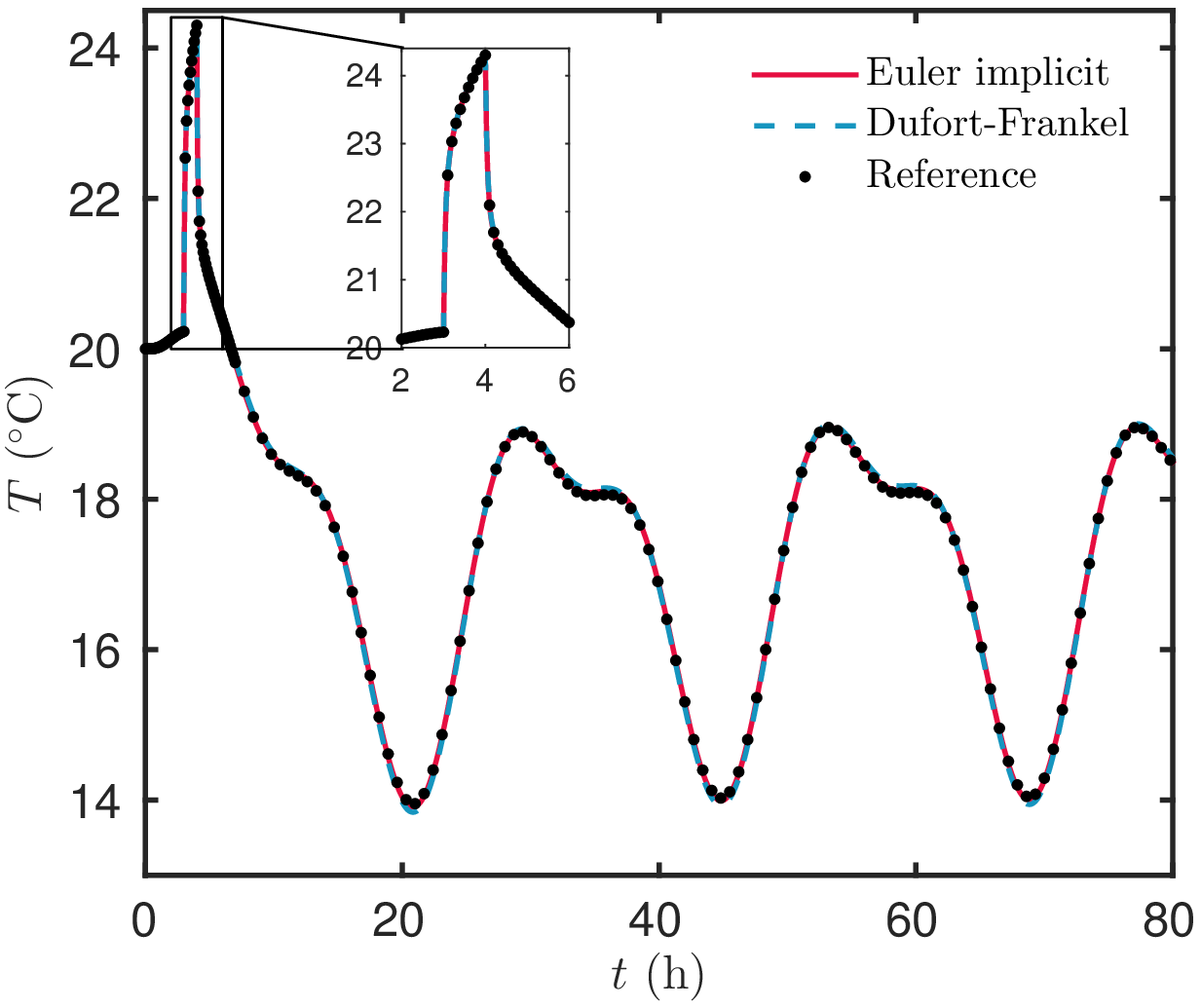}}
  \subfigure[][\label{fig_AN4:time_Phiz2}]{\includegraphics[width=.4\textwidth]{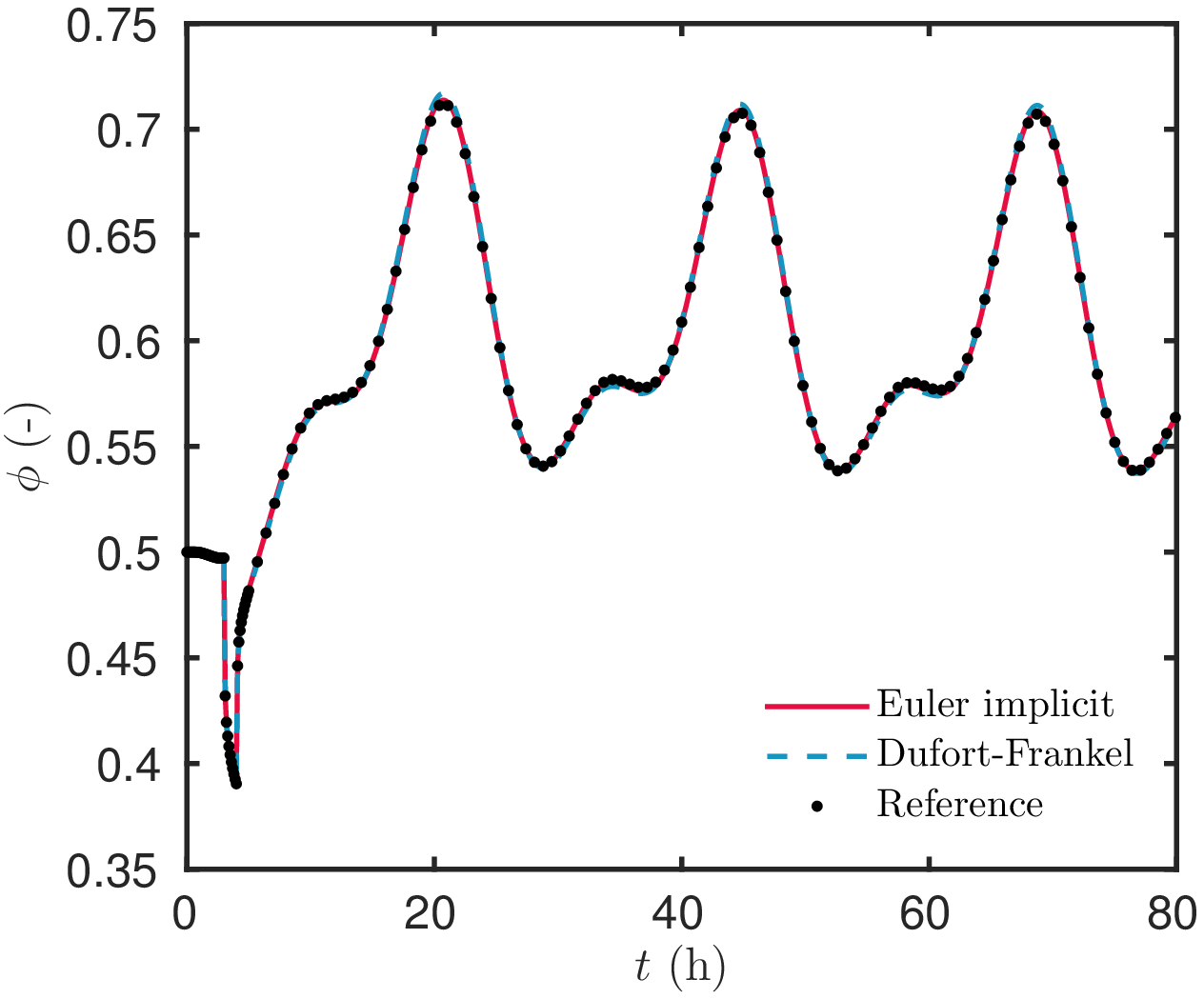}} 
  \caption{\small\em Evolution of the temperature and relative humidity for zone $1$ (a-b) and zone $2$ (c-d) for the nonlinear case.}
  \label{fig_AN4:time_fields_zone}
\end{center}
\end{figure}

\begin{figure}
\begin{center}
  \subfigure[][\label{fig_AN4:Erreur_fx_T}]{\includegraphics[width=.4\textwidth]{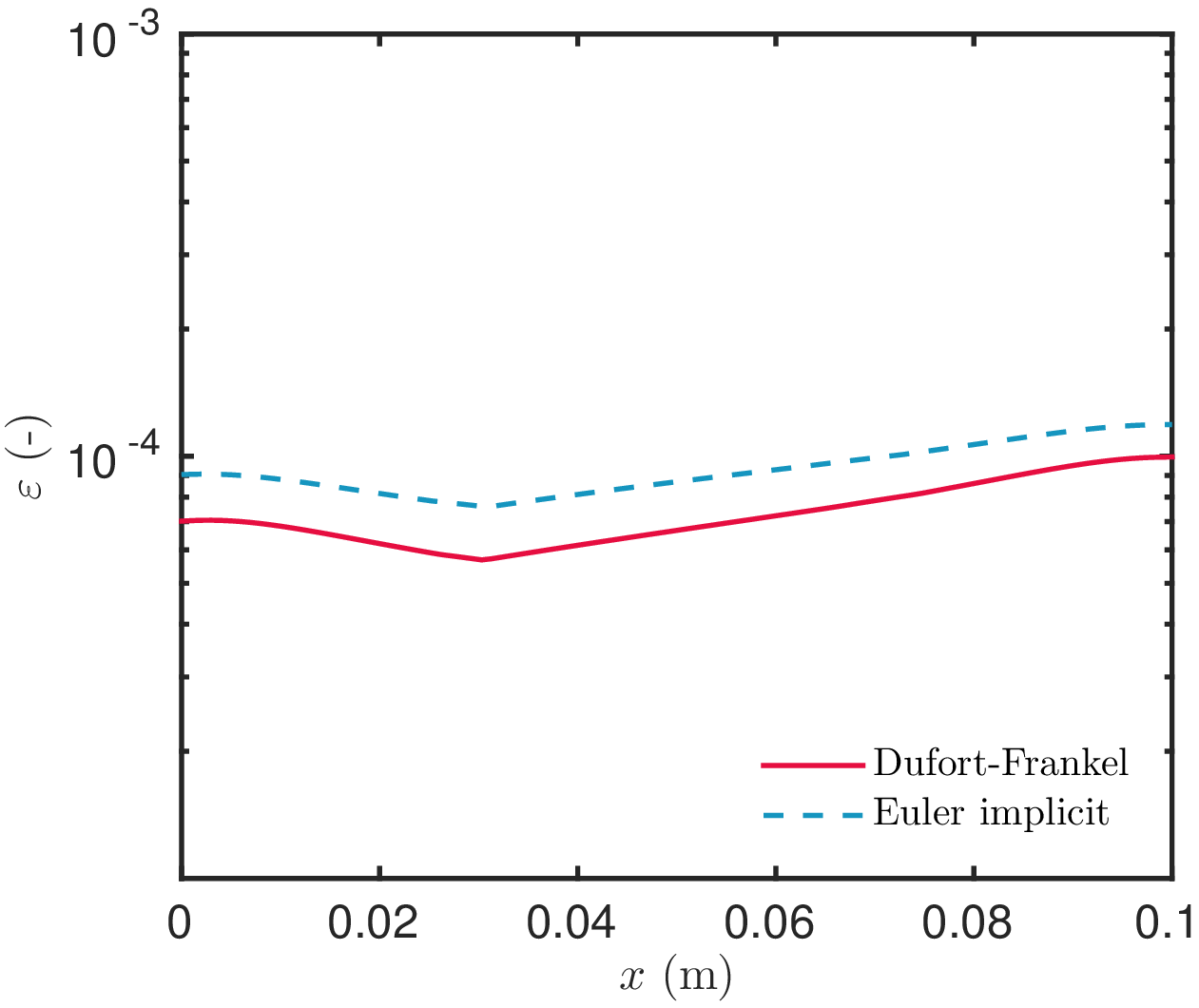}}
  \subfigure[][\label{fig_AN4:Erreur_ft_Tz}]{\includegraphics[width=.4\textwidth]{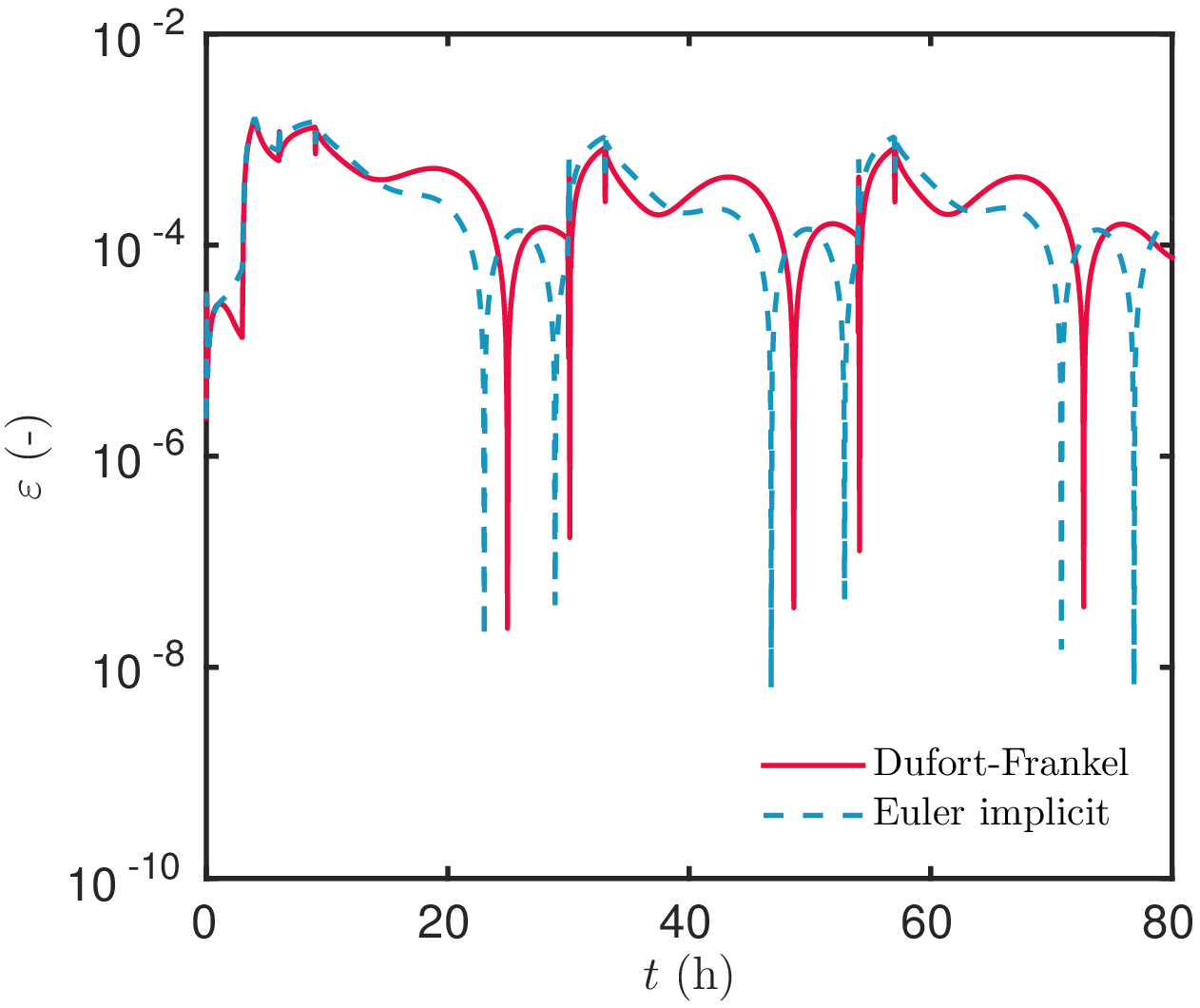}} 
  \caption{\small\em Error of the solutions computed with the \Eu ~implicit and \DF ~explicit schemes for the walls (a) and for the air zones (b) in the nonlinear case.}
  \label{fig_AN4:error_f_dt}
\end{center}
\end{figure}

Figures~\ref{fig_AN4:time_T1} and \ref{fig_AN4:time_Phi1} present the time evolution of the dependent variable fields for the wall $1$, showing a very good agreement among the three solutions. By comparing Figures~\ref{fig_AN3:time_T1}, \ref{fig_AN3:time_Phi1}, \ref{fig_AN4:time_T1} and \ref{fig_AN4:time_Phi1}, the effect of considering nonlinear material properties can be observed. Similar observations can be done by analysing Figures~\ref{fig_AN4:time_Tz1}, \ref{fig_AN4:time_Phiz1}, \ref{fig_AN4:time_Tz2} and \ref{fig_AN4:time_Phiz2}. The moisture production in zone $1$ is enhanced by an increase of the relative humidity from $6$ to $9 \ \mathsf{h}\,$. As the moisture generation only occurs in this zone, there is no increase of the relative humidity in zone $2$ due to this phenomena. Moreover, the heat production in zone $2$ is highlighted from $2$ to $6 \ \mathsf{h}\,$. The temperature reach $24^{\circ}\, \mathsf{C}\,$. A slight increase is observable in zone $1$ during this period due to the heat generation in zone $2$ and to the airflow between both zones. The error with the reference solution is of the order of $\O\,(10^{\,-4})$ for the wall and zone fields, as shown in Figures~\ref{fig_AN4:Erreur_fx_T} and \ref{fig_AN4:Erreur_ft_Tz}.

\begin{figure}
\begin{center}
  \includegraphics[width=.4\textwidth]{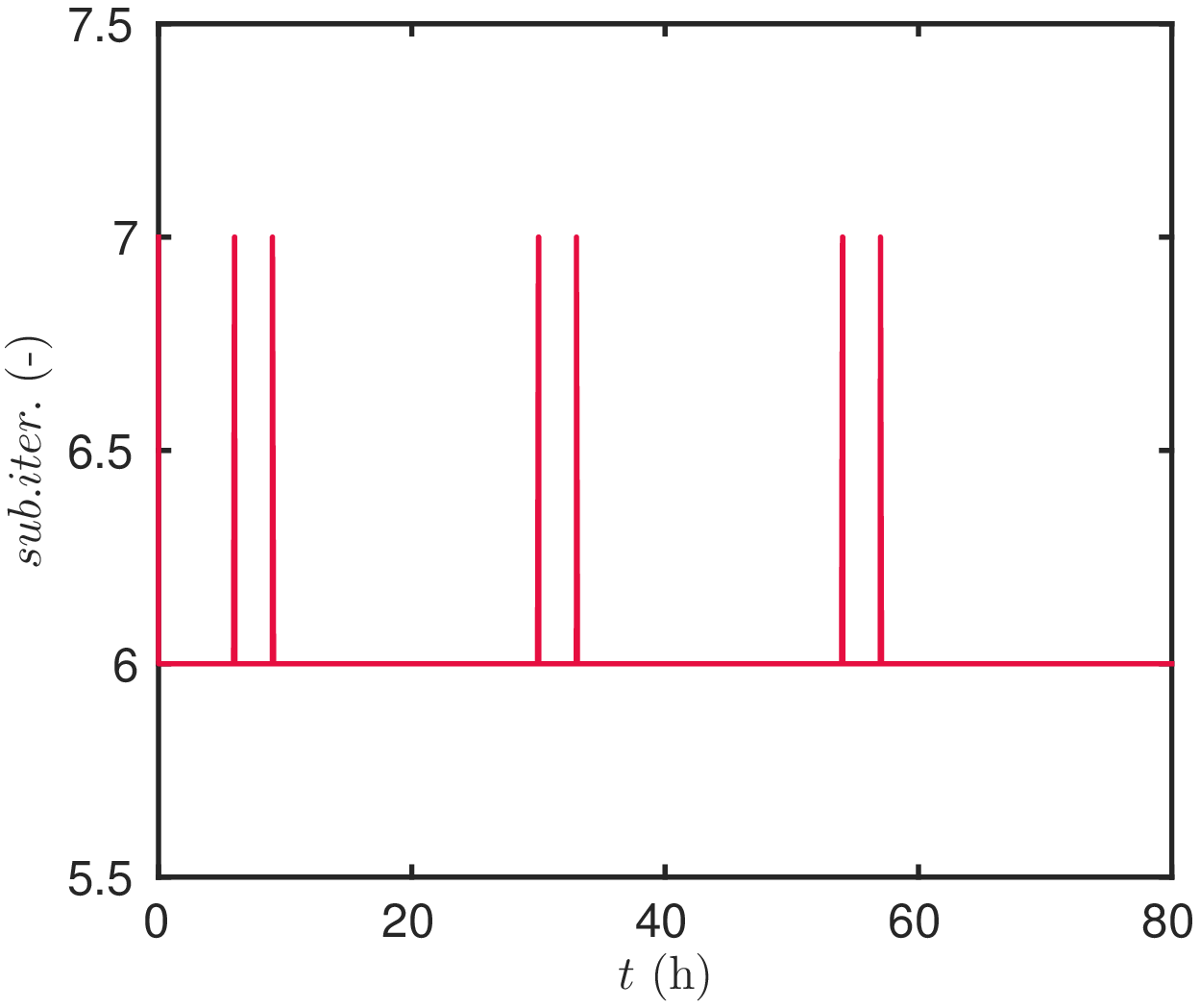}
  \caption{\small\em Sub-iterations required for the \Eu ~implicit scheme compute the solution of the whole-building energy model for the nonlinear case.}
  \label{fig_AN4:subiter_ft}
\end{center}
\end{figure}

By considering nonlinear wall material properties, the \Eu ~implicit scheme require more sub-iterations at each time step as illustrated in Figure~\ref{fig_AN4:subiter_ft}. In the previous linear case study, the algorithm required around $3$ iterations whereas for the present case, it needs at least $6$ to achieve the same accuracy. Therefore, the CPU time of the \Eu ~implicit scheme to compute the numerical solution increases, as reported in Table~\ref{tab:CPU_time}. As no sub-iterations are necessary for the \DF ~explicit scheme, the computation gain rises compared to the previous case study. It needs only $5 \%$ of the CPU time of the \Eu ~implicit scheme. These gains might considerably increase when considering highly nonlinear phenomena such as driving rain and  iteration with HVAC systems \cite{Barbosa2008}.

The purpose of this study was essentially to compare the numerical methods and the innovative \DF ~explicit scheme on a nonlinear case of whole building hygrothermal simulation. The computational time was measured for different simulation time horizon:

\begin{center}
\bigskip
\begin{tabular}{cccc}
  \hline \hline
  Simulation time & \DF & \Eu ~implicit  & Ratio (DF/Im)\\
  \hline \hline
  $80\ \mathsf{h}$ & $480\ \mathsf{s}$ & $8900\ \mathsf{s}$ & $5.4 \%$\\
  $160\ \mathsf{h}$ & $960\ \mathsf{s}$ & $17800\ \mathsf{s}$ & $5.4 \%$\\
  $800\ \mathsf{h}$ & $4750\ \mathsf{s}$ & $89000\ \mathsf{s}$ & $5.4 \%$\\
  \hline \hline
\end{tabular}
\bigskip
\end{center}

It can be noted that the increase of the CPU time is linear for both approaches. Therefore for a simulation time horizon of $1$ year, the CPU ratio between the \DF ~scheme and the \Eu ~implicit approach would be conserved. It is important to compare the ratio between the different approaches and not only the absolute value of CPU time with the classical building simulation programs. Indeed, the algorithms developed in this study are not optimised and the computing platforms may be different.

\begin{table}
\begin{center}
\setlength{\extrarowheight}{.3em}
\begin{tabular}[l]{@{} lccc}
\hline\hline
\multicolumn{4}{c}{\it Wall model with nonlinear wall material properties} \\
Numerical Scheme & CPU time ($\mathsf{s}$) & CPU time (\%) & Average number of iterations \\
\DF & 150 & 28 & -- \\
\Eu ~explicit & 164 & 31 & -- \\
\Eu ~implicit & 530 & 100 & 3 \\
\hline\hline
\multicolumn{4}{c}{\it One-zone model with linear wall material properties} \\
Numerical Scheme & CPU time ($\mathsf{s}$) & CPU time (\%) & Average number of iterations \\
\DF & 8.5 & 9 & -- \\
\Eu ~explicit & $\infty$ & -- & -- \\
\Eu ~implicit & 95 & 100 & 2.95 \\
\hline\hline
\multicolumn{4}{c}{\it Two-zones model with nonlinear wall material properties} \\
Numerical Scheme & CPU time ($\mathsf{s}$) & CPU time (\%) & Average number of iterations \\
\DF & 480 & 5.4 & -- \\
\Eu ~implicit & 8900 & 100 & 6 \\
\hline\hline
\end{tabular}
\bigskip
\caption{\small\em Computer run time required for the numerical schemes.}
\label{tab:CPU_time}
\end{center}
\end{table}


\section{Conclusion}

Implicit methods are extensively used in building simulation codes due to their stability conditions. However, they may require important extra computation when dealing with highly nonlinear problems, such as the combined heat and moisture transfer through porous building elements or when the whole-building is simulated, demanding perfect synchronism. In this way, this study aimed at exploring the use of the improved \DF ~explicit scheme. The advantages of this approach are studied here in whole-building hygrothermal simulations, with one- and two-zones building models.

The solution computed with the \DF ~scheme was compared to three other different solutions. Results have shown that the \DF ~scheme enables to compute an accurate solution, which is second-order accurate in time --- $\O\,(\Delta t^{\,2})\,$. It also enhanced that the \DF ~scheme can overcome some disadvantages of the \Eu ~implicit and explicit approaches.

First, contrarily to the \Eu ~explicit scheme, it is unconditionally stable, enabling to compute the solution without satisfying any stability condition, although, it does not mean that any value of $\Delta t$ can be used. When analysing the error of the \DF ~scheme as a function of the discretisation in the time domain --- $\Delta t\,$, three regions can be highlighted. The first one corresponds to small values of $\Delta t\,$, where the solution obtained with the \DF ~scheme reaches a constant error value, which is lower than the \Eu ~implicit one. The second region is where the error is proportional to $\O\,(\Delta t^{\,2})$ but higher than the \Eu ~implicit. The third and last region includes large values of $\Delta t \,$ where the \Eu ~implicit scheme seems to be more accurate. However, both schemes do not succeed in representing the physical phenomena. As mentioned in \citep{Gasparin2017}, the time step has to be chosen in accordance with the characteristic time of the problem. For the case studied, this condition is reached for a time discretisation up to $180-\mathsf{s}\,$, which is in accordance with the precision given in \cite{DosSantos2004} for whole building simulation.

Then, when dealing with nonlinearities, the scheme does not require any sub-iteration at each time step. The solution is directly computed, reducing consequently the computational cost of the algorithm. For the first case study, the \Eu ~implicit scheme required around $3$ sub-iterations, making it three times more costly than the \DF ~scheme. When coupling the wall and zone models using implicit schemes, a nonlinear system of equations has to be solved. By using the \DF ~explicit scheme, the system of equations becomes linear and no sub-iterations are necessary. Therefore, within the \DF ~approach, the algorithm requires only $9\%$ of the CPU time of the implicit approach. When considering the nonlinearities of the wall material properties and long-wave radiative heat transfer among the room surfaces, the computational savings rise to $95\%\,$.

These results encourage to apply the \DF ~approach in building simulation tools. The computational gains should increase with the number of rooms, walls, partitions, and furniture. The proposed method enable perfect synchronism for simulation and co-simulation, which can reduce even more the computation efforts.


\bigskip

\subsection*{Acknowledgments}
\addcontentsline{toc}{subsection}{Acknowledgments}

The authors acknowledge the \textsc{Brazilian} Agencies CAPES of the Ministry of Education and CNPQ of the Ministry of Science, Technology and Innovation, for the financial support.

\bigskip


\appendix
\section{The Dufort-Frankel scheme for weakly coupled equations}

This Appendix details the application of the \DF ~approach for two weakly coupled equations, as the ones describing the heat and moisture transfer in porous materials (Section~\ref{sec:HAM_transfer}). For this purpose, we consider a uniform discretisation as described in Section~\ref{sec:numerical_schemes} and the following linear coupled diffusion equation: 
\begin{subequations}\label{eq:coupled_eq}
\begin{align}
  \pd{v}{t} & \egal \FoM \, \pd{^{\,2}v}{x^{\,2}} \,, \label{eq:coupled_eq1} \\[3pt]
  \pd{u}{t} \plus \gamma \, \pd{v}{t} & \egal \FoTT \,  \pd{^{\,2}u}{x^{\,2}} \plus \gamma \, \FoTM  \,\pd{^{\,2}v}{x^{\,2}} \,. \label{eq:coupled_eq2}
\end{align}
\end{subequations}

The boundary conditions associated to Eq.~\eqref{eq:coupled_eq} are written as:
\begin{subequations}\label{eq:coupled_bc}
\begin{align}
  \pd{v}{x} & \egal \BiM \, \bigl(\, v - \vinf \,\bigr)  \,, \label{eq:coupled_bc1} \\[3pt]
  \FoTT \, \pd{u}{x} \plus \gamma \, \FoTM  \ \pd{v}{x} & \egal \BiTT \, \bigl(\, u- \uinf \,\bigr) \plus \BiTM \, \bigl(\, v - \vinf \,\bigr)  \,, \label{eq:coupled_bc2}
\end{align}
\end{subequations}

Considering Eq.~\eqref{eq:coupled_eq1} and the straightforward application of the \DF ~scheme described in Section~\ref{sec:DF_scheme}, we get: 
\begin{align}\label{eq:DF_eq1}
  & v_{\,j}^{\,n+1} \egal \frac{1 \moins \lambda}{1 \plus \lambda}\;v_{\,j}^{\,n-1}\ +\ \frac{\lambda}{1 \plus \lambda}\;\bigl(v_{\,j+1}^{\,n} \plus v_{\,j-1}^{\,n}\bigr) \,, &&  j\ =\ 1,\,\ldots,\,N\,, n\ \geqslant\ 1\,,
\end{align}
where:
\begin{align*}
  \lambda\ \eqdef\ 2\,\FoM\;\frac{\Delta t}{\Delta x^{\,2}} \,.
\end{align*}

For Eq.~\eqref{eq:coupled_eq2}, the numerical scheme is expressed as: 
\begin{align}\label{eq:DF_eq2}
  \frac{u_{\,j}^{\,n+1}\ -\ u_{\,j}^{\,n-1}}{2\,\Delta t} \plus \gamma \, \frac{v_{\,j}^{\,n+1}\ -\ v_{\,j}^{\,n-1}}{2\,\Delta t} & \egal  \FoTT\;\frac{u_{\,j-1}^{\,n}\ -\ \bigl(u_{\,j}^{\,n-1}\ +\ u_{\,j}^{\,n+1}\bigr)\ +\ u_{\,j+1}^{\,n}}{\Delta x^{\,2}} \nonumber \\
  & \plus \gamma \, \FoTM\;\frac{v_{\,j-1}^{\,n}\ -\ \bigl(v_{\,j}^{\,n-1}\ +\ v_{\,j}^{\,n+1}\bigr)\ +\ v_{\,j+1}^{\,n}}{\Delta x^{\,2}}\,, \nonumber \\
  &  j\ =\ 1,\,\ldots,\,N\,, n\ \geqslant\ 1\,,
\end{align}
With Eq.~\eqref{eq:DF_eq1} and rearranging the terms of Eq.~\eqref{eq:DF_eq2}, the numerical scheme is derived as follows for the field $u\,$:
\begin{align}\label{eq:DF_eq2b}
  u_{\,j}^{\,n+1} & \egal \frac{1 \moins \mu}{1 \plus \mu}\;u_{\,j}^{\,n-1}\ +\ \frac{\mu}{1 \plus \mu}\;\bigl(u_{\,j+1}^{\,n} \plus u_{\,j-1}^{\,n}\bigr) \nonumber \\
  & \plus \frac{\gamma \moins \beta}{1 \plus \mu}\;v_{\,j}^{\,n-1} \plus \frac{\beta}{1 \plus \mu}\;\bigl(v_{\,j+1}^{\,n} \plus v_{\,j-1}^{\,n}\bigr) \moins \frac{\gamma \plus \beta}{1 \plus \mu}\;v_{\,j}^{\,n+1}\
 \,, && n\ \geqslant\ 1\,,
\end{align}
where:
\begin{align*}
  \beta\ \eqdef\ 2\,\FoTM \, \gamma \;\frac{\Delta t}{\Delta x^{\,2}} && \text{and} && \mu \ \eqdef \ 2\,\FoTT \;\frac{\Delta t}{\Delta x^{\,2}} \,.
\end{align*}

For the boundary conditions, the application of the \DF ~scheme to Eq.~\eqref{eq:coupled_bc1} gives: 
\begin{align}\label{eq:bc_v}
  \frac{v_{\,2}^{\,n} \moins v_{\,0}^{\,n} }{2 \, \Delta x} \egal \BiM \, \Biggl(\, \frac{v_{\,1}^{\,n+1} \plus v_{\,1}^{\,n-1}}{2} \ \moins \vinf\, \Biggr)  \,.
\end{align}
Here, the node $j \egal 0$ is a ghost one located a distance $\Delta x$ from the node $j \egal 1$. From Eq.~\eqref{eq:bc_v}, we can deduce $v_{\,0}\,$:
\begin{align}\label{eq:ghost_node_v0}
  v_{\,0}^{\,n} \egal v_{\,2}^{\,n} \moins 2 \, \BiM \, \Delta x \, \biggl(\, \frac{v_{\,1}^{\,n+1} \plus v_{\,1}^{\,n-1}}{2} \moins \vinf \,\biggr) \,.
\end{align}
In a similar way for Eq.~\eqref{eq:coupled_bc2}, we get: 
\begin{align}\label{eq:ghost_node_u0}
  u_{\,0}^{\,n} \egal u_{\,2}^{\,n} \plus \frac{\FoTM}{\FoTT} \, \gamma \, \bigl(\, v_{\,2}^{\,n} \moins v_{\,0}^{\,n}  \,\bigr) \moins 2 \, \BiTT \, \Delta x \, \biggl(\, \frac{u_{\,1}^{\,n+1} \plus u_{\,1}^{\,n-1}}{2} \moins \uinf \,\biggr) \nonumber \\
  \moins 2 \, \BiTM \, \Delta x \, \biggl(\, \frac{v_{\,1}^{\,n+1} \plus v_{\,1}^{\,n-1}}{2} \moins \vinf \,\biggr) \,.
\end{align}
Using Eqs.~\eqref{eq:ghost_node_v0} and \eqref{eq:ghost_node_u0}, it is possible to compute Eqs~\eqref{eq:DF_eq1} and \eqref{eq:DF_eq2b} for $j \egal 1\,$. A similar approach is adopted for node $j \egal N\,$.


\section{Dimensionless numerical values}
\label{sec:appendix}

This Appendix provides the dimensionless values of the linear case study considered in this work.


\subsection{Whole-building model with linear material properties}

The dimensionless properties of the wall materials, considered in Section~\ref{sec:AN3}, are:
\begin{align*}
& \text{North Wall:}
&& \FoM \egalb 1.16 \dix{-1} \,, 
&& \FoTT \egalb 1.37 \dix{-1} \,,
&& \FoTM \egalb 3.76 \,,
&& \gamma \egalb 7.87 \dix{-3} \,, \\
& \text{South Wall:}
&& \FoM \egalb 8.9 \dix{-2} \,, 
&& \FoTT \egalb 2.36 \dix{-1} \,,
&& \FoTM \egalb 0.107 \,,
&& \gamma \egalb 3.07 \dix{-2} \,, \\
& \text{East and West Walls:}
&& \FoM \egalb 3.23 \dix{-2} \,, 
&& \FoTT \egalb 1.61 \dix{-1} \,,
&& \FoTM \egalb 1.08 \,,
&& \gamma \egalb 2.35 \dix{-2} \,,
\end{align*} 
and for all the walls:
\begin{align*}
\cMs \egal \kMs \egal \cTTs \egal \cTMs \egal \kTTs \egal \kTMs \egal 1 \,.
\end{align*} 
At $x \egal 0$, the \textsc{Biot} numbers are:
\begin{align*}
& \text{North Wall:}
&& \BiM \egal 3.39 \,, 
&& \BiTT \egal 1.7  \,,
&& \BiTM \egal 6.78 \dix{-1} \,, \\
& \text{South Wall:}
&& \BiM \egal 2.73  \,, 
&& \BiTT \egal 2.97 \,,
&& \BiTM \egal 9.48 \dix{-1} \,, \\
& \text{East and West Walls:}
&& \BiM \egal 7.31 \,, 
&& \BiTT \egal 3.1 \,,
&& \BiTM \egal 1.03 \,,
\end{align*} 
and at $x \egal 1$
\begin{align*}
& \text{North Wall:}
&& \BiM \egal 5.09 \dix{-1} \,, 
&& \BiTT \egal 2.72  \,,
&& \BiTM \egal 1.01 \dix{-1} \,, \\
& \text{South Wall:}
&& \BiM \egal 1.02 \dix{-1}  \,, 
&& \BiTT \egal 9.5 \dix{-1} \,,
&& \BiTM \egal 3.55 \dix{-2} \,, \\
& \text{East and West Walls:}
&& \BiM \egal 5.48 \dix{-1} \,, 
&& \BiTT \egal 2.06 \,,
&& \BiTM \egal 7.72 \dix{-2} \,,
\end{align*} 
For the zone model, the properties are $\caTTzero \egal 1$, $\caTTun \egal 1.43 \dix{-2}$ and the coupling parameter $\theta$:
\begin{align*}
& \text{North Wall:}
&& \thetaT \egal 3.45 \,, 
&& \thetaM \egal 111.1  \,, \\
& \text{South Wall:}
&& \thetaT \egal 9.89 \,, 
&& \thetaM \egal 55.3  \,, \\
& \text{East and West Walls:}
&& \thetaT \egal 2.27 \,, 
&& \thetaM \egal 5.18  \,, 
\end{align*}
The source term due to ventilation system equals: 
\begin{align*}
& q_{\,v,\,1}^{\,\star} \egal 7.1 \dix{-3} \,,
&& q_{\,v,\,2}^{\,\star} \egal 3.09 \dix{-2} \,,
&& g_{\,v}^{\,\star} \egal 0.5 \,.
\end{align*}
The source term due to moisture load equals: 
\begin{align*}
& q_{\,o}^{\,\star} \egal 3.8 \dix{-3} \plus \left. \begin{cases} 
6.1 \dix{-2} \,, & t \ \in \ \bigl[\,6 \,, 9 \,\bigl] \\
6.1 \dix{-2} \,, & t \ \in \ \bigl[\,6 \,, 9 \,\bigl] \plus 24  \\
6.1 \dix{-2} \,, & t \ \in \ \bigl[\,6 \,, 9 \,\bigl] \plus 48  \\
0  & \text{otherwise}
\end{cases} 
\ \right. 
&& \,,
\end{align*}
\begin{align*}
g_{\,o}^{\,\star} \egal 6.25 \dix{-2} \plus \left. \begin{cases} 
1\,, & t \ \in \ \bigl[\,6 \,, 9 \,\bigl] \\
1\,, & t \ \in \ \bigl[\,6 \,, 9 \,\bigl] \plus 24  \\
1\,, & t \ \in \ \bigl[\,6 \,, 9 \,\bigl] \plus 48  \\
0  & \text{otherwise}
\end{cases} 
\ \right. 
\end{align*}

The outside boundary conditions are expressed as: 
\begin{align*}
  & \uinf \egal 1 \moins 0.02 \ \sin \biggl(\, 2 \, \pi \, \frac{t}{24} \,\biggr)^{\,2} \,,
  && \vinf \egal 1 \plus 0.06 \ \sin \biggl(\, 2 \, \pi \, \frac{t}{24} \,\biggr) \,.
\end{align*} 
The final simulation time is fixed to $\tau^{\,\star} \egal 80\,$.


\section*{Nomenclature}

\begin{tabular*}{0.56\textwidth}{@{\extracolsep{\fill}} |@{} >{\scriptsize} c >{\scriptsize} l >{\scriptsize} l| }
\hline
\multicolumn{3}{|c|}{\emph{Latin letters}} \\
$A$ & surface & $\bigl[\mathsf{m^2}\bigr]$ \\
$c_{\,p,a}$ & air specific heat capacity & $\bigl[\unitfrac{J}{kg\cdot K}\bigr]$ \\
$c_{\,p,v}$ & vapour specific heat capacity & $\bigl[\unitfrac{J}{kg\cdot K}\bigr]$ \\
$\cz$ & material specific heat capacity & $\bigl[\unitfrac{J}{kg\cdot K}\bigr]$ \\
$\cw$ & liquid water specific heat capacity & $\bigl[\unitfrac{J}{kg\cdot K}\bigr]$ \\
$\cM$ & moisture storage coefficient & $\bigl[\unitfrac{s^2}{m^2}\bigr]$ \\
$\cTM$ & coupling storage coefficient & $\bigl[\unitfrac{W\cdot s^3}{kg\cdot m^2}\bigr]$ \\
$\cTT$ & energy storage coefficient & $\bigl[\unitfrac{W\cdot s}{m^3\cdot K}\bigr]$ \\
$\hM$ & convective vapour transfer coefficient & $\bigl[\unitfrac{s}{m}\bigr]$ \\
$\hT$ & convective heat transfer coefficient & $\bigl[\unitfrac{W}{m^2\cdot K}\bigr]$ \\
$g_{\,\inf}$ & liquid flow & $\bigl[\unitfrac{kg}{m^2\cdot s}\bigr]$ \\
$g$ & flow & $\bigl[\unitfrac{kg}{m^2\cdot s}\bigr]$ \\
$G$ & room moisture source term & $\bigl[\unitfrac{kg}{s}\bigr]$ \\
$\kl$ & liquid permeability & $\bigl[\mathsf{s}\bigr]$ \\
$\kM$ & moisture transf. coeff. under vap. press. grad. & $\bigl[\mathsf{s}\bigr]$ \\
\hline
\end{tabular*}

\begin{tabular*}{0.5\textwidth}{@{\extracolsep{\fill}} |@{} >{\scriptsize} c >{\scriptsize} l >{\scriptsize} l| }
\hline
& & \\
$\kTM$ & heat transf. coeff. under vap. press. grad.& $\bigl[\unitfrac{m^2}{s}\bigr]$ \\
$\kTT$ & heat transf. coeff. under temp. grad. & $\bigl[\unitfrac{W}{m\cdot K}\bigr]$ \\
$L$ & length & $\bigl[\mathsf{m}\bigr]$ \\
$\Lv$ & latent heat of evaporation & $\bigl[\unitfrac{J}{kg}\bigr]$ \\
$\Pa$ & air pressure & $\bigl[\mathsf{Pa}\bigr]$ \\
$\Pc$ & capillary pressure & $\bigl[\mathsf{Pa}\bigr]$ \\
$\Ps$ & saturation pressure & $\bigl[\mathsf{Pa}\bigr]$ \\
$\Pv$ & vapour pressure & $\bigl[\mathsf{Pa}\bigr]$ \\
$Q$ &  room heat source term & $\bigl[\unitfrac{W}{m^3} \bigr]$ \\
$q$ &  heat flux & $\bigl[\unitfrac{W}{m^2}\bigr]$ \\
$\Rv$ & water gas constant & $\bigl[\unitfrac{J}{kg\cdot K}\bigr]$\\
$T$ & temperature & $\bigl[\mathsf{K}\bigr]$ \\
$s$ &  view factor & $\bigl[\mathsf{-}\bigr]$ \\
$V$ & volume & $\bigl[\mathsf{m^3}\bigr]$ \\
$\wa$ & humidity ratio & $\bigl[\unitfrac{kg}{kg}\bigr]$ \\
\hline
\end{tabular*}

\vspace{0.3cm}

\begin{tabular*}{0.5\textwidth}{@{\extracolsep{\fill}} |@{} >{\scriptsize} c >{\scriptsize} l >{\scriptsize} l| }
\hline
\multicolumn{3}{|c|}{\emph{Greek letters}} \\
$\delta_{\,v}$ & permeability & $\bigl[\mathsf{s}\bigr]$ \\
$\lambda$ & thermal conductivity & $\bigl[ \unitfrac{W}{m \cdot K} \bigr]$ \\
$\sigma$ & \textsc{Stefan}--\textsc{Boltzmann} constant & $\bigl[-\bigr]$ \\
$\phi$ & relative humidity & $\bigl[-\bigr]$ \\
$\rho$ & specific mass & $\bigl[\unitfrac{kg}{m^3}\bigr]$ \\
\hline
\end{tabular*}

\vspace{0.3cm}

\begin{tabular*}{0.7\textwidth}{@{\extracolsep{\fill}} |@{} >{\scriptsize} c >{\scriptsize} l >{\scriptsize} l| }
\hline
\multicolumn{3}{|c|}{\emph{Parameters involved in the dimensionless representation}} \\
$\mathrm{Bi}$ & \textsc{Biot} number & $\bigl[-\bigr]$ \\
$c$, $\kappa$ & storage coefficient & $\bigl[-\bigr]$ \\
$\mathrm{Fo}$ & \textsc{Fourier} number & $\bigl[-\bigr]$ \\
$k$ & permeability coefficient & $\bigl[-\bigr]$ \\
$q$, $g$ & source terms & $\bigl[-\bigr]$ \\
$u$, $\ua$, $v$, $\va$ & field & $\bigl[-\bigr]$ \\
$\theta$ & weighted contribution & $\bigl[-\bigr]$ \\
$\nu$ & diffusion coefficient & $\bigl[-\bigr]$ \\
\hline
\end{tabular*}


\bigskip
\addcontentsline{toc}{section}{References}
\bibliographystyle{abbrv}
\bibliography{biblio}
\bigskip\bigskip

\end{document}